\documentclass[aps,prl,reprint,groupedaddress]{revtex4-1}
\usepackage{makecell}
\usepackage{amsfonts}
\usepackage{flafter}
\usepackage{float}
\usepackage{amssymb,graphics,graphicx,subfigure,rotating}

\usepackage{booktabs}
\usepackage{amsmath}
\allowdisplaybreaks[4]

\usepackage{latexsym}
\usepackage{amsmath}
\usepackage[colorlinks=True,linkcolor=blue,anchorcolor=blue,citecolor=blue,CJKbookmarks=true]{hyperref}
\usepackage{xcolor}
\usepackage{natbib}

\begin{document}
	
	\title{\textbf{
			Explicit Formulas for Estimating Trace of Reduced Density Matrix Powers via Single-Circuit Measurement Probabilities }}
	
	\author{Rui-Qi Zhang$^{1}$, Xiao-Qi Liu$^{1}$, Jing Wang$^{1}$}
	\author{Ming Li$^{1}$} \email{liming3737@163.com} 
	\author{Shu-Qian Shen$^{1}$}
	\author{Shao-Ming Fei$^{2}$}
	\affiliation{$^{1}$College of the Science, China University of Petroleum, 266580 Qingdao, China\\
		$^{2}$School of Mathematical Science, Capital Normal University, 100048, Beijing, China}

	\date{\today}
	
	\begin{abstract}
		In the fields of quantum mechanics and quantum information science, the traces of reduced density matrix powers play a crucial role in the study of quantum systems and have numerous important applications. In this paper, we propose a universal framework to simultaneously estimate the traces of the $2$nd to the $n$th powers of a reduced density matrix using a single quantum circuit with $n$ copies of the quantum state. Specifically, our approach leverages the controlled SWAP test and establishes explicit formulas connecting measurement probabilities to these traces. We further develop two algorithms: a purely quantum method and a hybrid quantum-classical approach combining Newton-Girard iteration. Rigorous analysis via Hoeffding inequality demonstrates the method's efficiency, requiring only $M=O\left(\frac{1}{\epsilon^2}\log(\frac{n}{\delta})\right)$ measurements to achieve precision $\epsilon$ with confidence $1-\delta$. Additionally, we explore various applications including the estimation of nonlinear functions and the representation of entanglement measures. Numerical simulations are conducted for two maximally entangled states, the GHZ state and the W state, to validate the proposed method.
	\end{abstract}
	\maketitle
	
	\section{I. Introduction}
	Quantum entanglement \cite{Horodecki,Vedral} is a crucial resource in quantum information science, essential for tasks like communication \cite{Barrett,Gisin,Bose}, cryptography \cite{Ribordy,Bernstein}, and computation \cite{DiVincenzo,Briegel}. Significant research has been dedicated to detecting and measuring entanglement due to its fundamental importance in these areas.\par
	The traces of reduced density matrix powers are essential in numerous applications such as characterizing the entanglement measure and detecting quantum entanglement. For example, researchers have introduced several entanglement measures based on the traces of reduced density matrix powers for both bipartite and multipartite quantum systems (see \cite{Horodecki1,Guhne,Szalay}) to quantify entanglement. These measures include concurrence \cite{Wootters,Fan}, entanglement of formation \cite{Wootters1,Kai}, relative entropy of entanglement \cite{Henderson,Moitra}, concentratable entanglements (CEs) \cite{Beckey,Cullen,Schatzki,Beckey1}, parametrized entanglement measures \cite{Li,Zhou}, and informationally complete entanglement measures (ICEMs) \cite{Jin}.
	Particularly, the CEs introduced by Beckey et al. \cite{Beckey}  can be realized via the parallel controlled SWAP tests \cite{Foulds,Zhang}. Subsequently, Zhang et al. \cite{Zhang} implemented the entropy criterion and provided a representation of entanglement measures based on the measured probabilities of the controlled SWAP test.
	More recently, Jin et al. \cite{Jin} proposed an  informationally complete entanglement measure that can be estimated on a quantum computer by leveraging the complete information of the reduced density matrix. These advancements highlight the increasing synergy between quantum information theory and the practical capabilities of quantum computing, paving the way for more refined and accurate entanglement quantification methods.\par
	Estimating the traces of reduced density matrix powers is exceptionally important, and several quantum algorithms have been proposed to compute ${\rm tr}(\rho_A^k)$ \cite{Ekert,Johri,Suba,Yirka,Shin,Huang,YangR} ($A$ denotes the subsystem of quantum state $\rho$). 
	However, these algorithms are primarily restricted to calculating the ${\rm tr}(\rho_A^k)$ for a specific number of copies of the circuit with $k$ copy quantum states. Computing other powers, like ${\rm tr}(\rho_A^{k'})$ with $k'\ne k$,  demands the design of a new quantum circuit customized for it. This limitation becomes acute in tasks requiring multiple $k$ such as entropy series or ICEM entanglement measures \cite{Jin}. A key question arises: Can a single quantum circuit estimate all $\left\lbrace {\rm tr}(\rho_A^k)\right\rbrace _{k=1}^n$ simultaneously?  \par
	 We answer affirmatively by introducing a unified circuit design inspired by controlled SWAP tests. In this paper,
	 we address the question of whether it is possible to simultaneously estimate arbitrary power traces of reduced density matrix using a single quantum circuit along with explicit and concrete representations. We propose two independent  algorithms, a pure quantum algorithm and a hybrid quantum algorithm that combines with a classical iterative method, for estimating the trace of powers of the reduced density matrix. This means that no matter which power trace of the reduced density matrix needs to be computed, it can be done in the same circuit framework without the need to redesign the circuit for different powers, which greatly improves the flexibility and efficiency of the computation and saves the time and resource cost of designing several different circuits. \par Based on the proposed quantum circuit inspired by \cite{Jin}, we rectify  inappropriate estimate of the traces of reduced density matrix powers presented in \cite{Jin}. We further develop two algorithms: a purely quantum method and a hybrid quantum-classical approach combining Newton-Girard iteration. Rigorous analysis via Hoeffding inequality demonstrates the method's efficiency, requiring only $M=O\left(\frac{1}{\epsilon^2}\log(\frac{n}{\delta})\right)$ measurements to achieve precision $\epsilon$ with confidence $1-\delta$. Additionally, we investigate the applications of these traces of powers, such as estimating the nonlinear functions and expressing entanglement measures via the measurement probabilities of quantum circuits. Finally for the two maximally entangled states, the GHZ state and the W state, we perform numerical simulations of quantum circuits on \emph{Qiskit}.\par 
  \section{II. Estimation of traces of reduced density matrix powers}
  In this section, we elaborate on our method for estimating the power trace of reduced density matrices through two subsections. The first subsection presents the mathematical foundations of circuit design, laying out the theoretical framework for constructing the quantum circuits. The second subsection details the implementation scheme of our quantum algorithm.\par 
  \subsection{A. Mathematical Foundations}
  To present the subsequent main result, we first introduce the following notations. 
	Let $\mathcal{S}=\left\lbrace 1,2,\dots,n-1\right\rbrace $ denote the set of labels corresponding to each qubit within the control qubits. Additionally, let $\mathcal{P}(\mathcal{S})=\left\lbrace   \varnothing,\left\lbrace 1\right\rbrace ,\cdots ,\left\lbrace n-1\right\rbrace ,\cdots,\left\lbrace      1,2,3,\cdots,n-1\right\rbrace \right\rbrace$ represent the power set of $\mathcal{S}$ comprising $2^{n-1}$ elements. Moreover, let $\emph{\textbf{z}}=z_{1} z_{2}\cdots z_{n-1}$ be a bitstring associated with the $n-1$ control qubits and $z_{i}\in\left\lbrace 0,1\right\rbrace ,i=1,2,\cdots,n-1$. Define $\mathcal{S}^\emph{\textbf{z}}=\left\lbrace i\in\mathcal{S}|z_{i}=1\right\rbrace $ and $c^\emph{\textbf{z}}_{ x}=\left|\mathcal{S}^\emph{\textbf{z}}\cap x\right|$ for any $x\in\mathcal{P}(\mathcal{S})$. The expression `` $c^\emph{\textbf{z}}_{x}\ is \ even$'' indicates that the number of qubits in $x$ which are in the state $|1\rangle$ is even. \par 
  To complete the proof of our main theorems, we state several useful lemmas. For the sake of readability, in the main text we focus on our main results and put both the proofs of lemmas in Appendix.\par
	\begin{figure}[!h]
		\includegraphics[width=1\linewidth]{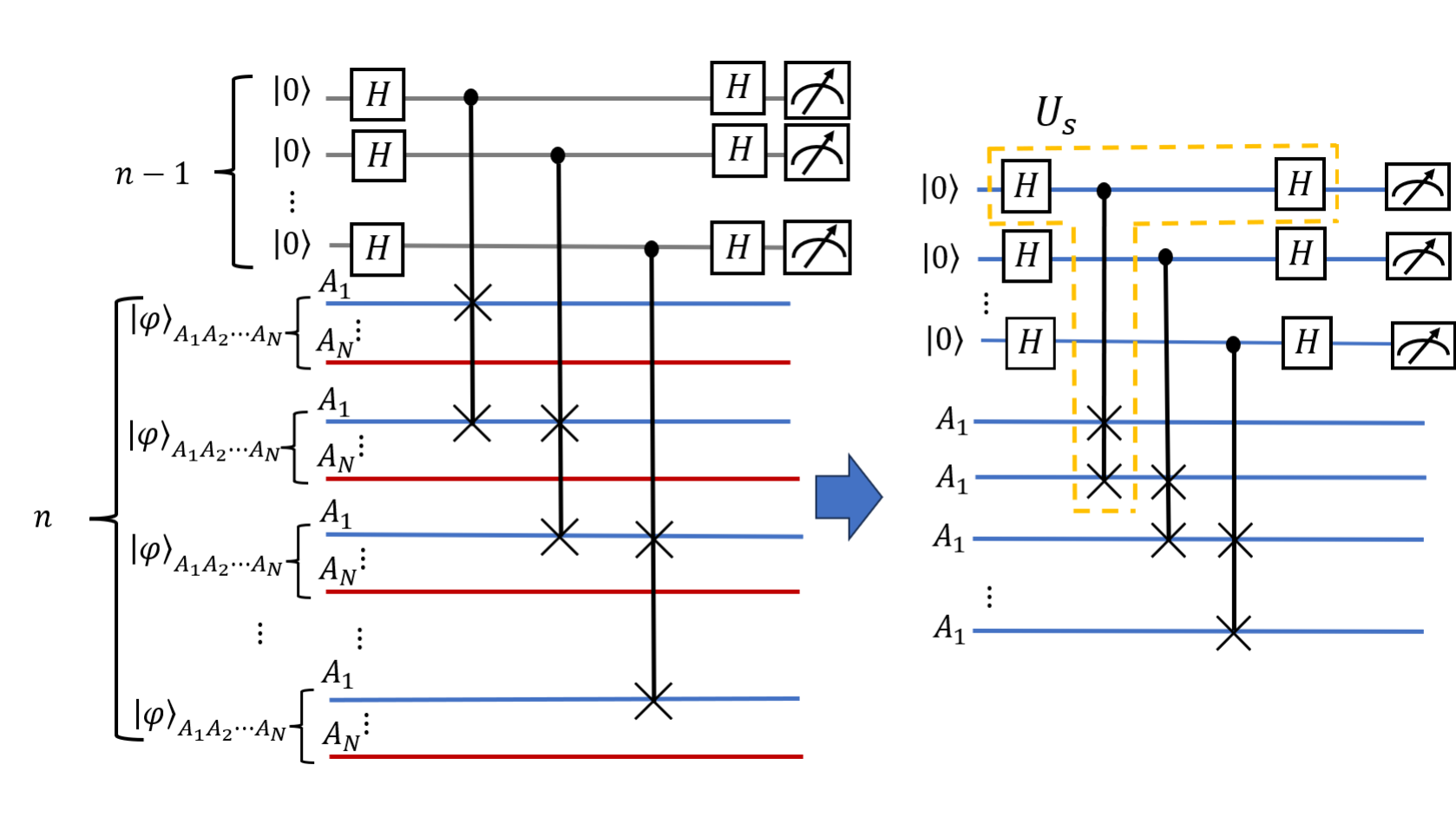}
		\caption{\textbf{The controlled SWAP test of $n$ copy quantum states about $A_1$ system.} Given $n-1$ control qubits and $n$ copy quantum states, each of the $n-1$ control qubits alternately controls the swap of the $n$ copy quantum states. $U_s$ represents the unitary matrix corresponding to the controlled SWAP test between two copy quantum states.
		}\label{fig1}
	\end{figure}
	
	First in order to give expressions for arbitrary traces of reduced density matrix powers, we consider an equivalent representation of the evolution of quantum circuits based on the kraus operator. Let $S$ denote the SWAP operator acting on a bipartite Hilbert
	space $\mathcal{H}\otimes\mathcal{H}$ with orthonormal product basis $B=\left\lbrace |i\rangle|i'\rangle\right\rbrace$, and the SWAP operator $S$ satisfies $$S|i\rangle|i'\rangle=|i'\rangle|i\rangle  \ \ \forall|i\rangle|i'\rangle \in B.$$
	It is important to note that $S$ is both unitary and Hermitian.
	As illustrated in Figure \ref{fig1}, for two Hilbert spaces $\mathcal{H} \otimes \mathcal{H}$ and $|\psi\rangle\in \mathcal{H}$, after passing through the unitary transformation $U_s$, the state $|0\rangle|\Psi\rangle =|0\rangle |\psi\rangle |\psi\rangle$ is transformed into the output state
	\begin{align*}
		U_s|0\rangle|\Psi\rangle = \frac{1}{2}\sum_{z=0}^{1}|z\rangle[I+(-1)^zS]|\Psi\rangle,
	\end{align*}
	where $I$ denotes the identity operator  on $\mathcal{H}\otimes \mathcal{H}$. If the control system is traced out, the evolution of the input state is given by a channel with kraus operators $K_{+}=\frac{1}{2}(I+S)$ and $K_{-}= \frac{1}{2}(I-S)$.
	For a $n$-partite system $\otimes_{i=1}^{n}\mathcal{H}_i,$ the evolution of the input states is characterized by a channel with kraus operators \cite{Jin}: 
	$$ K_\emph{\textbf{z}}=\frac{1}{2^{n-1}}\prod_{k=1}^{n-1} [I+(-1)^{z_k}S_k],$$
	where  $\emph{\textbf{z}}=z_1z_2\cdots z_{n-1}$ is the string corresponding to the measurement probability of the $n-1$ control qubits, $S_k\ (k=1,2,\cdots,n-1)$ denotes  $k$th SWAP test performing on the
	$k$th and $(k+1)$th states of Hilbert spaces $\mathcal{H}_k$ and $\mathcal{H}_{k+1}$, i.e.,
	$S_k|i_k\rangle|i_{k+1}\rangle=|i_{k+1}\rangle|i_k\rangle.$ Given that the control register is an $n$-qubit system, a Von Neumann measurement of the control register can result in a bitstring $\emph{\textbf{z}}\in \left\lbrace 0, 1\right\rbrace ^{n-1}$. The probability of obtaining bitstring $\emph{\textbf{z}}$ is then given by
	\begin{align*}
		&p(\emph{\textbf{z}})=\\&\frac{1}{2^{2n-2}}\langle\psi|(\prod_{k=1}^{n-1} [I+(-1)^{z_k}S_k])^{\dagger}
		 \prod_{k=1}^{n-1} [I+(-1)^{z_k}S_k]|\psi\rangle.
	\end{align*}
	
	\textbf{Remark 1.} We can readily observe that the multiplication of neighboring SWAP operators, $S_i$ and $S_{i+1}$, does not adhere to the commutative property due to the overlapping regions of the spaces on which these operators act, resulting in $K_\emph{\textbf{z}}^{\dagger}K_\emph{\textbf{z}} \neq K_\emph{\textbf{z}}$.  This explains the inappropriateness of the probability representation in \cite{Jin} since they wrongly assume that $K_\emph{\textbf{z}}^{\dagger}K_\emph{\textbf{z}} = K_\emph{\textbf{z}}$. Although the evolution of the kraus operator representation of quantum circuits is also considered in \cite{Beckey}, it is actually different from that of \cite{Jin}. In \cite{Beckey}, the condition $K_\emph{\textbf{z}}^{\dagger} K_\emph{\textbf{z}}= K_\emph{\textbf{z}}$ is satisfied by rearranging the $S_i,S_j$ operators, as neighboring SWAP operators $S_i,S_{i+1}$ act on distinct subsystems.\par 
	In order to assist us better understand the following lemmas, we will introduce a graphical method to describe the trace estimation.\par 
	We introduce the following graphical representation \cite{Liu}. A matrix is depicted as a box with open legs, which correspond to uncontracted
	indices. Those left-oriented legs and the right-oriented legs stand for row indices and column indices, respectively. Vectors and
	scalars are represented as boxes with legs of same orientation and boxes with no leg, respectively. The $k$ copy quantum states are represented by $k$ boxes that have $k$ pairs of legs. Each pair of legs represents the row and column indices of the corresponding part of the system. Connecting legs represents index contraction, such as matrix multiplication and taking the trace.\par 
	\begin{figure}[!h]
		\centering
		\includegraphics[width=1\linewidth]{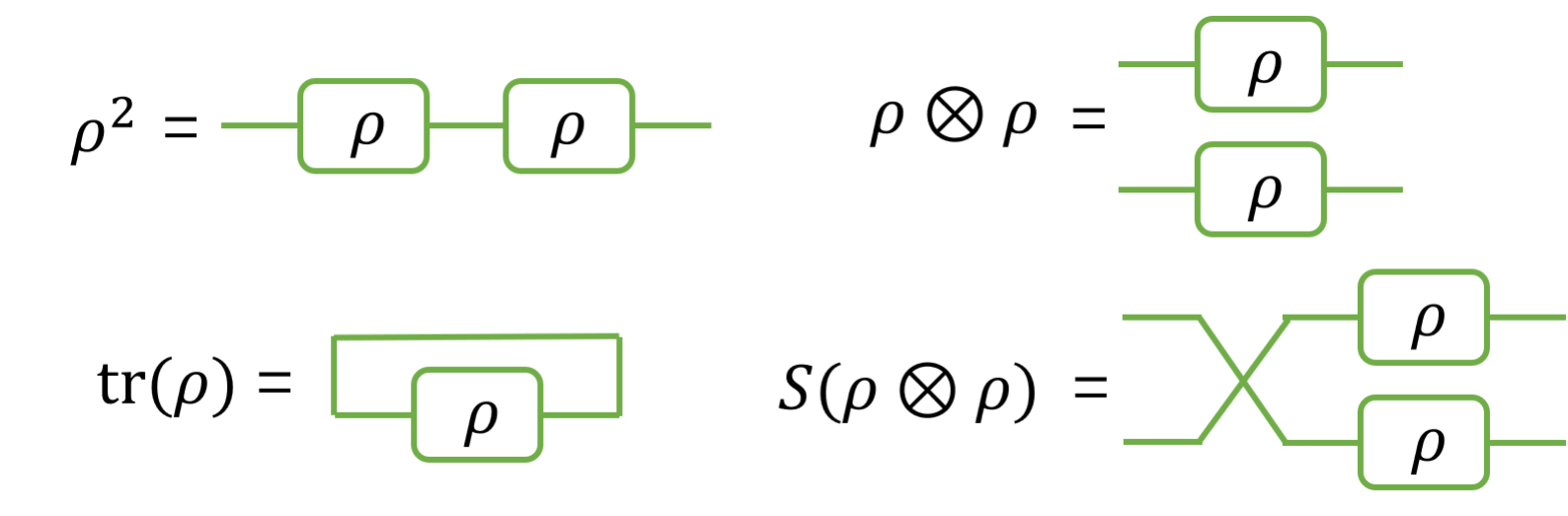}
		\caption{\small \textbf{Graphical representations of matrix multiplication, matrix tensor product, the trace of matrix, and the action of SWAP operator on matrix tensor product.}}
		\label{fig2}
	\end{figure}
	
	To generalize the result to the quantum circuit with multiple copy quantum states, we require the establishment of several key lemmas.
	\par 
	\textbf{Lemma 1.} \emph{ Consider $k$-partite quantum systems $\otimes_{i=1}^{k}\mathcal{H}_i$. Let $\rho_i\in \mathcal{H}_i \ (i=1,2,\cdots ,k)$ be $k$ identical quantum states, i.e., $\rho_i=\rho $ and let $S_i:\mathcal{H}_i\otimes\mathcal{H}_{i+1}\longrightarrow \mathcal{H}_i\otimes\mathcal{H}_{i+1}$  denote the SWAP operator acting on $\mathcal{H}_i$ and $\mathcal{H}_{i+1}$.
		Then we can obtain
		$${\rm tr}(S_{i_1}S_{i_2}\cdots S_{i_{k-1}}\otimes_{i=1}^{k}\rho_i)={\rm tr}( \prod_{i=1}^{k}\rho_i)={\rm tr}( \rho^k), $$
		where $i_1,i_2,\cdots,i_{k-1}$ is a rearrangement of $ 1,2,\cdots,k-1$.}\\\par
\begin{figure}[!ht]
		\centering
		\includegraphics[width=1\linewidth]{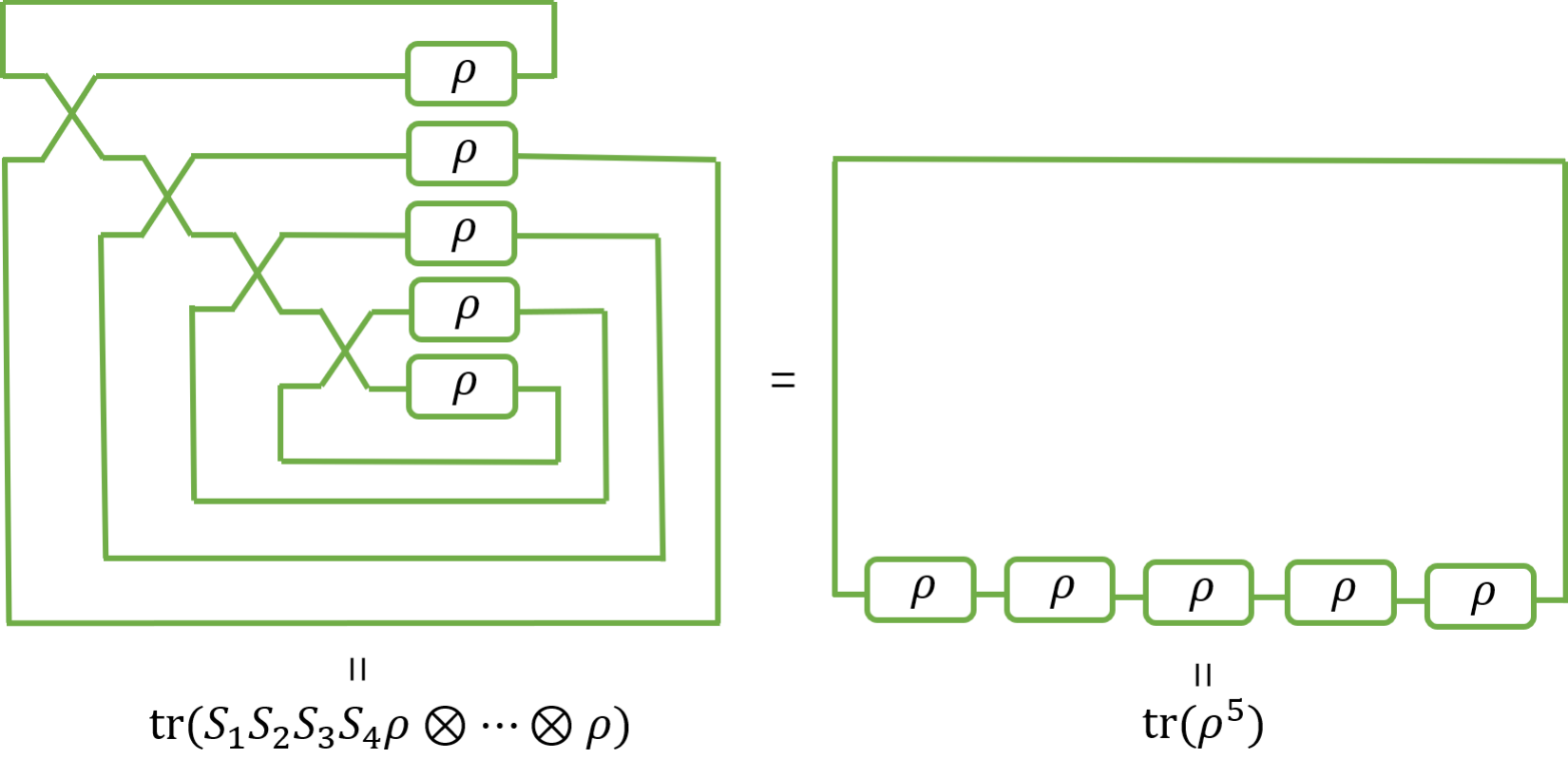}
		\caption{\small \textbf{The graphical representation corresponding to the trace of the operator $S_1S_2S_3S_4$ acting on the five copy quantum states.}}
		\label{fig3}
	\end{figure}

\begin{figure}[!ht]
		\centering
		\includegraphics[width=1\linewidth]{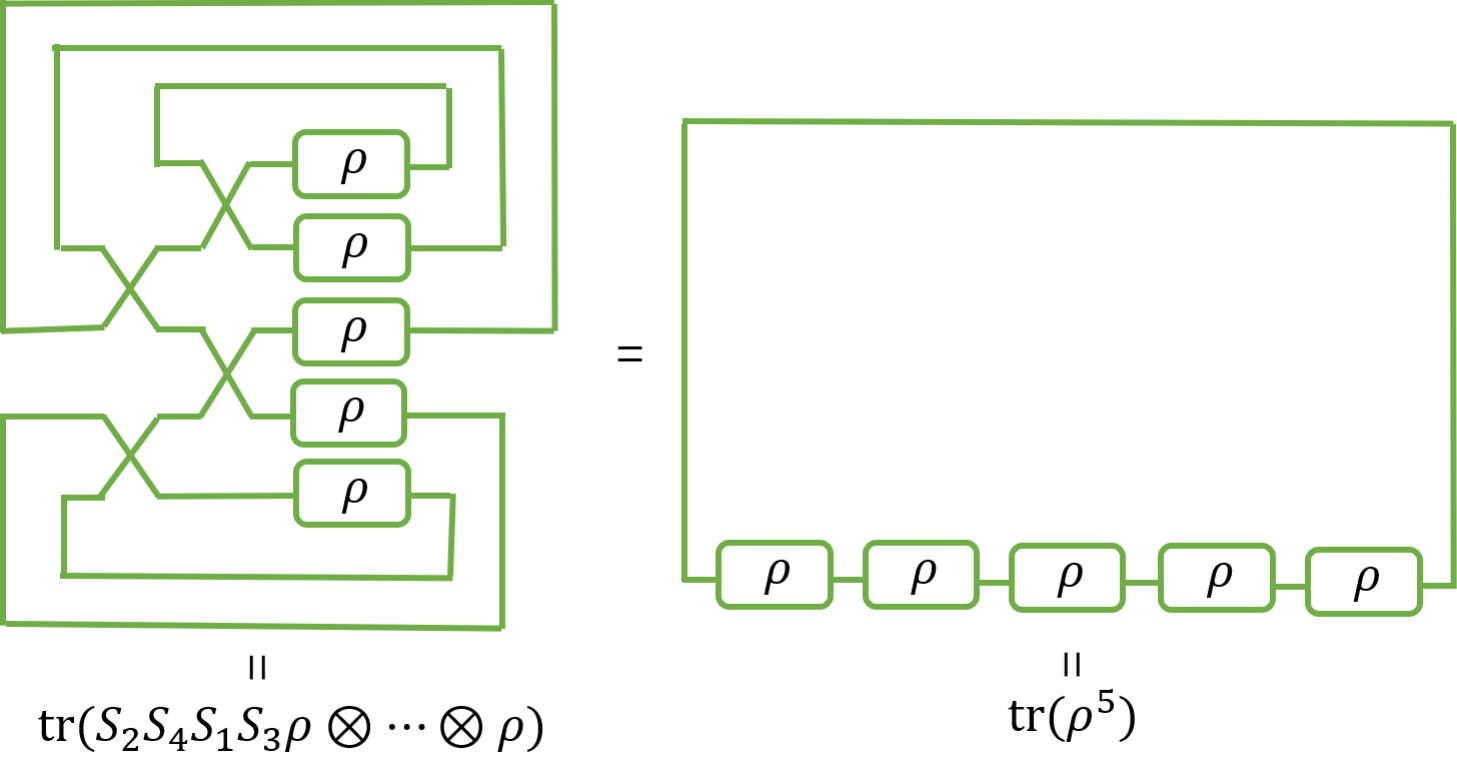}
		\caption{\small \textbf{ The graphical representation corresponding to the trace of the operator $S_2S_4S_1S_3$ acting on the five copy quantum states.}}
		\label{fig4}
	\end{figure}

 
	To make it easier to understand Lemma 1, we will use graphical representations to illustrate and verify it. Considering $k=5$, we present the graphical representations of the traces taken by the operators $S_1S_2S_3S_4$ and $S_2S_4S_1S_3$ act on five copy quantum states respectively, as shown in Figure \ref{fig3} and Figure \ref{fig4}.\par 

	\textbf{Lemma 2.} \emph{ Consider $k$-partite quantum systems $\otimes_{i=1}^{k}\mathcal{H}_i$. Let $S_i$ and $S_j$ be two nonadjacent SWAP operators $(i,j\in\left\lbrace 1,2,\cdots,k-1\right\rbrace ,i<j,$ and $i\ne j-1)$ and $\rho_i=\rho \in \mathcal{H}_i, (i=1,2,\cdots,k)$. We can obtain 
		\begin{align*}
			{\rm tr}(S_iS_j\otimes_{i=1}^{k}\rho_i)=({\rm tr}\rho^2)^2.
		\end{align*}}\\\par
\begin{figure}[!ht]
		\centering
		\includegraphics[width=1\linewidth]{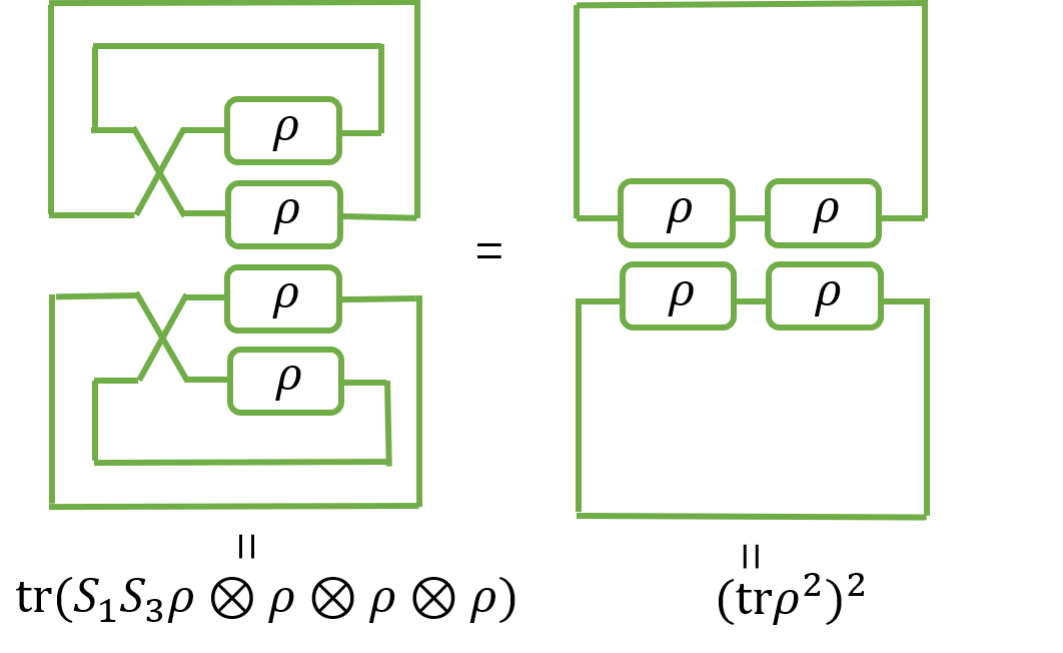}
		\caption{\small \textbf{ The graphical representation corresponding to the trace of the operator $S_1S_3$ acting on the four copy quantum states.}}
		\label{fig5}
	\end{figure}

	Lemma 2 shows that in a $k$-partite quantum system, due to the action of two non-adjacent SWAP operators on the corresponding copy states, the cross term $({\rm tr}\rho^2)^2$ will appear. As shown in Figure \ref{fig5}, taking the four copy quantum states as an example, our graphical representation verifies the establishment of Lemma 2.\par 
	
	The result of Lemma 2 explains the reason why cross terms appear in the measurement probabilities of quantum circuits with four copy quantum states.\par  
	\textbf{Lemma 3.} \emph{For $k$-partite quantum systems $\otimes_{i=1}^{k}\mathcal{H}_i$, if there are $k$ copies of the same quantum state $\rho=\rho_i\in\mathcal{H}_{i},(i=1,2,\cdots,k)$, then we have 
		\begin{align*}
			&{\rm tr}(\bar{S}_1 \bar{S}_2\cdots\bar{S}_t\otimes_{i=1}^{k}\rho_i)={\rm tr}\rho^{s_1+1}{\rm tr}\rho^{s_2+1}\cdots{\rm tr}\rho^{s_t+1},
		\end{align*}
		where $\bar{S}_1=S_{j_1^{(1)}}S_{j_2^{(1)}}\cdots S_{j_{s_1}^{(1)}}$, $\bar{S}_2=S_{j_1^{(2)}}S_{j_2^{(2)}}\cdots S_{j_{s_2}^{(2)}}$, $\cdots$, $\bar{S}_t=S_{j_1^{(t)}}S_{j_2^{(t)}}\cdots S_{j_{s_t}^{(t)}}$ are a series of operators composed of the products of adjacent SWAP operators, and the SWAP operators within operators $\bar{S}_m \ (m=1,2,\cdots, t)$ and those within the other operators  $\bar{S}_n \ (n=1,2,\cdots, t, \ n\ne m)$ have no intersection and are not adjacent. And $s_1+s_2+\cdots+s_t\le k-t$, where $s_m$ denotes the number of all SWAP operators in $\bar{S}_m$.}\\

	Lemma 3 gives an explanation for the appearance of Various types of cross terms in the representation of the probability results with the number of quantum copies of quantum circuit increases. \par 
	In the following we define several operators $\hat{S}_i \ (i=1,2,3,4)$ which are composed of several SWAP operators $S_i$. It should be noted that when an $S_i$ appears twice within $\hat{S}$, the effects of action of the repeated SWAP operator will cancel out. \\\par 
	\textbf{Lemma 4.} \emph{Consider $k$-partite quantum systems  $\otimes_{i=1}^{k}\mathcal{H}_i$, let 
		\begin{align*}
			&\hat{S}_1=S_{i_1}S_{i_2}\cdots S_{i_m}S_{i_1},\\
			&\hat{S}_2=S_{i_1}\cdots S_{i_{s-1}}S_{i_1}S_{i_{s}}\cdots S_{i_m},\notag\\
			&\hat{S}_3=S_{i_1}\cdots S_{i_{s-1}}S_{i_{s}}\cdots S_{i_m}S_{i_{s}},\\
			&\hat{S}_4=S_{i_1}\cdots S_{i_{s-1}}S_{i_{s}}\cdots S_{i_{t-1}}S_{i_{s}}S_{i_{t}}\cdots S_{i_m}
		\end{align*}
		be operators composed of some sequentially and continuously adjacent SWAP operators $S_j$ and only one SWAP operator appears twice are repeated in these operators and $m\le k-1$. The indices $i_1, i_2, \cdots, i_m$ denote a sequence of consecutive integers corresponding to adjacent SWAP operators, satisfying $1 \leq i_1 < i_2 < \cdots < i_m \leq k-1$.  If there are $k$ copies of the same quantum state $\rho=\rho_i\in\mathcal{H}_{i},(i=1,2,\cdots,k)$, then 
		\begin{align*}
			&{\rm tr}(\hat{S}_1\otimes_{i=1}^{k}\rho_i)= {\rm tr}(\hat{S}_2\otimes_{i=1}^{k}\rho_i)={\rm tr}(\hat{S}_3\otimes_{i=1}^{k}\rho_i)\\
			=&{\rm tr}(\hat{S}_4\otimes_{i=1}^{k}\rho_i)={\rm tr}\rho^m.
		\end{align*}
	}\\

	Lemma 4 illustrates that in a specific quantum system, if there are certain operators composed of sequentially and continuously adjacent SWAP operators where some SWAP operators appear twice, the trace of these operators acting on the tensor product of the same quantum state is equal to the trace of the corresponding power of that quantum state. These repeated SWAP operators do not affect the representation of the trace of the power of the quantum state. In other words, the actions of these repeated SWAP operators cancel each other out, which is equivalent to having no effect. \par
	In fact, for the term in the expansion that satisfies $K_\emph{\textbf{z}}^{\dagger }K_\emph{\textbf{z}}$, even if a certain sequential structure is not satisfied,  we can still obtain the result that the action of the repeated SWAP operator will cancel each other out.
	For example, for $k$-partite quantum systems  $\otimes_{i=1}^{k}\mathcal{H}_i$  $(k>5)$, there are $k$ copies of the same quantum state $\rho=\rho_i\in\mathcal{H}_{i},(i=1,2,\cdots,k)$. If we consider the following operators:
	\begin{align*}
		&\bar{S}_1=S_1S_3S_4S_5S_2S_1,\\
		&\bar{S}_2=S_2S_3S_5S_4S_2S_1,\\
		&\bar{S}_3=S_1S_2S_4S_5S_3S_2,\\
		&\bar{S}_4=S_2S_3S_4S_5S_4S_1,
	\end{align*}
	we can verify the following conclusion by means of spectral decomposition in Lemma 4: 
	\begin{align*}
		&{\rm tr} (\bar{S}_1\otimes_{i=1}^{k}\rho_i)={\rm tr}(\bar{S}_2\otimes_{i=1}^{k}\rho_i)\\
		=&{\rm tr} (\bar{S}_3\otimes_{i=1}^{k}\rho_i)={\rm tr} (\bar{S}_4\otimes_{i=1}^{k}\rho_i)={\rm tr}\rho^5.
	\end{align*}
	After performing spectral decomposition on the copy quantum states and then successively applying the SWAP operators in $\bar{S}_i \ (i=1,2,3,4)$, we will notice that the term ${\rm tr}(\sum_{j_x}a_{j_x}|j_x\rangle\langle j_x|)$ in the proof will appear in a certain subsystem after the swapping, which is due to the presence of the repeated operator and thus has no effect.\par 
	In particular, if there are some terms in the expansion for $K_\emph{\textbf{z}}^{\dagger }K_\emph{\textbf{z}}$ that contain nonadjacent SWAP operators, i.e., as follows:
	\begin{align*}
		&\tilde{S}_1=S_1S_3S_5S_4S_1,\\
		&\tilde{S}_2=S_2S_4S_5S_2S_1,\\
		&\tilde{S}_3=S_1S_2S_4S_5S_2,\\
		&\tilde{S}_4=S_1S_3S_5S_3S_2,\\
	\end{align*}
	we have 
	\begin{align*}
		&{\rm tr}(\tilde{S}_1\otimes_{i=1}^{k}\rho_i)={\rm tr}\rho^4\\
		&{\rm tr}(\tilde{S}_2\otimes_{i=1}^{k}\rho_i)={\rm tr}\rho^2{\rm tr}\rho^3\\
		&{\rm tr}(\tilde{S}_3\otimes_{i=1}^{k}\rho_i)={\rm tr}\rho^2{\rm tr}\rho^3\\
		&{\rm tr}(\tilde{S}_4\otimes_{i=1}^{k}\rho_i)={\rm tr}\rho^2{\rm tr}\rho^3.
	\end{align*}
	Here we will explain the reasons for the differences in the representations of the above results. 
	Although there are repeated SWAP operators in these operators $\tilde{S}_i$, their overall structures are different. When they act on the tensor product of quantum states, this difference cause differences in the exchange and combination patterns of quantum states. Consequently, the expressions for the final trace turn out to be different. However, essentially, they all demonstrate the property that the effects of repeated SWAP operators can cancel each other out.
	\par
\begin{figure}[htbp]
		\centering
		\includegraphics[width=1\linewidth]{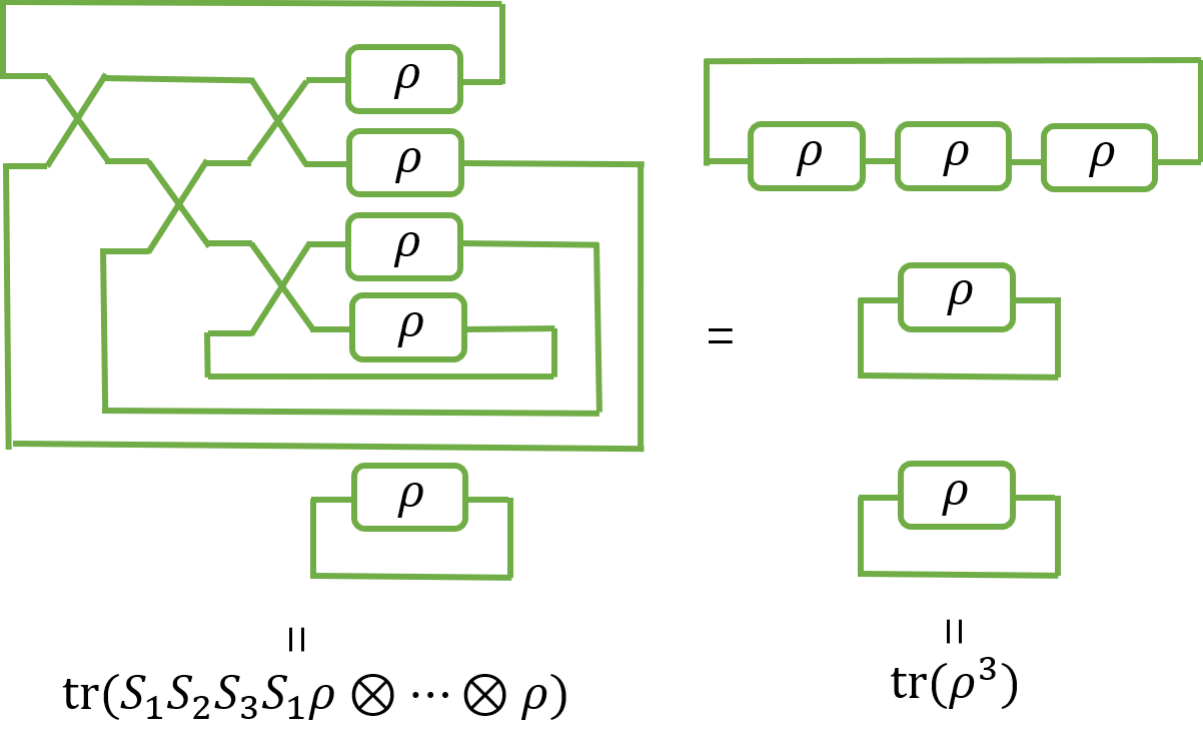}
		\caption{\small \textbf{ The graphical representation corresponding to the trace of the operator $S_1S_2S_3S_1$ acting on the five copy quantum states.}}
		\label{fig6}
	\end{figure}

\begin{figure}[htbp]
		\centering
		\includegraphics[width=1\linewidth]{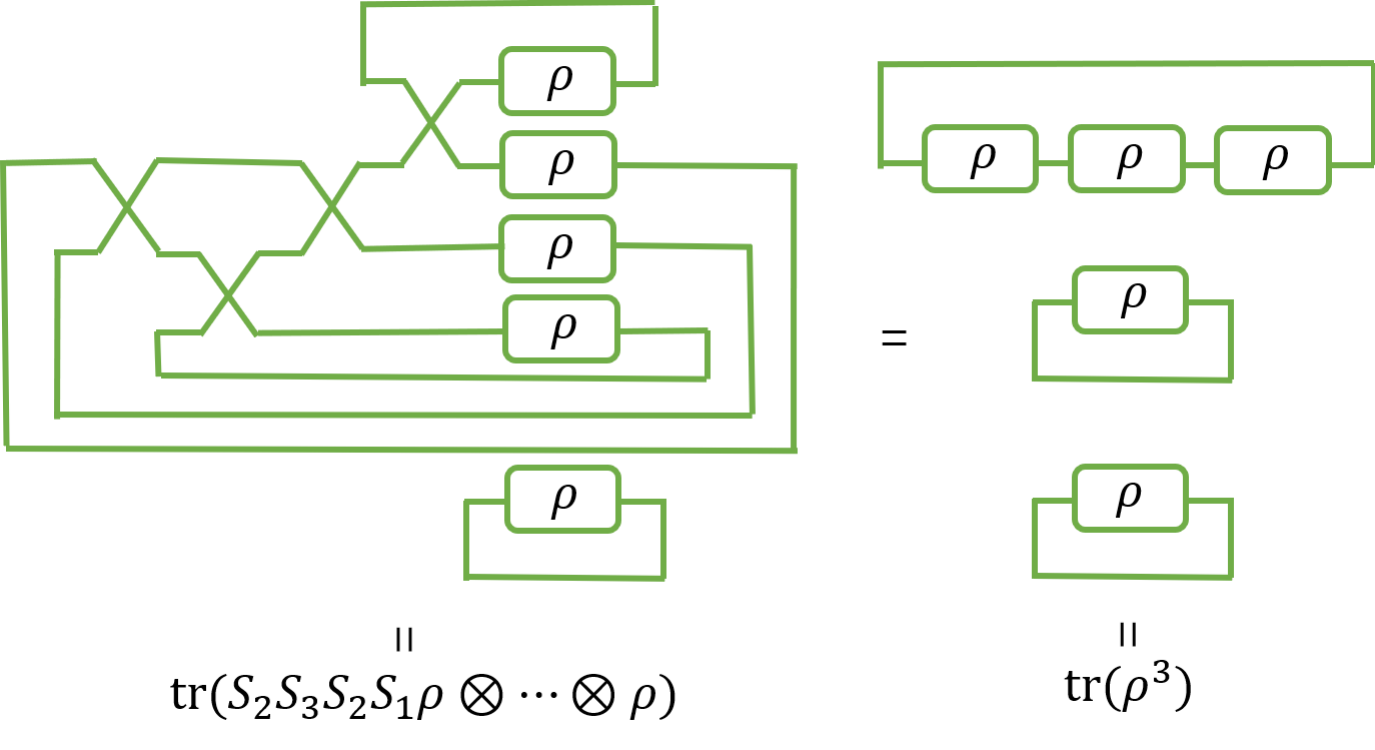}
		\caption{\small \textbf{ The graphical representation corresponding to the trace of the operator $S_2S_3S_2S_1$ acting on the five copy quantum states.}}
		\label{fig7}
	\end{figure}

\begin{figure}[htbp]
		\centering
		\includegraphics[width=1\linewidth]{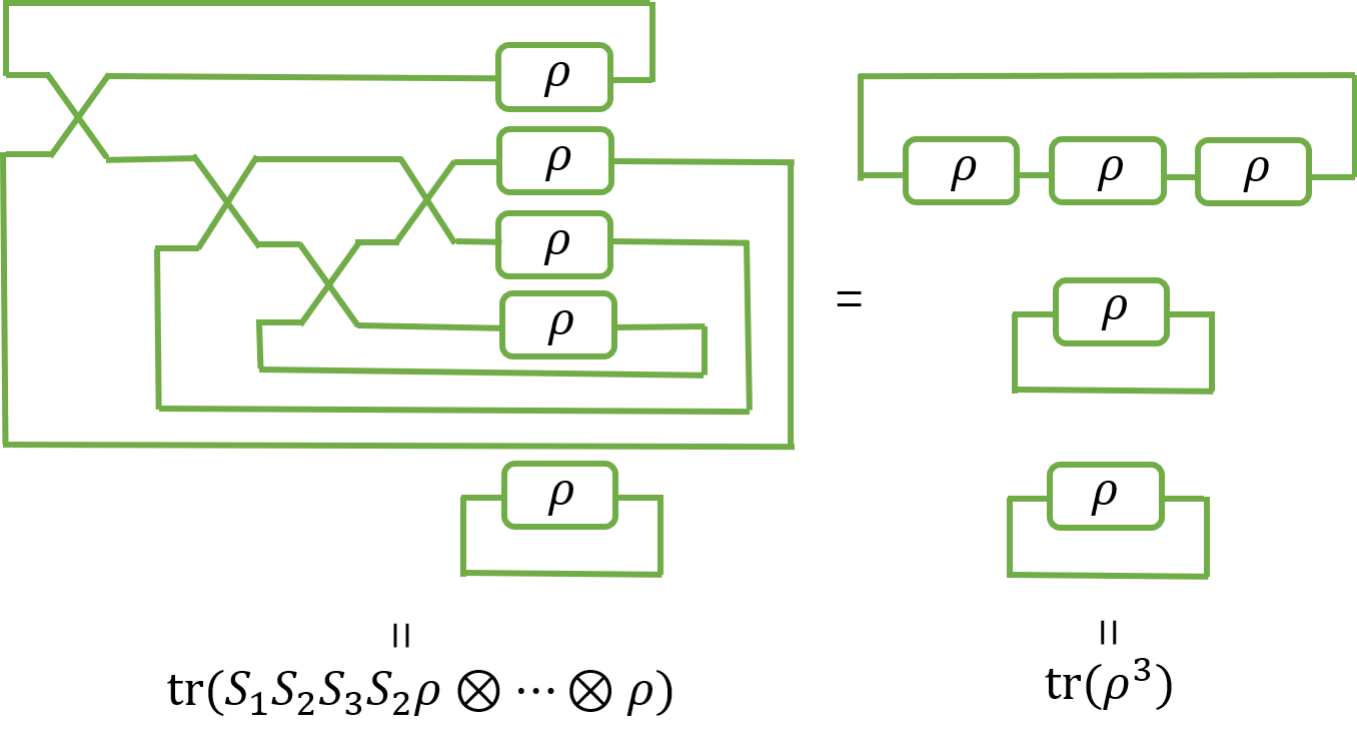}
		\caption{\small \textbf{ The graphical representation corresponding to the trace of the operator $S_1S_2S_3S_2$ acting on the five copy quantum states.}}
		\label{fig8}
	\end{figure}

\begin{figure}[htbp]
		\centering
		\includegraphics[width=1\linewidth]{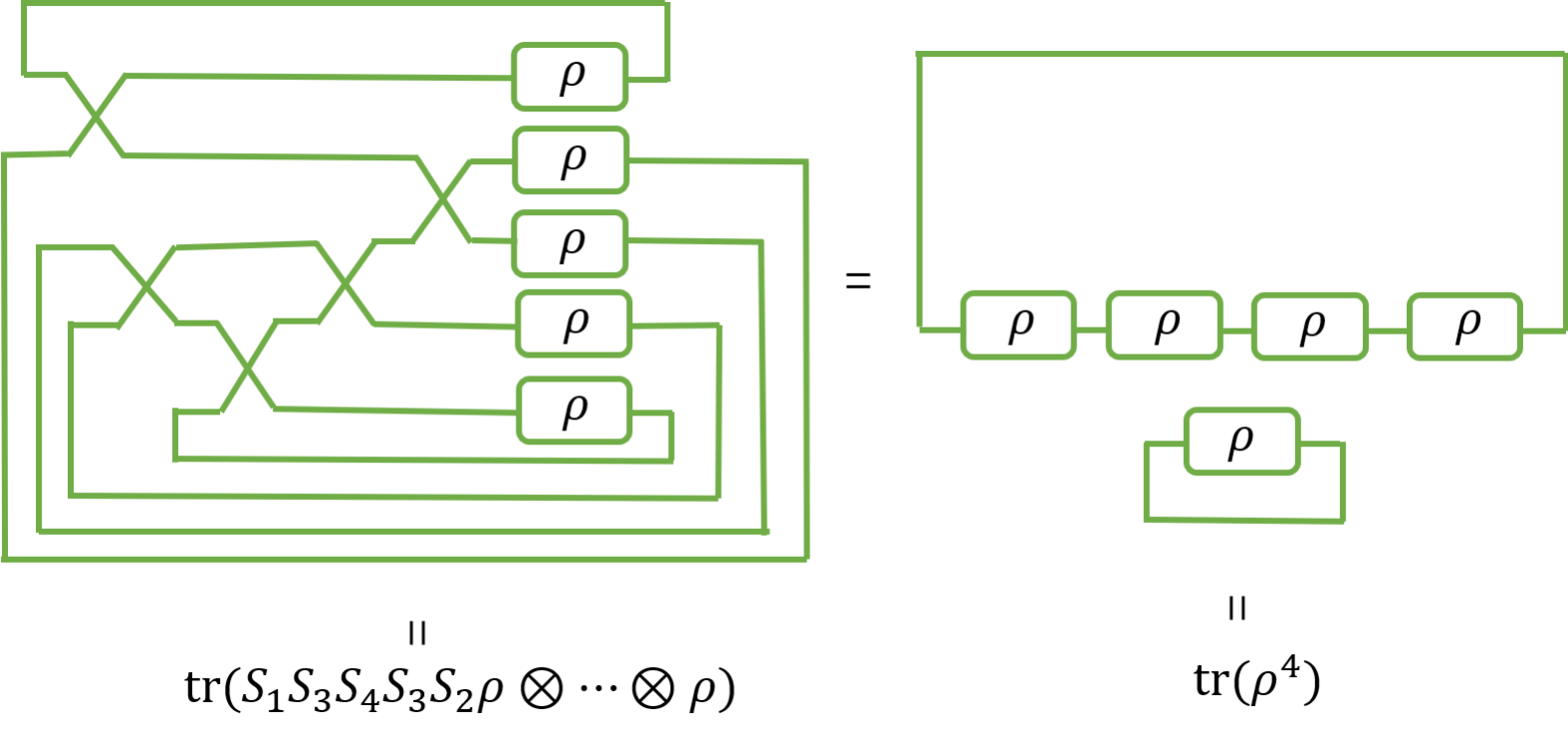}
		\caption{\small \textbf{ The graphical representation corresponding to the trace of the operator $S_1S_3S_4S_3S_2$ acting on the five copy quantum states.}}
		\label{fig9}
	\end{figure}

 
	To visually verify the validity of Lemma 4 and to explain the above phenomenon, we consider five copy quantum states and present four cases corresponding to Figure \ref{fig6}, Figure \ref{fig7}, Figure \ref{fig8} and Figure \ref{fig9}.

	In order to give the circuit evolution process, the following lemma is necessary.\\\par 
	
	\textbf{Lemma 5.} \emph{ Any $N$-partite quantum state $\rho \in \mathbb{C}^{d_1}\otimes\mathbb{C}^{d_2}\otimes\cdots\otimes\mathbb{C}^{d_N}$ has the following expression as
		$$	\rho=\sum_{i}p_{i}|\phi_{i}^1\rangle\langle\phi_{i}^1|\otimes|\phi_{i}^2\rangle\langle\phi_{i}^2|\otimes\cdots\otimes|\phi_{i}^N\rangle\langle\phi_{i}^N|,  $$
		where $ p_{i}\in\mathbb{R}$,  $\sum_{i}p_{i}=1,$ and $|\phi_{i}^j\rangle \ (j=1,2,\cdots,n)$ denote the pure states in the $jth$ subsystem.	}\\
	
	Lemma 5 demonstrates that any multipartite state can be expressed as a linear combination of pure separable states, with the possibility of negative coefficients in the combination. \par
	Based on these facts, we can turn to stating our main result in next subsection. \\\par 
	
	\begin{figure}[htbp]
		\centering
		\includegraphics[width=1\linewidth]{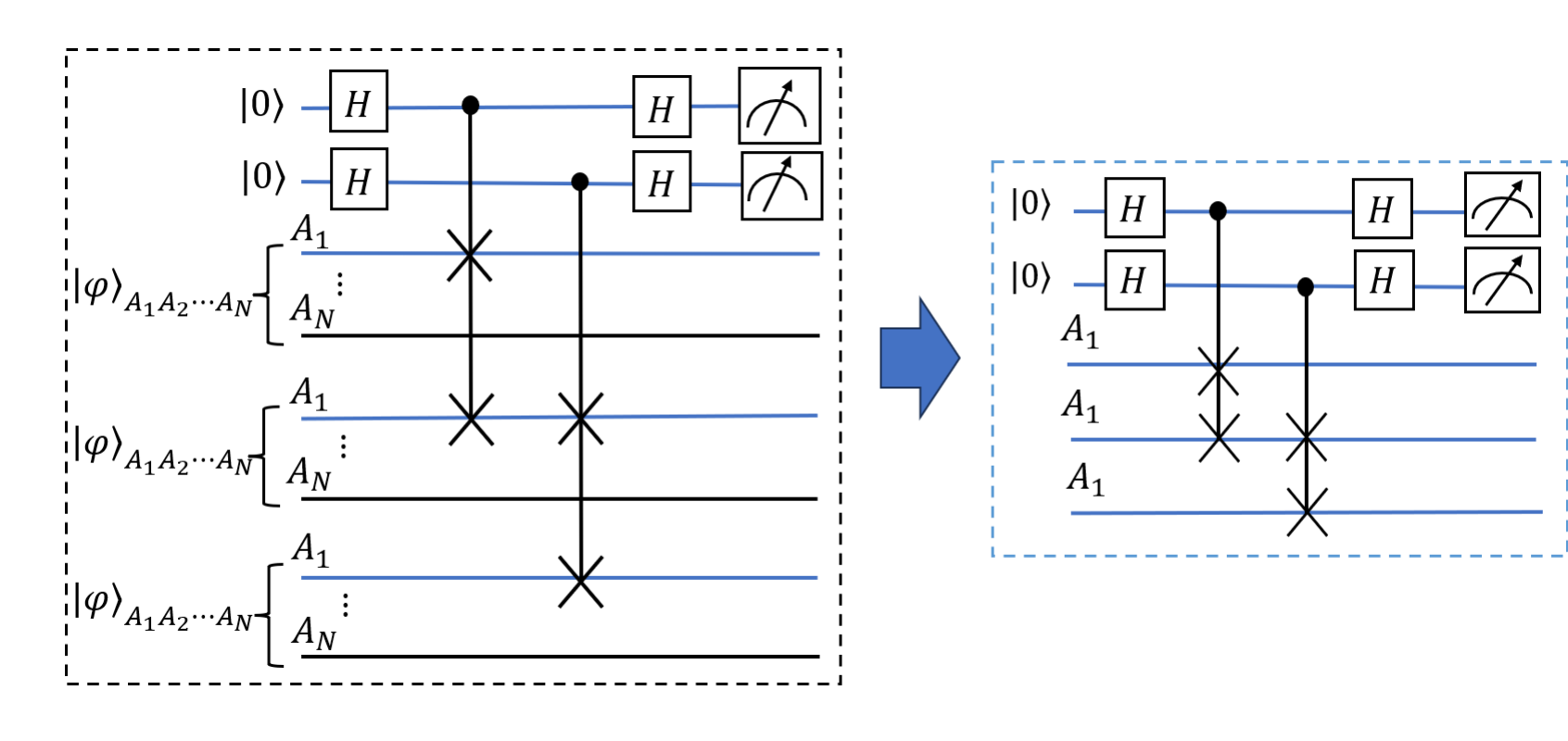}
		\caption{\textbf{The controlled SWAP test for the $A_1$ subsystem with three copies of the quantum state.} Given three copies of a quantum state and two control qubits, apply the first controlled SWAP gate operation to the first two copies of the quantum state, which is controlled by the first control qubit. Then, apply the second controlled SWAP gate operation to the last two copies of the quantum state, which is controlled by the second control qubit. }
		\label{fig10}
	\end{figure}
	\subsection{B. Quantum Algorithm Implementation} 
	In this section, we mainly present the representations of the traces of reduced density matrix powers. For the detailed analysis of circuit implementation and the derivation of results, please refer to the lemma and theorem proofs in the Appendix.\par 
	In order to give expressions for arbitrary traces of reduced density matrix powers, we consider an equivalent representation of the evolution of quantum circuits based on the kraus operator. This approach allows us to characterize the results to quantum circuits with multiple copies of quantum states, facilitating the computation of ${\rm tr}\rho_{A_1}^k$ for $k=2,3,\cdots,n$ in the controlled SWAP test for $n$ copy states.  \par 
    In the following, we will first consider the quantum circuits for three copy quantum states and four copy quantum states. According to the Lemmas 1-4, for quantum circuits with three copy states and four copy states, we can give a representation of ${\rm tr}\rho_{A_1}^k,(k=2,3,4)$ based on the kraus operators.\\\par
	
	\textbf{Theorem 1.} \emph{Let $\rho\in \mathbb{C}^{d_{A_1}}\otimes\mathbb{C}^{d_{A_2}}\otimes\cdots\otimes\mathbb{C}^{d_{A_N}}$ be an arbitrary $N$-partite quantum state, and let $S_1$ and $S_2$ denote the SWAP operators acting on the $A_1$ subspace of the first two copies and the last two copies in the quantum circuit of Figure \ref{fig10}, respectively. Passing through the quantum circuit in Figure \ref{fig10}, we obtain the following expressions: 
		\begin{align*}
			&{\rm tr}\rho_{A_1}^2=2\sum_{c^\textbf{z}_{1}\ is\ even } p(\textbf{z})-1=2\left( p(00)+p(01)\right) -1,\notag\\
			&{\rm tr}\rho_{A_1}^3=2\sum_{c^\textbf{z}_{1,2}\ is\ even } p(\textbf{z})-1=2\left( p(00)+p(11)\right) -1,
		\end{align*}
		where  $\rho_{A_1}$ is the reduced density matrix with $A_1$ subsystem.}
	\\
	
	\begin{figure}[!t]
		\centering
		\includegraphics[width=0.8\linewidth]{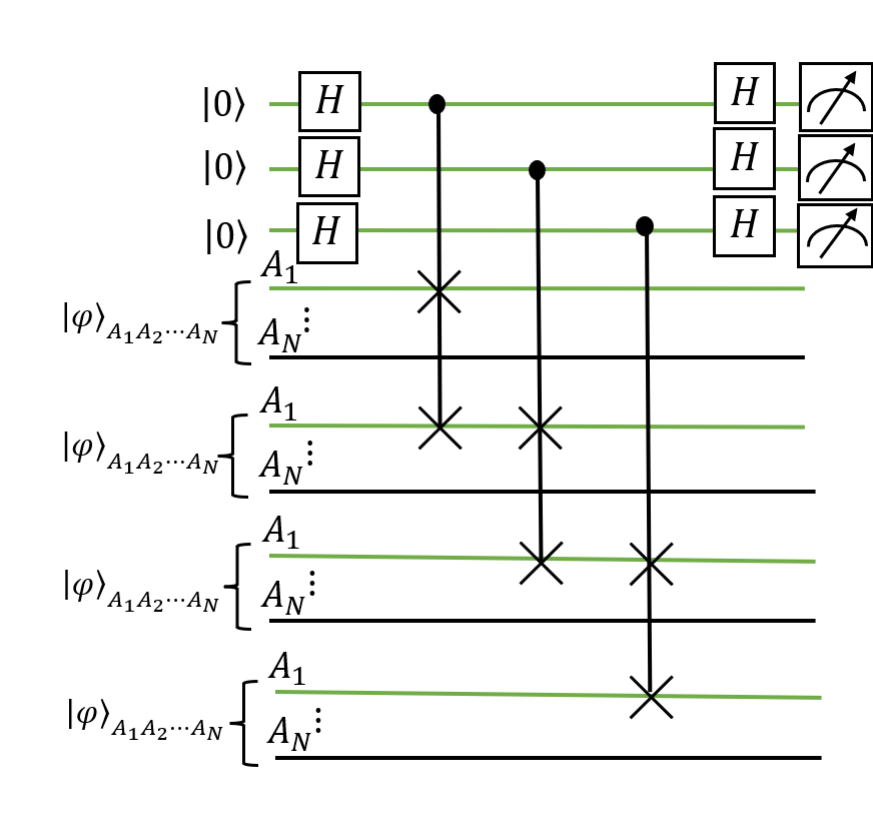}
		\caption{\textbf{The controlled SWAP test of four copy quantum states about $A_1$ system.} Given three control qubits and four copy quantum states, each of the three control qubits alternately controls the swap of the four copy quantum states. }
		\label{fig11}
	\end{figure}
	\textbf{Theorem 2.} \emph{Let $\rho$ in $ \mathbb{C}^{d_{A_1}}\otimes\mathbb{C}^{d_{A_2}}\otimes\cdots\otimes\mathbb{C}^{d_{A_N}}$ be any $N$-partite quantum state, and $S_i\ (i=1,2,3)$  denotes the SWAP operator acting on the subsystem $A_1$ in the space where the $i${\rm th} and $(i+1)${\rm th} copy states are located in Figure \ref{fig11}. Passing the quantum circuit in Figure \ref{fig11}, we can also get 
		\begin{align*}
			&{\rm tr}\rho_{A_1}^2=2\sum_{c^\textbf{z}_{1}\ is\ even } p(\textbf{z})-1\\=&2[ p(000)+p(001)+p(010)+p(011)] -1,\notag\\
			&{\rm tr}\rho_{A_1}^3=2\sum_{c^\textbf{z}_{1,2}\ is\ even } p(\textbf{z})-1\\=&2[ p(000)+p(001)+p(110)+p(111)] -1,\notag\\
			&{\rm tr}\rho_{A_1}^4=2\sum_{c^\textbf{z}_{1,2,3}\ is\ even } p(\textbf{z})-1\\=&2[ p(000)+p(011)+p(101)+p(110)] -1.
		\end{align*}
	}
	
	\textbf{Remark 2.} The proof of the Theorem 2 clearly show that cross terms like $({\rm tr}\rho_{A_1}^2)^2$ emerge in the measurement probabilities. This is exactly owing to the action of non-adjacent SWAP operators, as stated in Lemma 2, which results in the emergence of cross terms. In fact, as the number of quantum copy states increases, when a circuit contains multiple quantum copy states, Lemma 3 indicates that cross terms in the form of ${\rm tr}\rho^{s_1+1}{\rm tr}\rho^{s_2+1}\cdots{\rm tr}\rho^{s_t+1}$ may appear in the measurement probabilities. Since in \cite{Jin}, they omitted these cross terms when considering the measurement probabilities, it further demonstrates the inappropriateness of the results they gave when characterizing the entanglement measure ICEMs through quantum circuits.
	\par 
	And the proof of above theorems reveal that only the terms where $``c^\emph{\textbf{z}}_{1,\cdots,s}\ is\ even$'' $(s=1,2,\cdots,n-1)$ are relevant and affect the outcome, while the terms in other cases, i.e., those not satisfying the $``c^\emph{\textbf{z}}_{1,\cdots,s}\ is\ even$'' $(s=1,2,\cdots,n-1)$ condition, cancel out during the summation process. This cancellation occurs due to the parity characteristics of these terms. For instance, certain terms contain summation forms of $z_i$ that differ from those in the terms meeting the condition. When performing the summation operation, the combined value of these terms amounts to 0, thereby having no substantial contribution to the final result. Consequently, certain terms in the expressions for $K_\emph{\textbf{z}}^{\dagger}K_\emph{\textbf{z}}$ and $p(\emph{\textbf{z}})$ are irrelevant to the proof's process and result. Given that the expansion of $K_\emph{\textbf{z}}^{\dagger}K_\emph{\textbf{z}}$ can reach up to $2^{2n-2}$ terms, it is impractical and unnecessarily complex to specify an exact general expression for this expansion. For the sake of simplicity, we will focus on the essential terms, and for those that are insignificant, we will denote them with the symbol $\mathcal{S}_{waste}$, as they do not influence the proof's process and outcome. Next we consider circuit of $n$ copy quantum states via kraus operator representation. And we can obtain the representation of the trace of the reduced density matrix powers for the controlled SWAP test with $n$ copy states.  \\\par
	\textbf{Theorem 3.} \emph{For any $N$-partite quantum state $\rho_{A_1A_2\cdots A_N}$, consider $n$ copy states, and passing through the controlled SWAP test with $n-1$ control qubits as shown in Figure \ref{fig1}, for any $k=2,3,\cdots,n,$  we obtain 
		\begin{align*}
			{\rm tr}\rho_{A_1}^k=2\sum_{c^\textbf{z}_{1,2\cdots,k-1}\ is \ even }p(\textbf{z})-1.
	\end{align*}}
\\\par 

  This result shows that through a single quantum circuit containing $n$ copy quantum states, the estimates of the traces of reduced density matrix powers ranging from the second power to the $n$th power can be obtained simultaneously.\par 
  We remark that an expression for the traces of reduced density matrix powers of a mixed state can also be provided. Although the quantum circuit presented in Figure \ref{fig1} is designed for pure states, it is a well-known fact in quantum mechanics that any quantum state can be expressed as a convex combination of pure states through ensemble decomposition.  And it is clear that the proof process of the theorem primarily relies on the trace operation. Due to linear property of the trace operation, by performing a weighted summation of the calculation results of each pure state component based on their probabilities, the expression for the traces of the reduced density matrix  powers for the mixed state can be  obtained. 
   \par 
 
  We note that the the occurrence of cross terms in the proof may be due to the action of non-adjacent SWAP operators on the corresponding copy states. The combined effect of these SWAP operators will cause a variation in the correlations among the copy quantum states within the entire system, resulting in the appearance of cross terms during the measurement process.
  \par 
  Furthermore, we note that due to the sum of probabilities being 1 and some probabilities being equal by symmetry, in fact all representations of entanglement probabilities for the traces of powers of reduced density matrices share a single set of measurement data.\\\par 
  \textbf{Corollary 1.}\emph{ In the quantum circuit with $n$ copy states shown in Figure \ref{fig1}, the measurement probability terms in the expression for ${\rm tr}\rho_{A_1}^2$ can be used to represent ${\rm tr}\rho_{A_1}^k$ for $k=3,\cdots,n.$}\\\par 
   Corollary 1 implies that for different powers $k$, there is no need to redesign the quantum circuit or acquire new measurement data. Instead, the traces of multiple power can be represented through the same set of measurement probability terms, therefore significantly enhancing computational efficiency and reducing resource consumption.\par 
   	Actually the above discussion is equivalent to evolution of quantum circuit.  Subsequently, we present the evolution process of the mixed state quantum circuit, and prove that the result is consistent with that based on the kraus operator representation. And the following results reproduce the above results based on the kraus operator representation.
   \par
   \textbf{Theorem 4.} \emph{Let $\rho$ in $ \mathbb{C}^{d_{A_1}}\otimes\mathbb{C}^{d_{A_2}}\otimes\cdots\otimes\mathbb{C}^{d_{A_N}}$ be any quantum state, and let the initial state be $|0\rangle\langle 0|\otimes|0\rangle\langle 0|\otimes\rho\otimes\rho\otimes\rho$.  After the quantum circuit in Figure \ref{fig10} is applied, we have 
   	\begin{align*}
   		&	{\rm tr}\rho_{A_1}^2=2\left( p(00)+p(01)\right) -1,\\
   		&	{\rm tr}\rho_{A_1}^3=2\left( p(00)+p(11)\right) -1.
   	\end{align*}
   }\\\par 
\textbf{Theorem 5. }\emph{ Let $\rho$ in $ \mathbb{C}^{d_{A_1}}\otimes\mathbb{C}^{d_{A_2}}\otimes\cdots\otimes\mathbb{C}^{d_{A_N}}$ be any $N$-partite quantum state, and consider the initial state $|0\rangle\langle 0|\otimes|0\rangle\langle 0|\otimes|0\rangle\langle 0|\otimes\rho\otimes\rho\otimes\rho\otimes\rho$. After passing through the quantum circuit in Figure \ref{fig11}, we can obtain 
	\begin{align*}
		{\rm tr}\rho_{A_1}^2=2\left( p(000)+p(001)+p(010)+p(011)\right) -1,\\
		{\rm tr}\rho_{A_1}^3=2\left( p(000)+p(001)+p(110)+p(111)\right) -1.\\
		{\rm tr}\rho_{A_1}^4=2\left( p(000)+p(011)+p(101)+p(110)\right) -1.
	\end{align*}
}\par 
We can obtain an expression for ${\rm tr}\rho_{x}^k,( k=2,3,\cdots n)$ for any $x\in\mathcal{P}(\mathcal{S})\setminus \left\lbrace \varnothing \right\rbrace$ via the circuit in Figure \ref{fig1}. This requires us to consider applying the SWAP operators to the subsystem $x$. In particular, when $x = \mathcal{S}$, we can give the representation of ${\rm tr}\rho^k,( k=2,3,\cdots n).$\par

	As shown in Figure \ref{fig1}, the number of qubits required to calculate all ${\rm tr}\rho_{x}^k,( k=2,3,\cdots,n)$ is $ O(n{\rm log}(d)+n-1)$, where $d$ is the size of the Hilbert space  on which $\rho$ is defined, and the number of quantum gates required for a single measurement is $ O(3n-3)$. 
	
	\begin{figure}[!t]
		\centering
		\includegraphics[width=1\linewidth]{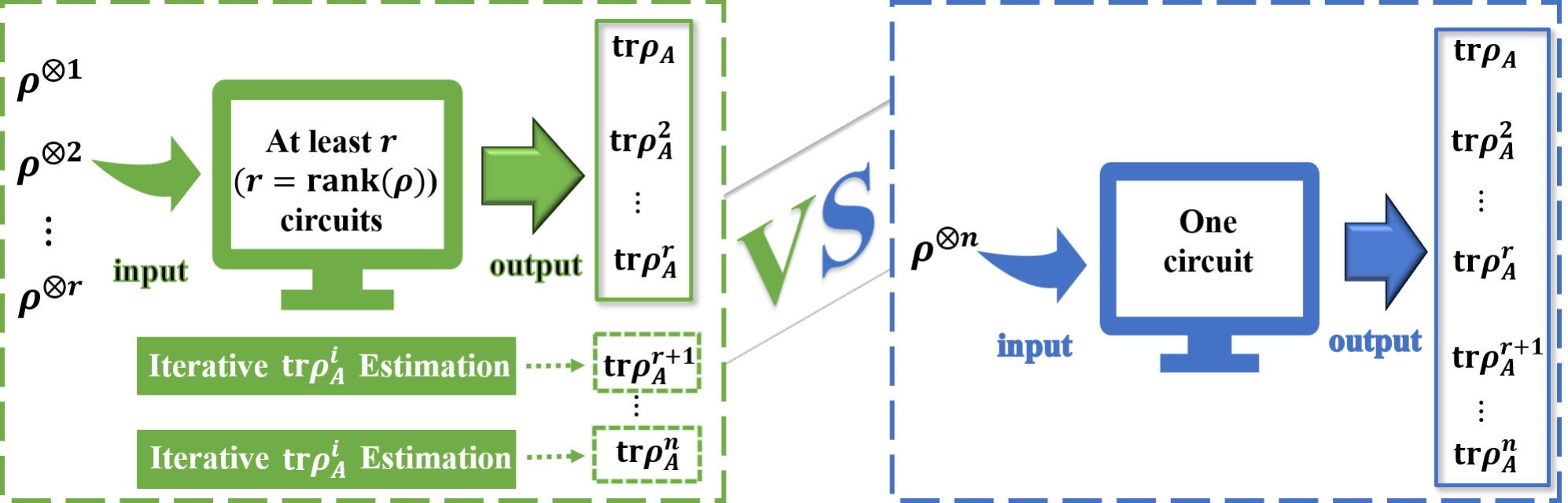}
		\caption{\textbf{Comparison of existing algorithms for estimating power traces of reduced density matrices with the algorithm in this paper.} The left figure illustrates the process of obtaining the traces of reduced density matrix powers through the method in \cite{Shin}. The right figure shows the process of obtaining the traces of reduced density matrix powers by using the method proposed in this Letter. Comparison of existing algorithms (left) requiring at least $r$ quantum circuits \cite{Shin} versus our single-circuit approach (right), highlighting reduced resource consumption and improved efficiency for estimating multiple power traces simultaneously}.
		\label{fig12}
	\end{figure}
\section{III. Improved quantum algorithm for estimating traces of multiple powers of a reduced density matrix}
    If we want to apply the existing quantum circuits to obtain all traces of reduced density matrix powers ${\rm tr}(\rho_A^k), (k=1,2,\cdots,n)$, we need at least $r={\rm rank}(\rho_A)$ quantum circuits, then iteratively estimate through classical calculations \cite{Shin},
   which is shown on the left figure of Figure \ref{fig12}.
   In contrast, our quantum algorithm requires only one quantum circuit to estimate all the traces of reduced density matrice powers simultaneously, as shown on the right in Figure \ref{fig12}.\par
\begin{figure}[!t]
		\centering
		\includegraphics[width=1\linewidth]{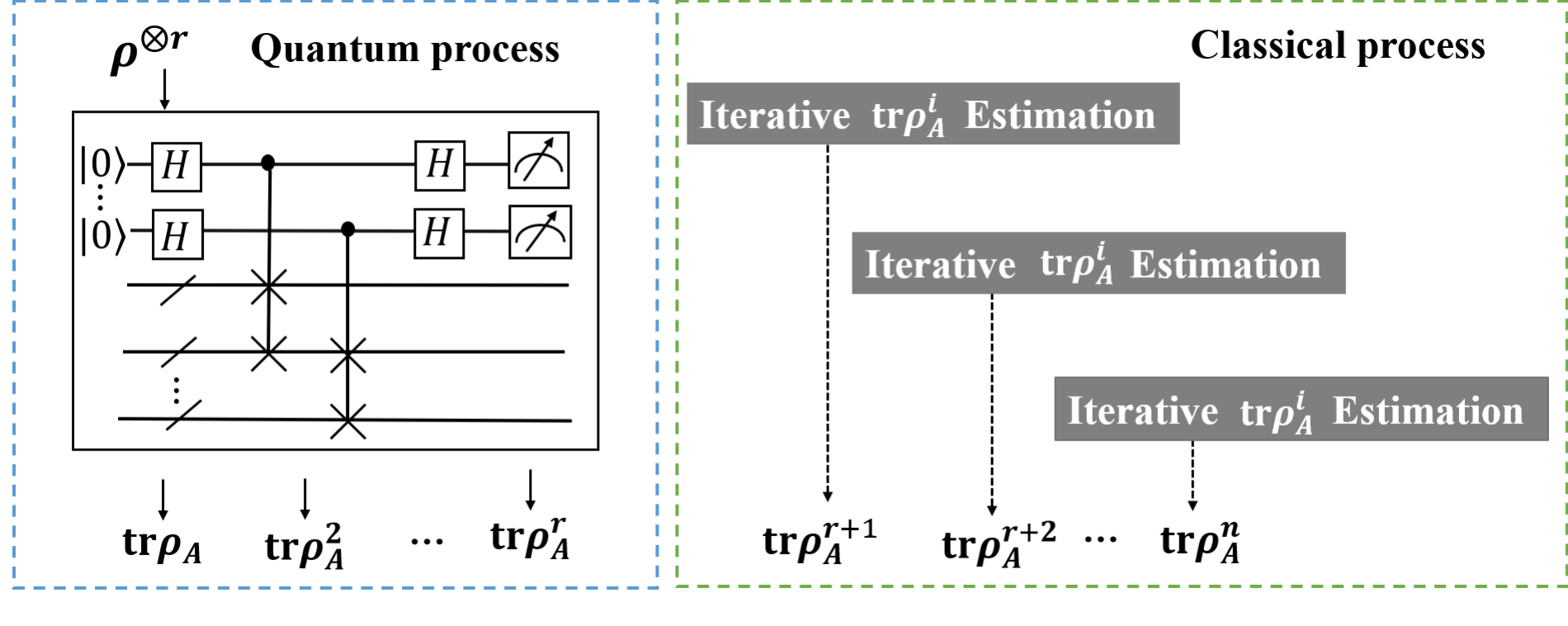}
		\caption{\textbf{Schematic diagram of the improved quantum algorithm for estimating traces of multiple powers of a reduced density matrix.} The black box indicates the quantum process and the gray box indicates the part where the classical iteration is performed. The quantum algorithm is realized by a quantum circuit with $r$ quantum copy states, and the classical process is realized by classical iteration.} 
		\label{fig13}
	\end{figure}

 
  To further optimize the above proposed algorithm, we can propose an improved quantum algorithm for estimating traces of multiple powers of a reduced density matrix, i.e., $\left\lbrace {\rm tr}(\rho_A^i)\right\rbrace_{i=1}^{n} $, which combines quantum circuit with classical iteration, which is inspired by \cite{Shin}. We need at most one quantum circuit containing $r$ quantum copies to estimate all $\left\lbrace {\rm tr}(\rho_A^i)\right\rbrace_{i=1}^{n} $, and the specific process is shown in Figure \ref{fig13}.\par 
	The classical iteration is realized by iterating the classical Newton-Girard method \cite{Shin,Neven,Natalini}. In the following, we briefly describe this method.\par

	Let $r={\rm rank}(\rho)$ denote the rank of the quantum state $\rho$ and the nonzero eigenvalue of the quantum state $\rho$ be $\lambda_{i},\ (i=1,2,\cdots,r)$.
	Consider a polynomial with these eigenvalues as roots:
	\begin{align*}
		F(\lambda)= & \prod_{i=1}^{r}\left(\textcolor{red}{\lambda}-\lambda_{i}\right)
		=\sum_{k=0}^r(-1)^ka_k\textcolor{red}{\lambda}^{r-k},
	\end{align*}
	where $a_k$ is the elementary symmetric polynomial expressed as the sum of all distinct products of $k$ distinct variables. There is a one-to-one correspondence between the coefficients of the characteristic polynomials $\left\lbrace a_1, a_2,\cdots, a_r \right\rbrace $ and the traces of the quantum state powers $\left\lbrace {\rm tr}\rho, {\rm tr}\rho^2
	,\cdots, {\rm tr}\rho^r
	\right\rbrace $ as follows:
	\begin{align*}
		a_{k}=\frac{1}{k}\sum_{i=1}^{k}(-1)^{i-1}a_{k-i}{\rm tr}\rho^{i},
	\end{align*}
	for any integer $1\le k\le r$.\par 
	Then we have the following relationship between the eigenvalues
	\begin{align*}
		\lambda_{i}^{r+l}=\sum_{j=1}^r(-1)^{j-1}a_j\lambda_i^{r+l-j},
	\end{align*}
	where $1\le i\le r$ and $ l\ge 1$ are both integers. Further we can get 
	\begin{align*}
		{\rm tr}\rho^{r+l}=\sum_{j=1}^r(-1)^{j-1}a_j{\rm tr}\rho^{r+l-j}.
	\end{align*}
	Then we can give an estimate to an arbitrary power trace of the reduced density matrix.\\\par 
\section{IV. Statistical analysis of traces of Reduced Density Matrix powers}
	In this section, we discuss the accuracy and precision with high probability of the estimation of trace of reduced density matrix powers via quantum circuit in Figure \ref{fig1} as presented in Theorem 3. Note that we do not need to store all $2^n$  probability values in estimating ${\rm tr}(\rho_A^k)$, we only need to count the number of bitstrings $\emph{\textbf{z}}$ that satisfy specific condition ``$c^\emph{\textbf{z}}_{1,2\cdots,k-1}\ is \ even $''. We have
	the following Theorems, see proof in Appendix. \\\par

\textbf{Theorem 6.}\emph{ For any $N$-qubit quantum state $\rho$ with reduced density matrix $\rho_A$ on subsystem $A$, consider performing a controlled SWAP test in Figure \ref{fig1} with $n$ copies of $\rho$, yielding measurement results $\emph{\textbf{z}} \in \left\lbrace 0,1\right\rbrace^{n-1}$. Let $\left\lbrace {\rm tr}(\rho_A^k)\right\rbrace_{k=2}^n$ denote the set of power traces to be estimated. The following statistical guarantees hold:\\ 
(1) There exist random variables $\left\lbrace \hat{T}_k\right\rbrace_{k=2}^n$ such that with $M = O\left(\frac{1}{\epsilon^2}\log(\frac{n}{\delta})\right)$ independent experiments, the following holds simultaneously:
\begin{align*}
	{\rm P}\left(\bigcap_{k=2}^n \left|\hat{T}_k - {\rm tr}(\rho_A^k)\right| \leq \epsilon \right) \geq 1-\delta,
\end{align*}
where $\epsilon>0$ is the maximum allowable error and $\delta \in (0,1)$ is the confidence level. \\
(2) For any $ k \in \left\lbrace 2, 3, \ldots, n\right\rbrace $, the variance of the estimator $\hat{T}_k$ satisfies 
\begin{align*}
	{\rm Var}(\hat{T}_k)=\frac{4}{M}p_k(1 - p_k),
\end{align*}
where $ p_k = \frac{{\rm tr}(\rho_A^k)+1}{2}$.} \\\par 
 Theorem 6 unifies the analysis of sample complexity guaranteed by the Hoeffding inequality \cite{Hoeffding} and variance to characterize the estimator efficiency.
The variance of $\hat{T}_k$ depends on the purity of the reduced density matrix $\rho_A$. For nearly pure states, i.e., ${\rm tr}(\rho_A^k)\approx 1$, the variance diminishes, allowing fewer samples to achieve the same precision. And for nearly pure states, the estimator becomes highly accurate as the variance tends to zero.
This indicates that the algorithm exhibits higher computational efficiency for quantum states with higher purity.
The unification of these statistical guarantees provides a rigorous foundation for employing the controlled SWAP test in practical quantum state analysis. \par 
To demonstrate the advantages of our approach, we present a comparison with some existing method in in certain aspects for estimating  Table \ref{comparision} contrasts our framework with some existing method in terms of quantum circuit requirement, simultaneous estimation capability, and measurement complexity, aligning with the theoretical analysis in Theorem 6 and the circuit design in Figure 1.

	\begin{table*}
		\renewcommand{\arraystretch}{1.5}
		\begin{tabular*}{0.9\linewidth}{c c c c}
			\hline
			Methods & Quantum Circuit & Simultaneous Estimation & Measurement Complexity \\
			\hline
			Existing Method \cite{Ekert,Johri,Suba,Yirka,Shin} &  At least $r $ circuits & A ${\rm tr}(\rho_A^k)$ with a circuit &  $O\left(\frac{r^2}{\epsilon^2}\log(\frac{1}{\delta})\right)$ \cite{Shin} \\
			Our method & A single unified circuit &  $\left\lbrace {\rm tr}(\rho_A^i)\right\rbrace_{i=1}^{n} $ & $O\left(\frac{1}{\epsilon^2}\log(\frac{n}{\delta})\right)$ \\
			\hline
		\end{tabular*}
		\caption{Comparison of our framework with some methods for estimating $\left\lbrace {\rm tr}(\rho_A^i)\right\rbrace_{i=1}^{n} $ within the error margin of $\epsilon$ and the confidence level $\delta$.}
		\label{comparision}
	\end{table*}

 \section{V. Applications}
 	Below we will give some applications for the above method.
    \subsection{A. Estimation of nonlinear functions based on quantum circuits}
	In quantum mechanics, many nonlinear functions have crucial applications. So being able to give estimates of nonlinear functions is an important task. Below we explore the estimation of several nonlinear functions based on the methods given in this Letter.\par 
	The exponential function $e^A$ ($A$ is the density matrix) is a quantity of great significance. However, estimating it directly through quantum circuits is a challenging task, as the computation of the exponential function is not easy to realize directly in the quantum domain. We can estimate the the trace of exponential function $e^A$ based on the quantum circuit proposed in this Letter.\par
	First we can expand ${\rm tr}(e^A)$ using a Taylor series into ${\rm tr}(e^A)={\rm tr}(\sum_{n=0}^{\infty }\frac{A^n}{n!})$. It can be approximated as ${\rm tr}(e^A)\approx \sum_{n=0}^{N }\frac{{\rm tr}(A^n)}{n!}$, where $N$ is a finite number of truncation terms. By choosing a suitable truncation term $N$, $ {\rm tr}(e^A)$ can be approximated within a certain accuracy $\epsilon$. We define $f(A)=\sum_{n=0}^{N }\frac{{\rm tr}(A^n)}{n!}$. With the specific expression of the traces of density matrice powers,  this approximation can be obtained by designing a quantum circuit with $N$ copies state. By using the measurement probability of this circuit, we can give the estimate 
	\begin{align*}
		f(A)&= {\rm tr}(I)+{\rm tr}(A)+\sum_{n=2}^{N }\frac{1}{n!}\left( 2\sum_{c^\emph{\textbf{z}}_{1,2,\cdots,n-1}\ is\ even }p(\emph{\textbf{z}})-1\right)  .
	\end{align*}
    In particular, based on the estimation of the trace of this exponential function, we can estimate ${\rm tr}(e^{\beta\rho})$\ ($\beta$ is a constant), which has many important applications in thermodynamics \cite{Rotter,Mohseni} and other fields.\par

	In addition we discuss another nonlinear function $\rho \ln \rho$ that has an important place in quantum information theory. Von Neumann entropy \cite{Boes} $-{\rm tr}(\rho \ln\rho)$ is an important tool used to quantify the degree of entanglement and purity of a quantum state. When it comes to multipartite quantum systems, such as two entangled qubit systems, the entanglement between them can be determined by calculating the Von Neumann entropy. We can then estimate the Von Neumann entropy based on the given expression for ${\rm tr}(\rho ^n)$.
	First we can expand ${\rm tr}(\rho \ln\rho)$ using a Taylor series into 
	\begin{align*}
	 \text{tr}(\rho \ln \rho) = {\rm tr} \left[ \rho \sum_{n = 1}^{\infty} \frac{(-1)^{n + 1}}{n} (\rho - I)^n \right],
	\end{align*}
where we suppose the matrix $\rho$ satisfies $\|\rho - I\| < 1$ (where $\|\cdot\|$ is a suitable matrix norm).\par 
	It can be approximated as ${\rm tr}(\rho \ln\rho)\approx \sum_{n=1}^{N }\frac{(-1)^{n+1}}{n} {\rm tr}(\rho(\rho-I)^n)$, where $N$ is a finite number of truncation terms. By choosing a suitable truncation term $N$, ${\rm tr}(\rho \ln\rho)$ can be approximated within a certain accuracy $\epsilon$. Since 
	\begin{align*}
		&\sum_{n=1}^{N}\frac{(-1)^{n+1}}{n}{\rm tr}( \rho(\rho-I)^n)\\=&\sum_{n=1}^{N}\frac{(-1)^{n+1}}{n}\sum_{k=0}^{n}C_{n}^k(-1)^{n-k}{\rm tr}(\rho^{k+1})
	\end{align*}
	 can be expressed as some terms containing different powers of ${\rm tr}(\rho ^k)$, specific estimates can be given.\par 
	
	Preparation of quantum Gibbs states \cite{wang1,Consiglio} is an important part of quantum computing and is used in various applications such as quantum simulation, optimization problems in quantum computing,  and quantum machine learning. The truncated Taylor series
	\begin{align*}
		S_k(\rho)=&\sum_{k=1}^{K}\frac{(-1)^k}{k}{\rm tr}((\rho-I)^k\rho)\\=&\sum_{k=1}^{K}\frac{(-1)^k}{k}\sum_{s=0}^{k}C_k^s(-1)^{k-s}{\rm tr}(\rho^{s+1})
	\end{align*}
	can be used as a cost function for the preparation of variational quantum Gibbs states. This cost function can be calculated by estimates based on using $\left\lbrace {\rm tr}(\rho^i)\right\rbrace_{i=1}^{K+1}$ in this Letter. \par 

\subsection{B. Efficient estimation of entanglement measures}
In the subsequent part, we will show Theorem can be employed to characterize entanglement measures via a single quantum circuit.\par 
The bipartite concurrence, as described in \cite{Wootters}, represents a crucial entanglement measure for bipartite quantum states.
\par 
	\textbf{Definition 1: Bipartite Concurrence \cite{Wootters}.} \emph{For bipartite pure state $|\psi\rangle\in\mathcal{H}_A\otimes\mathcal{H}_B$, entanglement concurrence is defined as  
		$C_E(|\psi\rangle)=\sqrt{2(1-{\rm tr}\rho_A^2)},$
		where $\rho_A={\rm tr}_B(|\psi\rangle\langle\psi|).$}\par
	The expression of the concurrence is related to the subsystem purity. When considering the quantum circuit of Figure \ref{fig1} with two copy quantum states, the subsystem purity is given by ${\rm tr}\rho_A^2=2p(0)-1.$ We can calculate the concurrence from the circuit, expressed as
	$C_E(|\psi\rangle)=2\sqrt{1-p(0)}.$\par 
   Jin et al. \cite{Jin} introduced a family of entanglement measures known as ICEMs which could be estimated in a quantum circuit (Figure 2 in \cite{Jin}). \par
	\textbf{Definition 2: Informationally Complete Entanglement Measures (ICEMs) \cite{Jin}.} \emph{For the bipartite state $|\psi\rangle_{AB}$ with Schmidt rank ${\rm R}(|\psi\rangle_{AB})=R+1$, the ICEMs is defined by
		$E^C(|\psi\rangle_{AB})=1-\frac{1}{2^R}\sum_{i=0}^{R}C_R^i {\rm tr}(\rho_{A}^{i+1}),$
		which is a well-defined measure of quantum entanglement.}\par 
	In \cite{Jin}, the authors give a representation of ICEMs based on circuit measurement probabilities. However, due to the non-exchangeability of neighboring SWAP operators, this representation is not suitable. In this Letter, we have taken this into account, and give the correct representation in the following:
	\begin{align*}
		&E^C(|\psi\rangle_{AB})=1-\frac{1}{2^R}\sum_{i=0}^{R}C_R^i {\rm tr}(\rho_{A}^{i+1})\notag\\
		=&2-\frac{1}{2^{R-1}}\sum_{i=1}^{R}C_R^i\sum_{c^\emph{\textbf{z}}_{1,2,\cdots,i}\ is\ even }p(\emph{\textbf{z}})-\frac{1}{2^{R-1}}.
	\end{align*}\par 
	\textbf{Definition 3: Tsallis-$q$ entanglement \cite{SanKim,Luo}.} \emph{For a bipartite pure state $|\psi\rangle_{AB}$ and each $q>0,$ Tsallis-$q$ entanglement is
		$\mathcal{T}_q(|\psi\rangle_{AB}):=T_q(\rho_A)=\frac{1}{q-1}(1-{\rm tr}\rho_{A}^q).$
		}\par 
	We can estimate the probabilistic representation of Tsallis-$q$ entanglement by a quantum circuit, i.e., 
	$$\mathcal{T}_q(|\psi\rangle_{AB}):=T_q(\rho_A)=\frac{2\sum_{c^\textbf{z}_{1,2\cdots,q-1}\ is \ even }p(\emph{\textbf{z}})}{q-1}.  $$\par 
	\textbf{Definition 4: $q$-concurrence \cite{Yang} .} \emph{For an arbitrary bipartite pure state $|\psi\rangle_{AB}$ on Hilbert space $\mathcal{H}_A\otimes\mathcal{H}_B$, the $q$-concurrence is defined as
	$C_q(|\psi\rangle_{AB})=1-{\rm tr}\rho_{A}^q,$
		for any $q\ge2$.}\par
	For $q$-concurrence, we also can estimate the probabilistic representation of $q$-concurrence by quantum circuit, i.e., 
	$$C_q(|\psi\rangle_{AB})=2\sum_{c^\textbf{z}_{1,2\cdots,q-1}\ is \ even }p(\emph{\textbf{z}}).$$\\\par 
	\section{VI. Numerical simulation}
\begin{figure}[!t]
		\centering
		\includegraphics[width=0.7\linewidth]{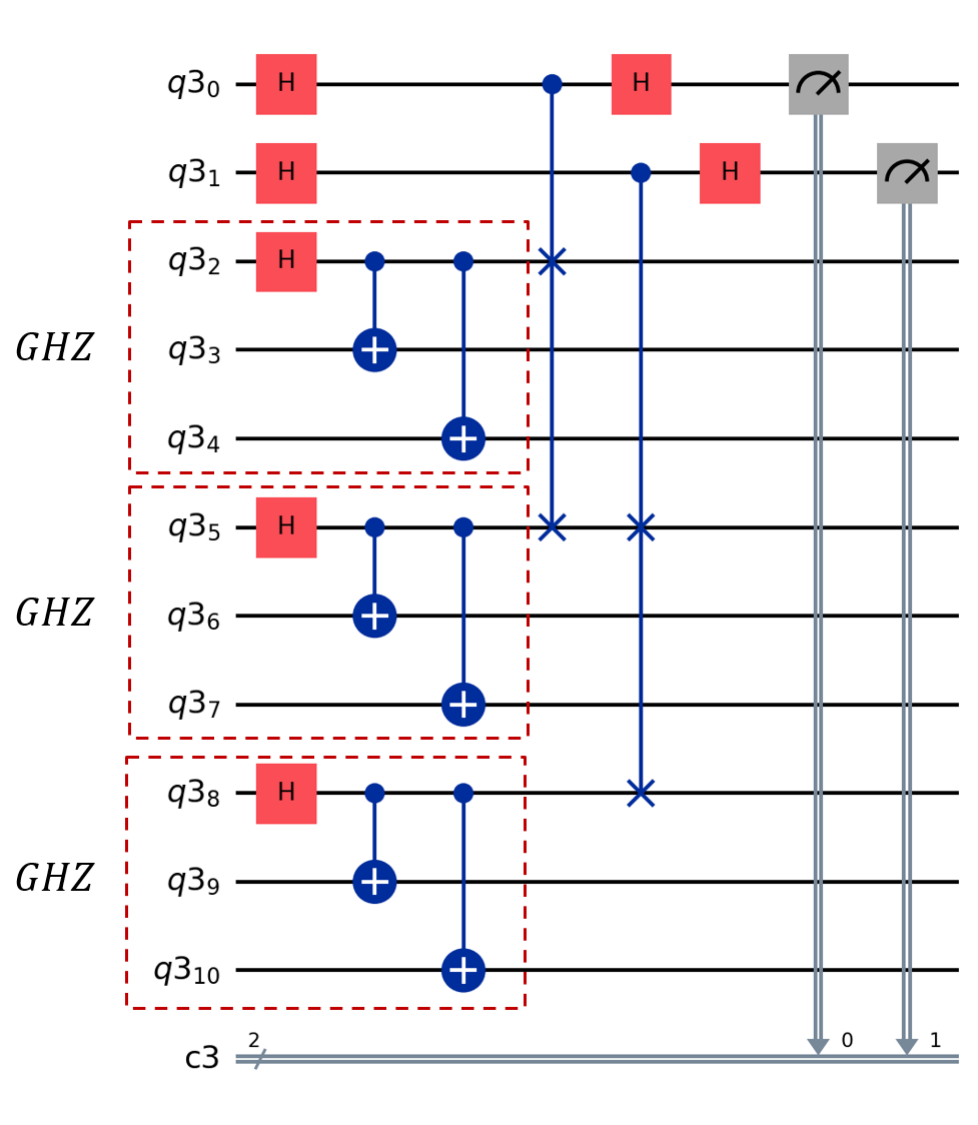}
		\caption{\textbf{The controlled SWAP test of three copies of the GHZ state in \emph{Qiskit}.} The red dashed box denotes the generation of the GHZ state.}
		\label{fig14}
	\end{figure}

\begin{figure}[!t]
		\centering
		\includegraphics[width=0.7\linewidth]{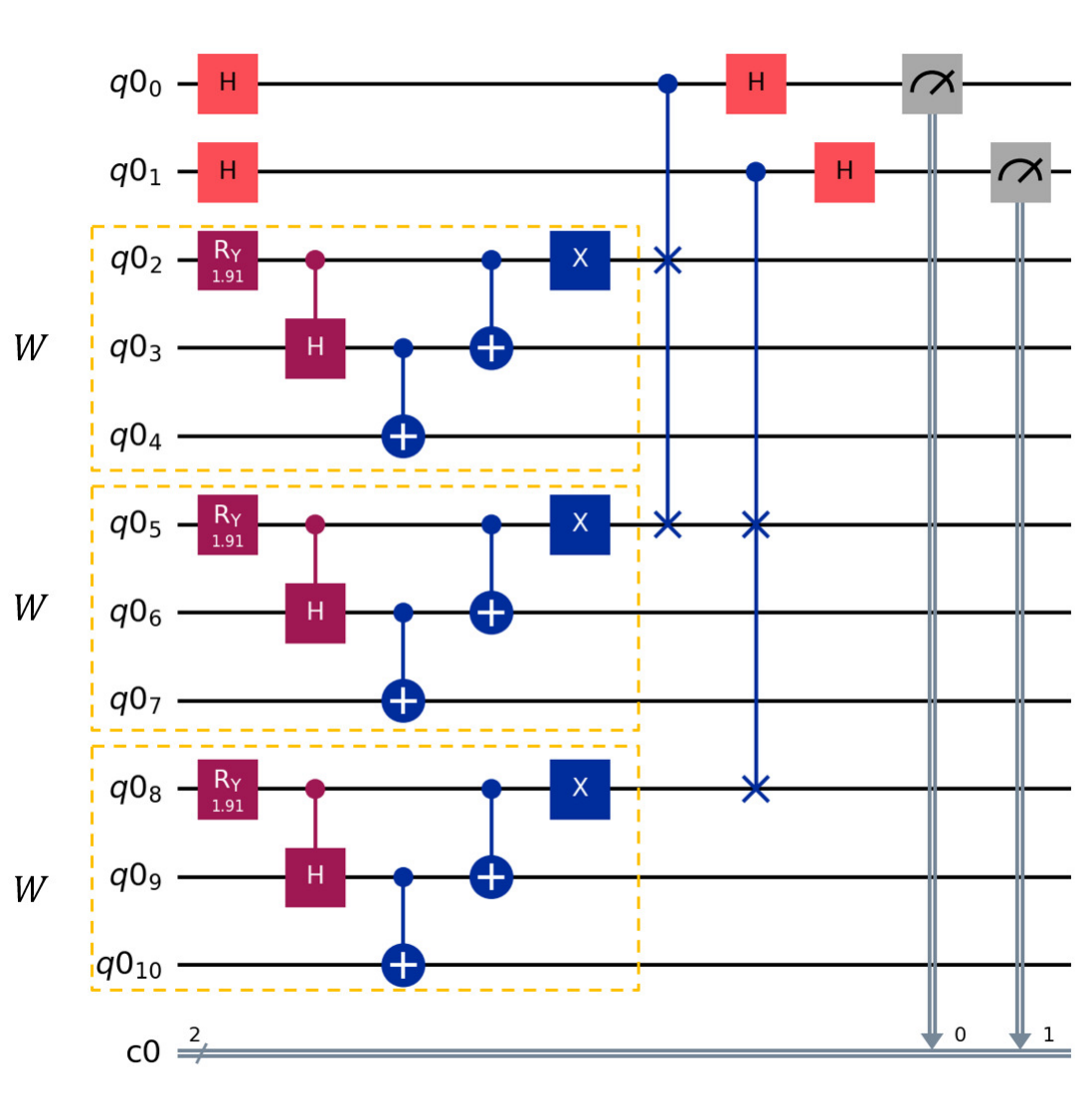}
		\caption{\textbf{The controlled SWAP test of three copies of the W state in \emph{Qiskit}. } The orange dashed box denotes the generation of the W state.}
		\label{fig15}
	\end{figure}

\begin{figure*}[htbp]
	\centering
	\subfigure[]{
		\includegraphics[width=0.45\linewidth]{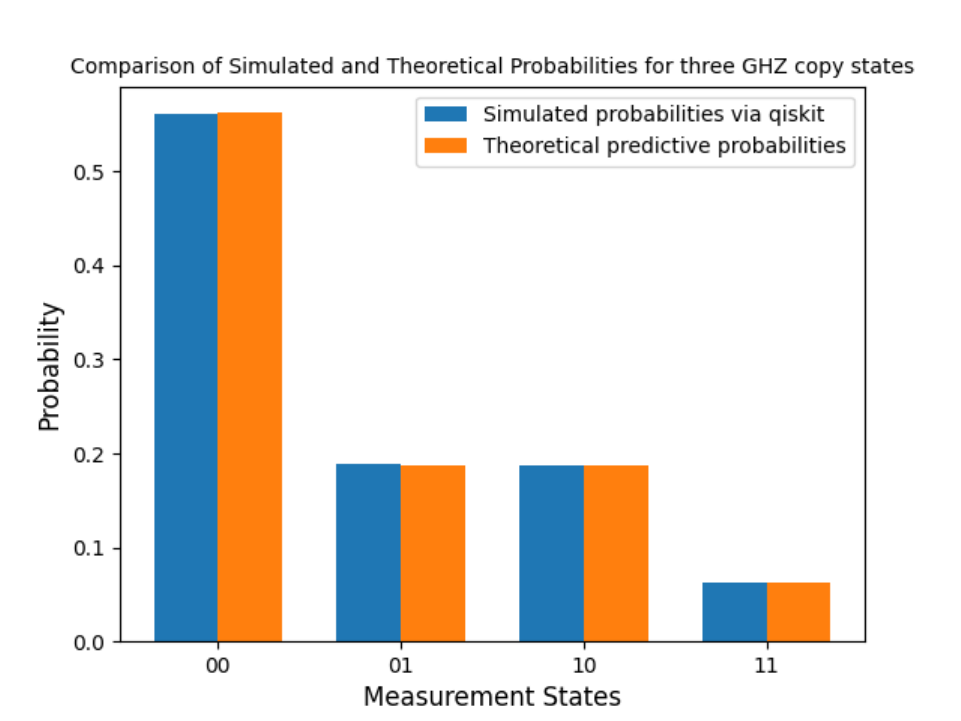}
		\label{fig16a}
	}
	\subfigure[]{
		\includegraphics[width=0.45\linewidth]{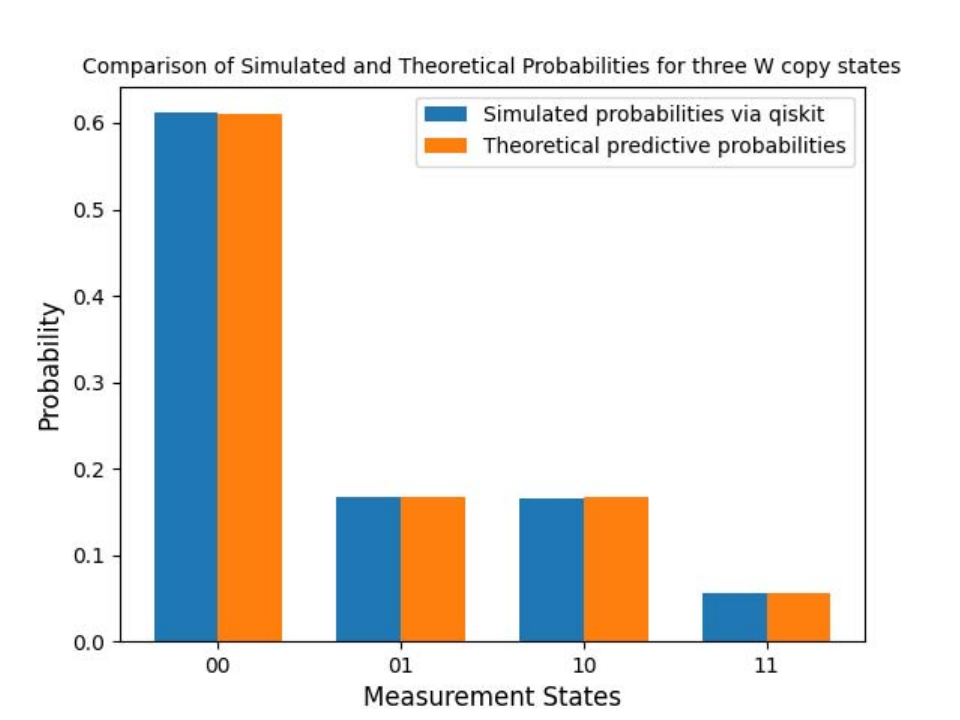}
		\label{fig16b}
	}
	\subfigure[]{
		\includegraphics[width=0.45\linewidth]{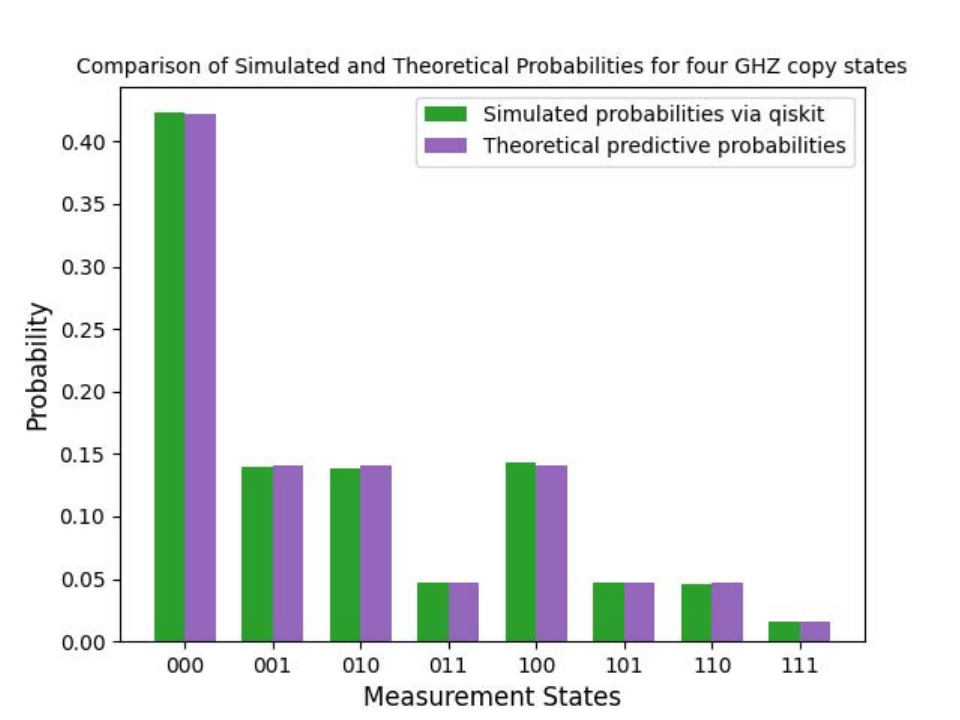}
		\label{fig16c}
	}
	\subfigure[]{
		\includegraphics[width=0.45\linewidth]{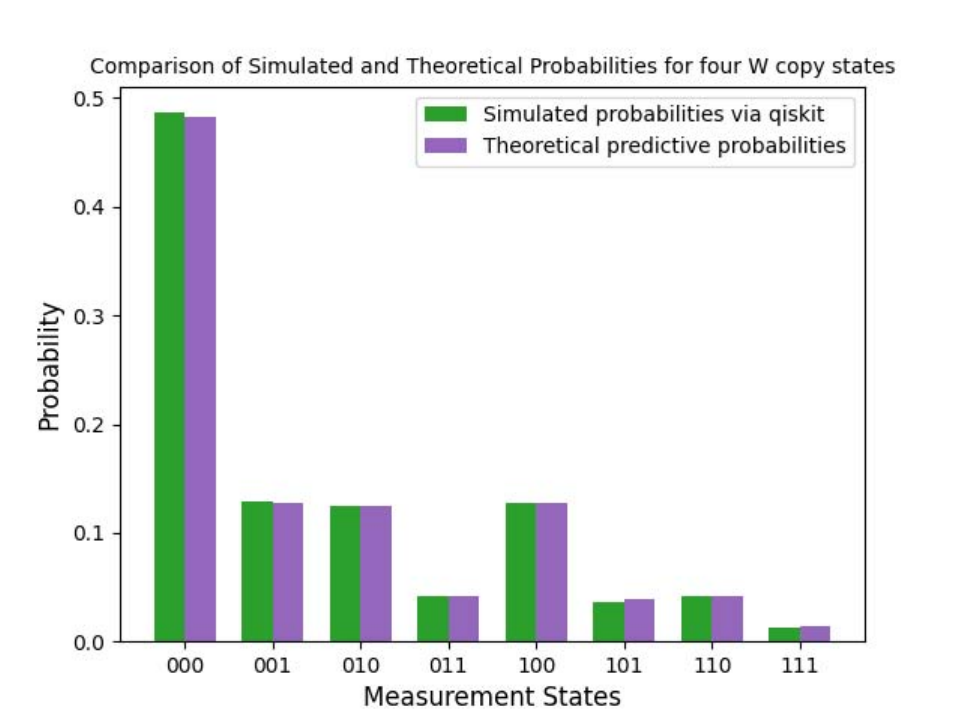}
		\label{fig16d}
	}
	\subfigure[]{
		\includegraphics[width=0.45\linewidth]{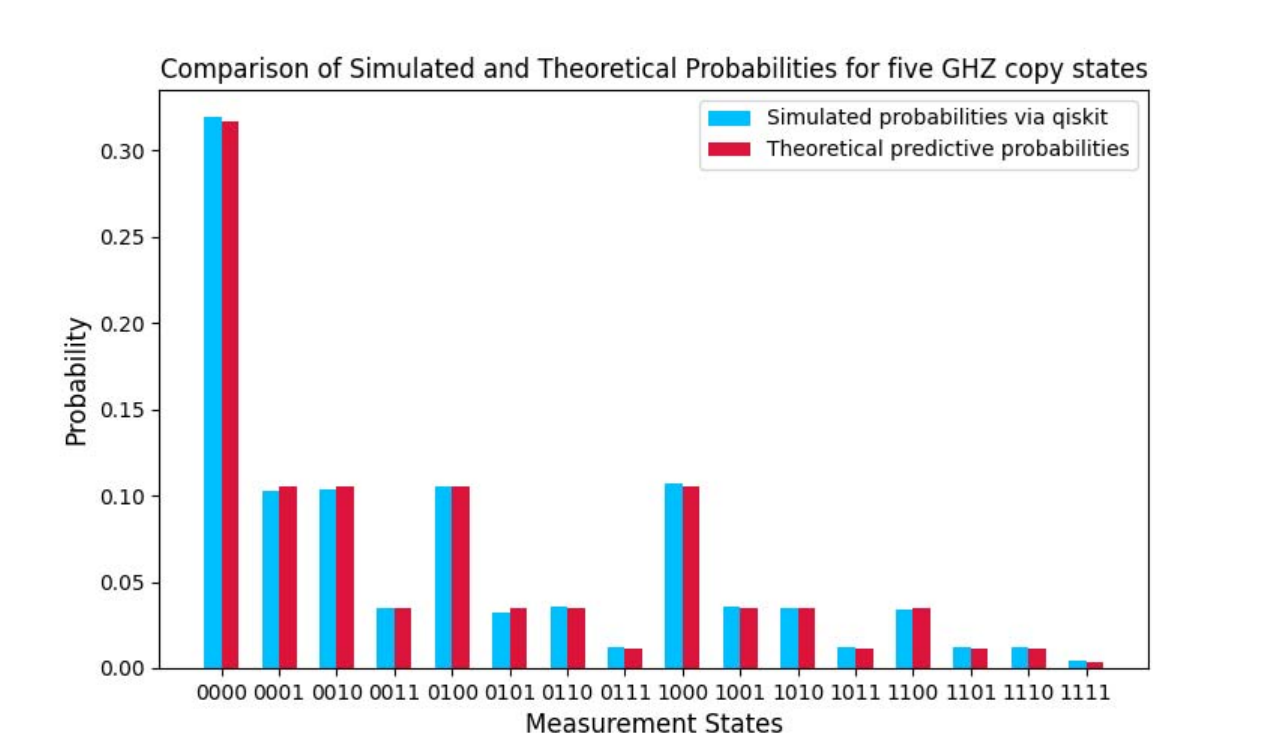}
		\label{fig16e}
	}
	\subfigure[]{
		\includegraphics[width=0.45\linewidth]{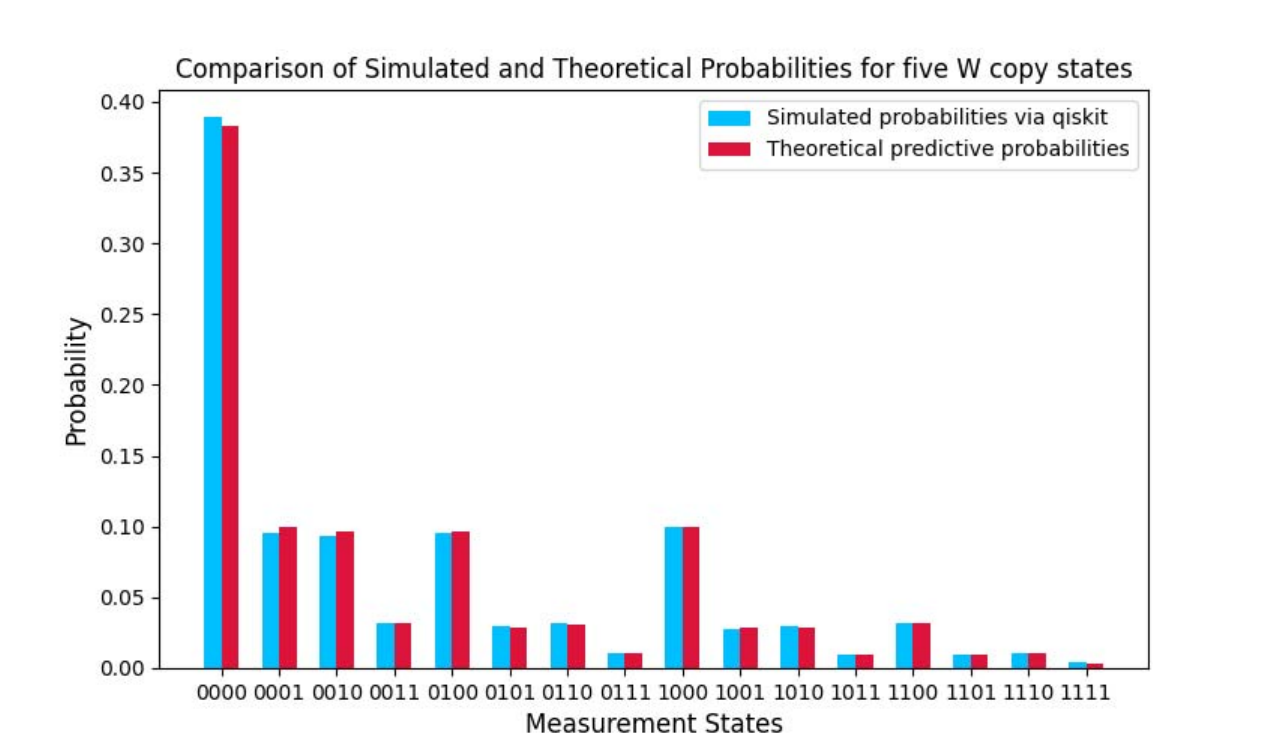}
		\label{fig16f}
	}
	\caption{\textbf{Representation of the simulated and theoretical probabilities corresponding to GHZ states and W states under the quantum circuit with three, four and five copy states, respectively.} The blue bars are the measured probabilities through the qiskit simulated circuit, and the orange bars indicate the theoretical predicted probabilities for three copy states. The green bars are the measured probabilities through the qiskit simulated circuit, and the purple bars indicate the theoretical predicted probabilities for four copy states. The cyan bars are the measured probabilities through the qiskit simulated circuit, and the pink bars indicate the theoretical predicted probabilities for five copy states.}
\end{figure*}

\begin{figure*}
	\centering
	\subfigure[]{
	\includegraphics[width=1\linewidth]{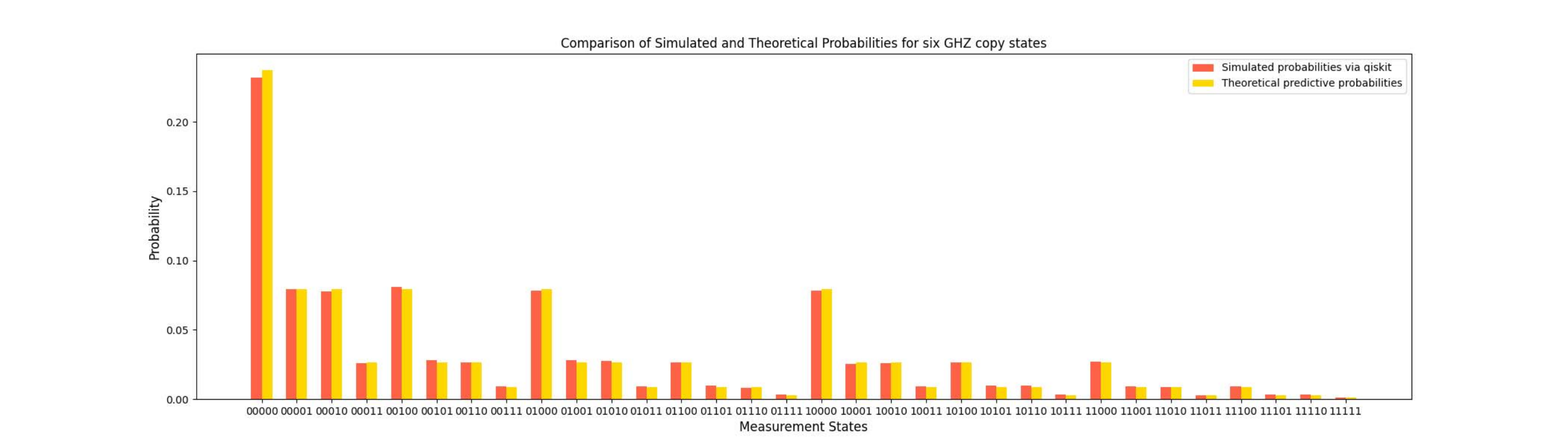}
	\label{fig17a}
}
\subfigure[]{
	\includegraphics[width=1\linewidth]{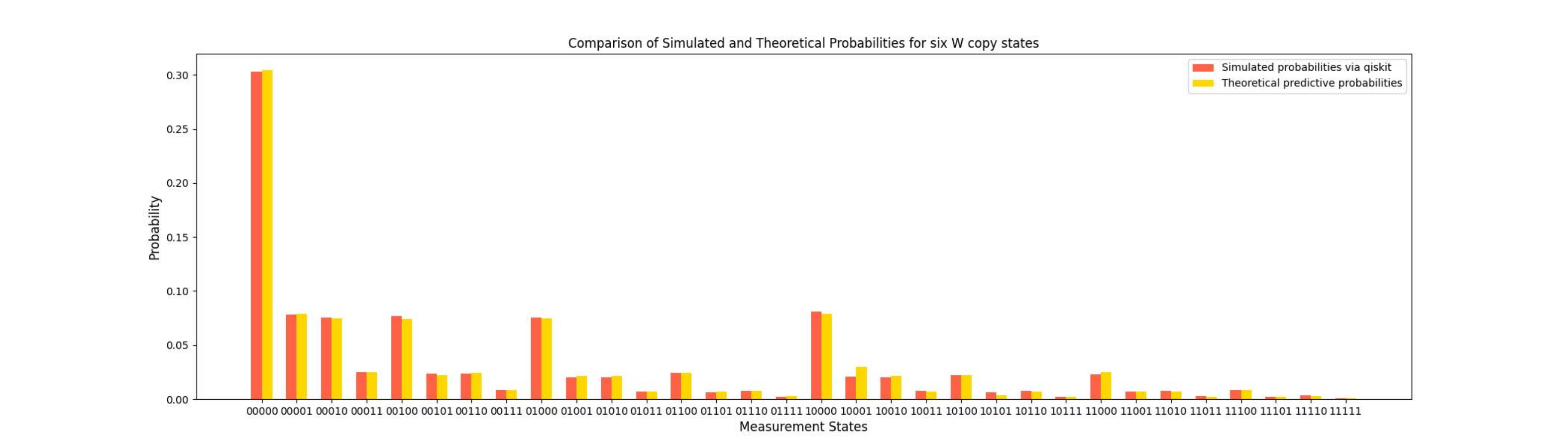}
	\label{fig17b}
}
	\caption{\textbf{Representation of the simulated and theoretical probabilities corresponding to GHZ states and W states under the quantum circuit with six copy states.} The red bar is the measured probabilities through the qiskit simulated circuit with six copy states, and the yellow bar indicates the theoretical predicted probabilities for six copy states.}
\label{six copy states}
\end{figure*}

	We numerically simulated the quantum circuit in the backend of the \emph{qasm\underline{ }simulator} in the \emph{Qiskit} package. We consider two important maximally entangled pure states, GHZ state and W state:
	\begin{align*}
	   &|\psi\rangle_{GHZ}=\frac{1}{\sqrt{2}}(|000\rangle+|111\rangle),\\
	   &|\phi\rangle_{W}=\frac{1}{\sqrt{3}}(|001\rangle+|010\rangle+|100\rangle).
	\end{align*}
	For three copies of the GHZ state and W state, we construct the quantum circuits shown in Figures \ref{fig14} and \ref{fig15} in \emph{Qiskit}, where the red dashed box denotes the generation of the W state and the orange dashed box indicates the generation of the GHZ state. And the controlled SWAP test of additional numbers of state copies can be constructed in the same manner.
	In the circuit simulation process, we set the number of measurements as $2^{15}$. Specifically, for both the GHZ and W states, we simulated circuits with three copies of the quantum state, and the measurement probabilities are presented in Figures \ref{fig16a} and \ref{fig16b}, respectively. Additionally, we simulated circuits with four, five and six copies of the quantum state for these two states, and
	 the corresponding measurement probabilities are shown in Figures \ref{fig16c}, \ref{fig16d}, \ref{fig16e},\ref{fig16f}, \ref{fig17a} and \ref{fig17b}, respectively.
	\par 
	In this simulation, we consider the mean square error method to measure the error between the simulated and theoretical probabilities.  
	It should be noted that, as derived from the previous theorems and lemmas, the power trace representation of the reduced density matrix can be expressed through measurement probabilities. Correspondingly, measurement probabilities can be obtained from the power traces of the reduced density matrix and the products of these power traces, enabling us to determine the theoretical probabilities accordingly.
    The mean square error is given by
	\begin{align*}
		MSE=\frac{1}{m}\sum_{i=1}^{m}\left( p_{sim}(|i\rangle)-p_{theo}(|i\rangle)\right) ^2,
	\end{align*}
	where $p_{sim}(|i\rangle)$ denotes the simulated probability of measuring state $|i\rangle$, $p_{theo}(|i\rangle)$ denotes the theoretical probability of measuring state $|i\rangle$, and $m$ denotes the total number of measuring states. We have performed an error analysis of the quantum circuit simulations. Table \ref{tab} mainly shows the errors in the simulations using quantum circuits with different number of copies (three copies, four copies, five copies and six copies) in GHZ and W states. The estimation of the simulation probabilities allows us to proceed to estimate the power traces of the reduced density matrix. \par 
	Through the error analysis table (Table \ref{comparision}) and Figures \ref{fig16a},\ref{fig16b},\ref{fig16c},\ref{fig16d},\ref{fig16e},\ref{fig16f},\ref{fig17a} and \ref{fig17b}, we systematically compared the differences between the measurement results of quantum circuits and the theoretical probability predictions based on the power traces of reduced density matrices for 3 to 6 copies of GHZ states and W states. The data shows that the relative errors remain at extremely low levels across all tested numbers of copies. This finding not only strongly validates the mathematical reliability of the power trace estimation formula for reduced density matrices proposed in this paper but also fully demonstrates the high precision of the controlled SWAP test quantum circuit in practical applications.
	

\begin{table*}[t]
	\centering
	\label{tab}
	\renewcommand{\arraystretch}{1.5}
	\begin{tabular}{|c|c|c|c|c|} 
		\hline
		\textbf{Error} & \textbf{Three copies} & \textbf{Four copies} & \textbf{Five copies} & \textbf{Six copies} \\
		\hline
		\textbf{GHZ state} & $3.2843\times10^{-6}$ & $1.3299\times10^{-6}$ & $2.1505\times10^{-6}$ & $1.6844\times10^{-6}$ \\ 
		\hline 
		\textbf{W state} & 
		$2.8379\times10^{-6}$ & $2.6062\times10^{-6}$ & $3.7108\times10^{-6}$ & $3.8313\times10^{-6}$ \\
		\hline
	\end{tabular}
	\caption{\textbf{Errors of GHZ and W states under circuit simulations of three-copy, four-copy, five-copy and six-copy} quantum states, respectively.}
\end{table*}

\begin{figure*}
	\centering
	\subfigure[]{
	\includegraphics[width=0.45\linewidth]{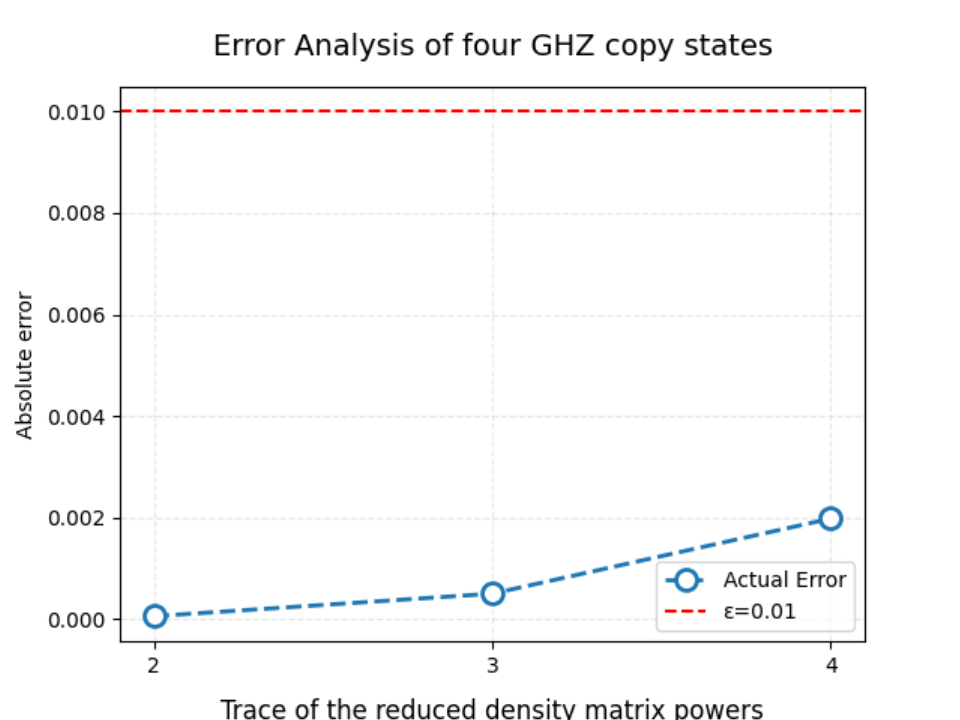}
		\label{fig18a}
	}
    \subfigure[]{
   \includegraphics[width=0.5\linewidth]{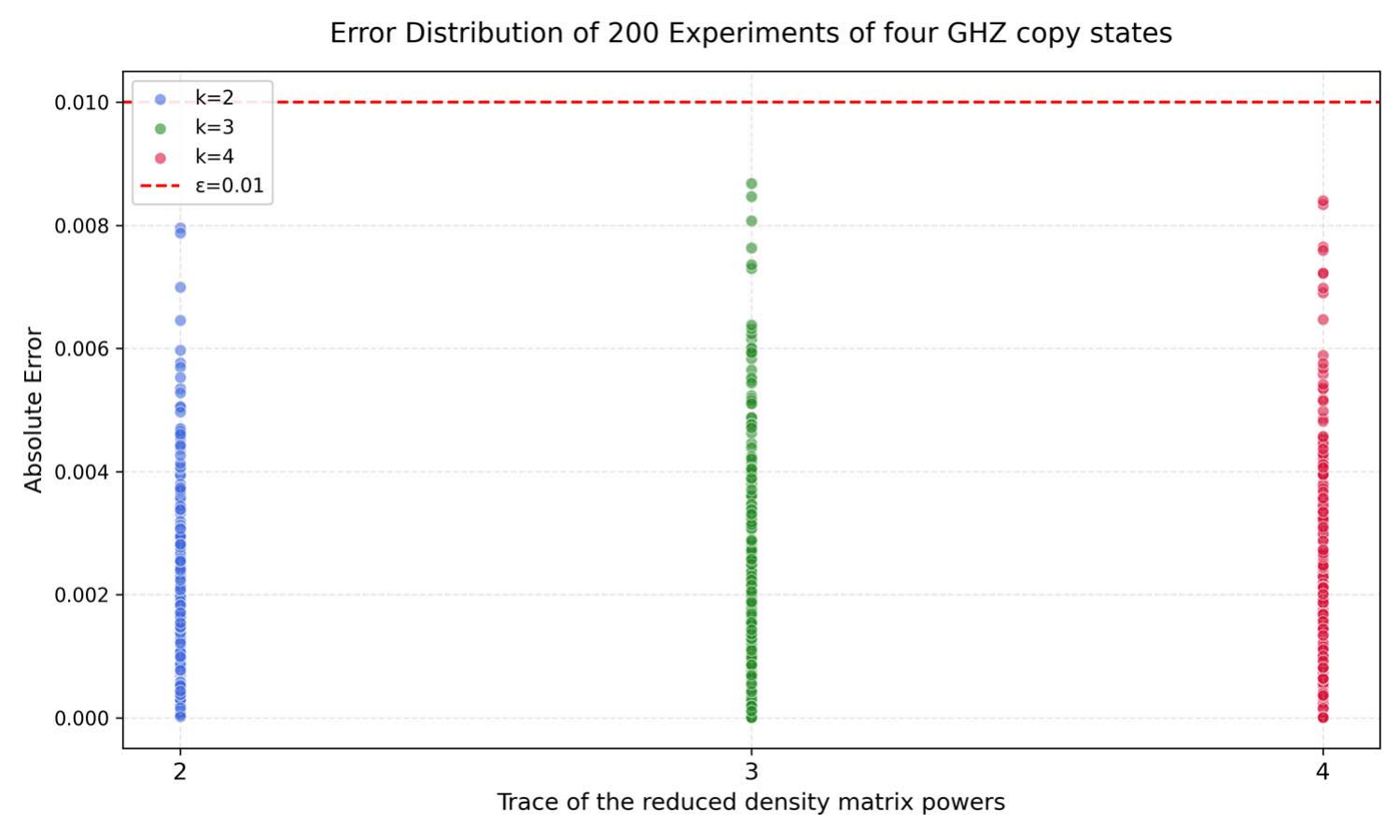}
     	\label{fig18b}
     }
 \subfigure[]{
 	\includegraphics[width=0.45\linewidth]{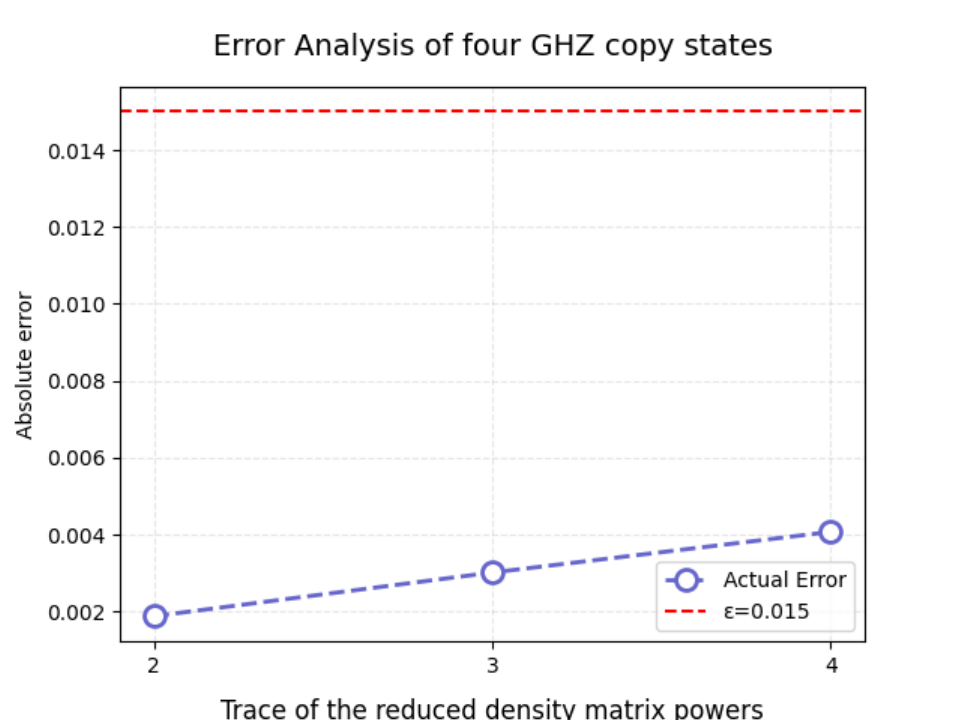}
 	\label{fig18c}
 }
 \subfigure[]{
 	\includegraphics[width=0.5\linewidth]{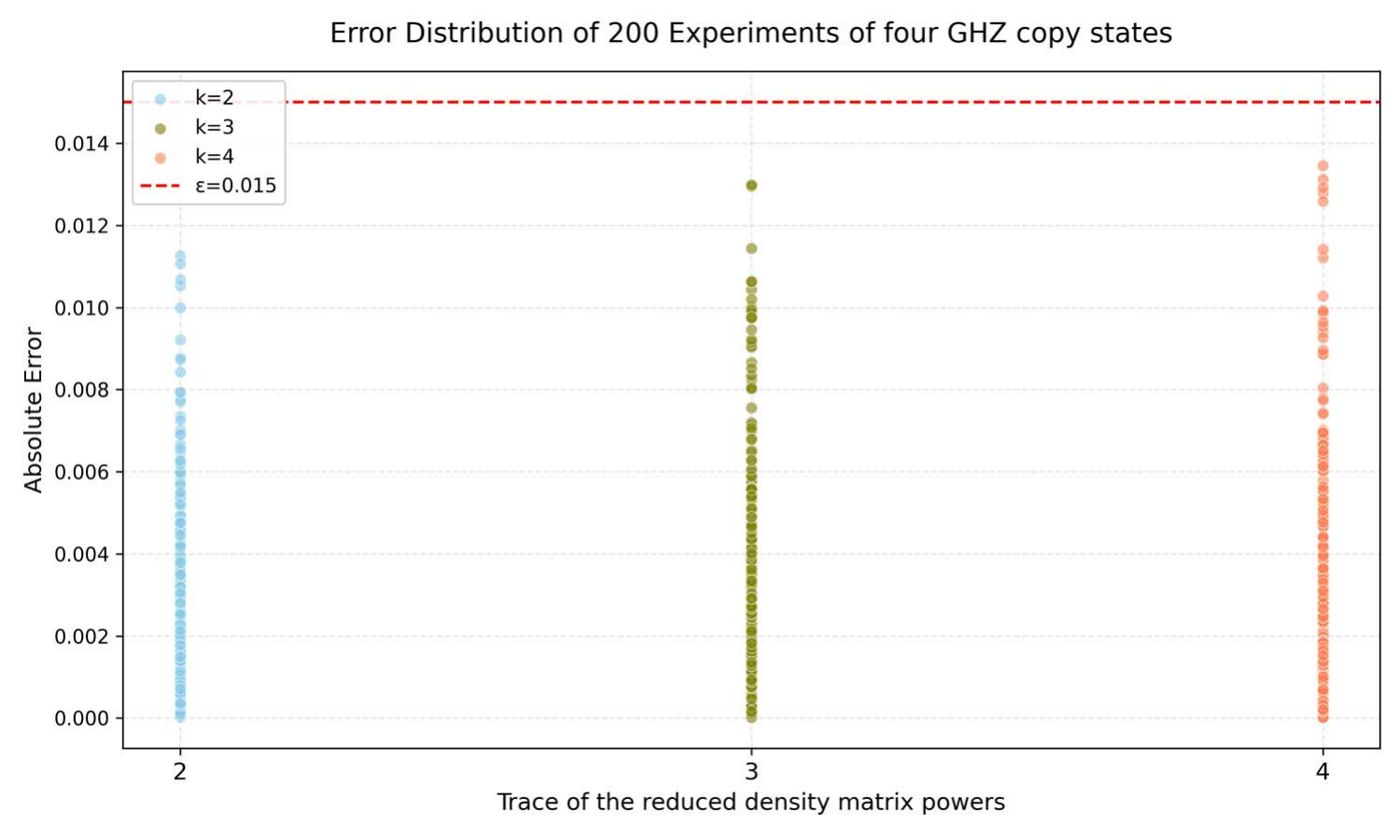}
 	\label{fig18d}
 }
\subfigure[]{
	\includegraphics[width=0.45\linewidth]{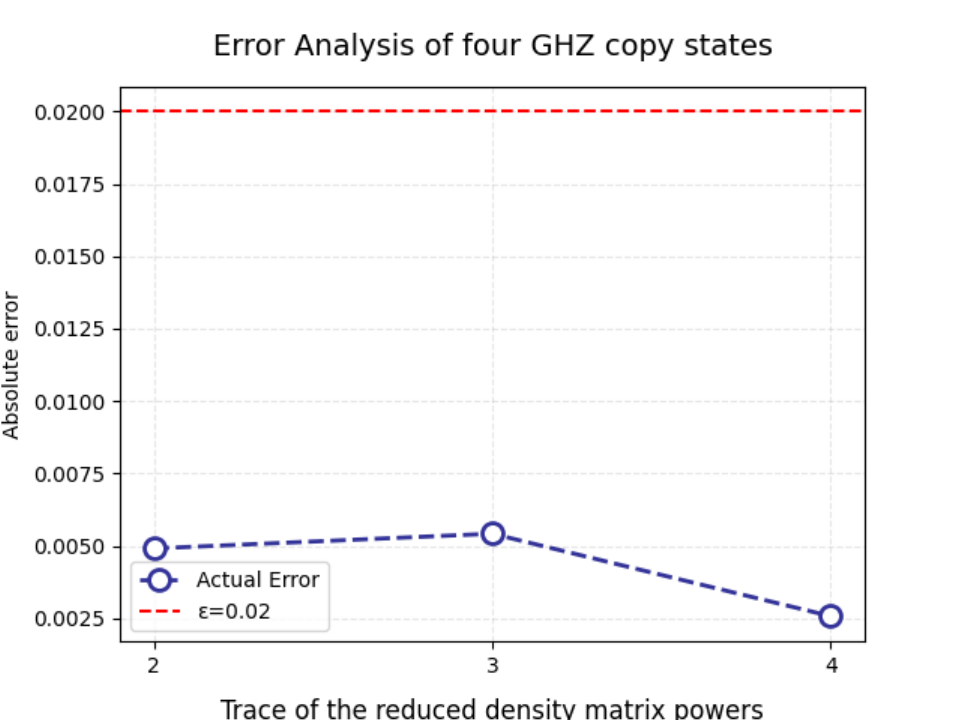}
	\label{fig18e}
}
\subfigure[]{
	\includegraphics[width=0.5\linewidth]{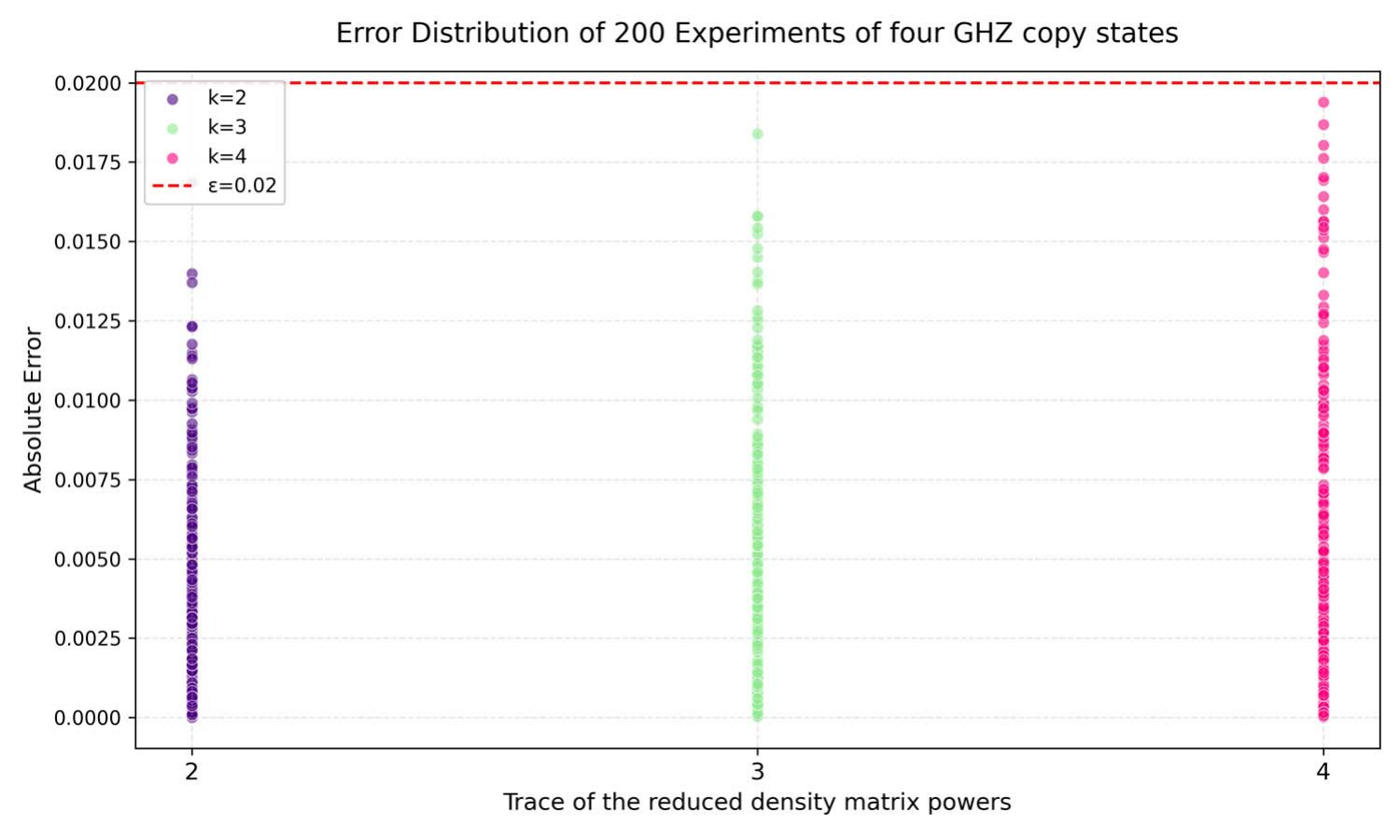}
	\label{fig18f}
}
	\caption{\textbf{Error analysis of the estimation for trace of reduced density matrix of four copy GHZ states under 
		different settings of maximum allowable error and confidence levels.} The red line represents the set maximum allowable error. (a)
	The absolute errors of estimation for trace of reduced density matrix powers when the circuit is run $\frac{2}{\epsilon^2} \ln\left(\frac{6}{\delta}\right)$ times under settings of maximum allowable error $\epsilon=0.01$ and confidence level $\delta=0.05$.  (b) Error distribution of estimation for trace of reduced density matrix powers in 200 experiments with each experiment run $\frac{2}{\epsilon^2} \ln\left(\frac{6}{\delta}\right)$ times. (c) The absolute errors of estimation for trace of reduced density matrix powers when the circuit is run $\frac{2}{\epsilon^2} \ln\left(\frac{6}{\delta}\right)$ times under settings of maximum allowable error $\epsilon=0.015$ and confidence level $\delta=0.06$. (d) Error distribution of estimation for trace of reduced density matrix powers in 200 experiments with each experiment run $\frac{2}{\epsilon^2} \ln\left(\frac{6}{\delta}\right)$ times. (e) The absolute errors of estimation for trace of reduced density matrix powers when the circuit is run $\frac{2}{\epsilon^2} \ln\left(\frac{6}{\delta}\right)$ times under settings of maximum allowable error $\epsilon=0.02$ and confidence level $\delta=0.08$. (f) Error distribution of estimation for trace of reduced density matrix powers in 200 experiments with each experiment run $\frac{2}{\epsilon^2} \ln\left(\frac{6}{\delta}\right)$ times.}
\end{figure*}

\begin{figure*}
	\subfigure[]{
	\includegraphics[width=0.45\linewidth]{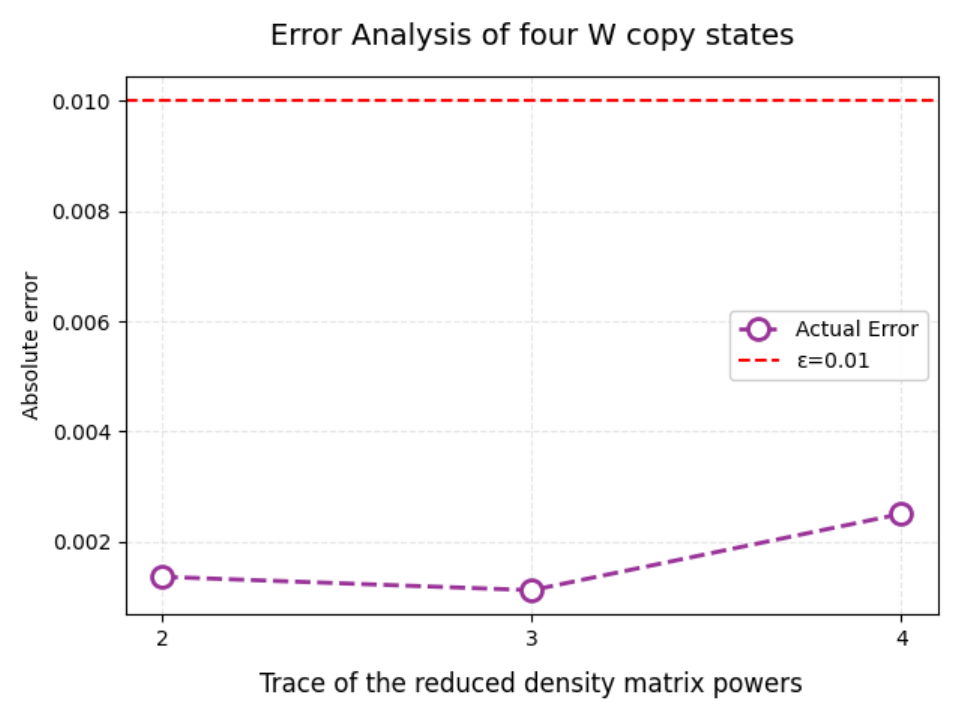}
		\label{fig19a}
}
\subfigure[]{
	\includegraphics[width=0.5\linewidth]{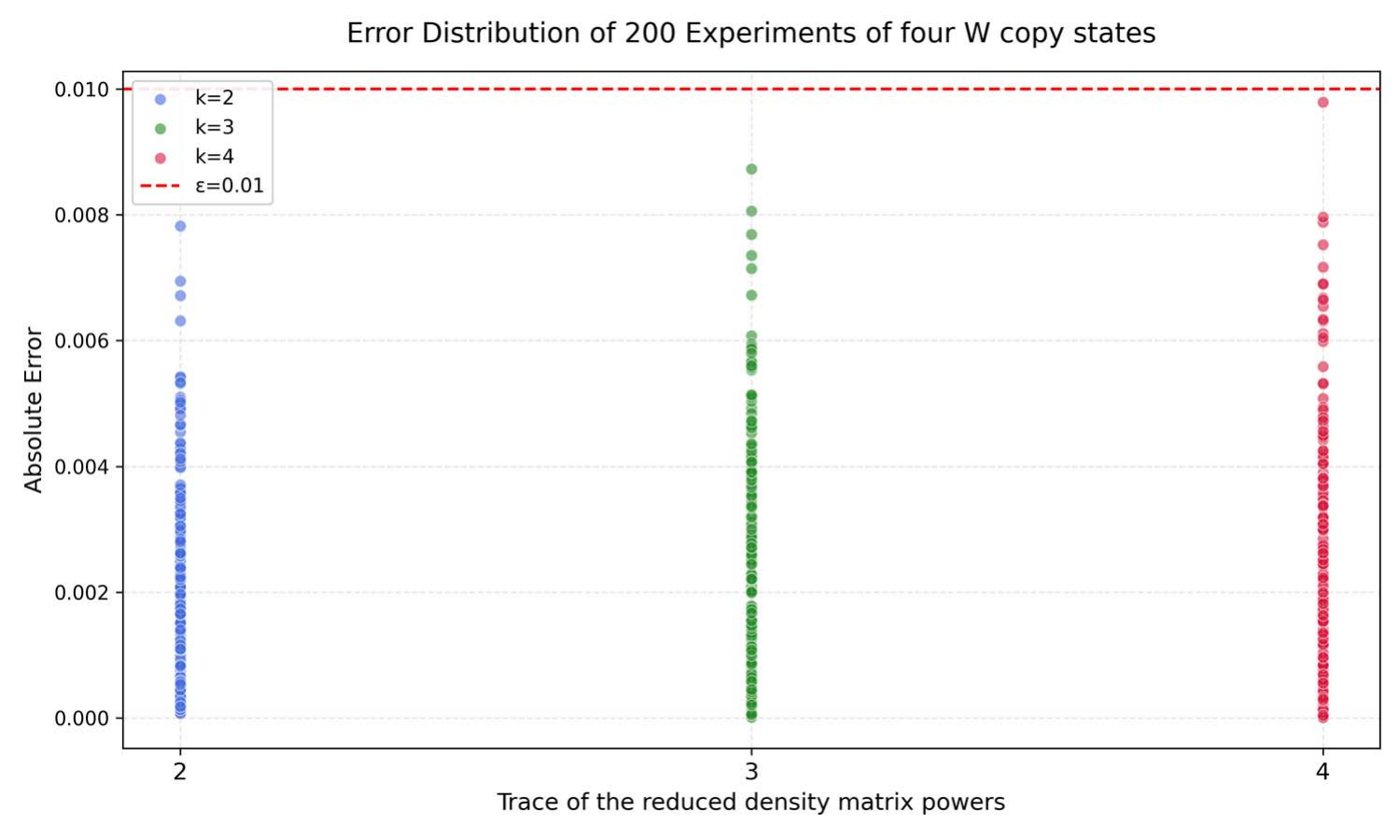}
		\label{fig19b}
}
\subfigure[]{
	\includegraphics[width=0.45\linewidth]{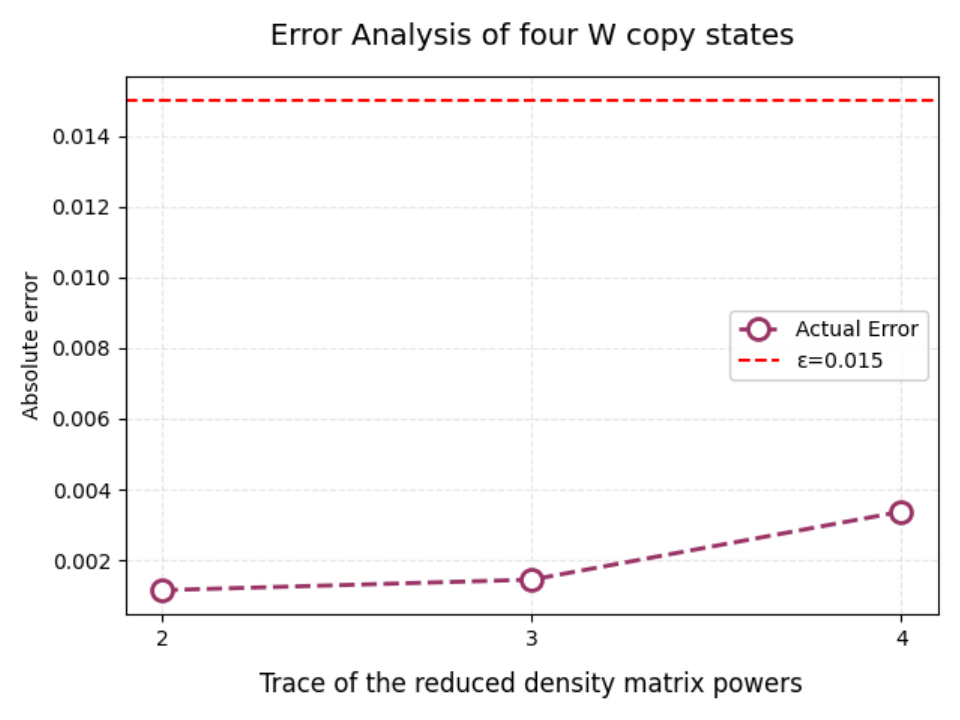}
	\label{fig19c}
}
\subfigure[]{
	\includegraphics[width=0.5\linewidth]{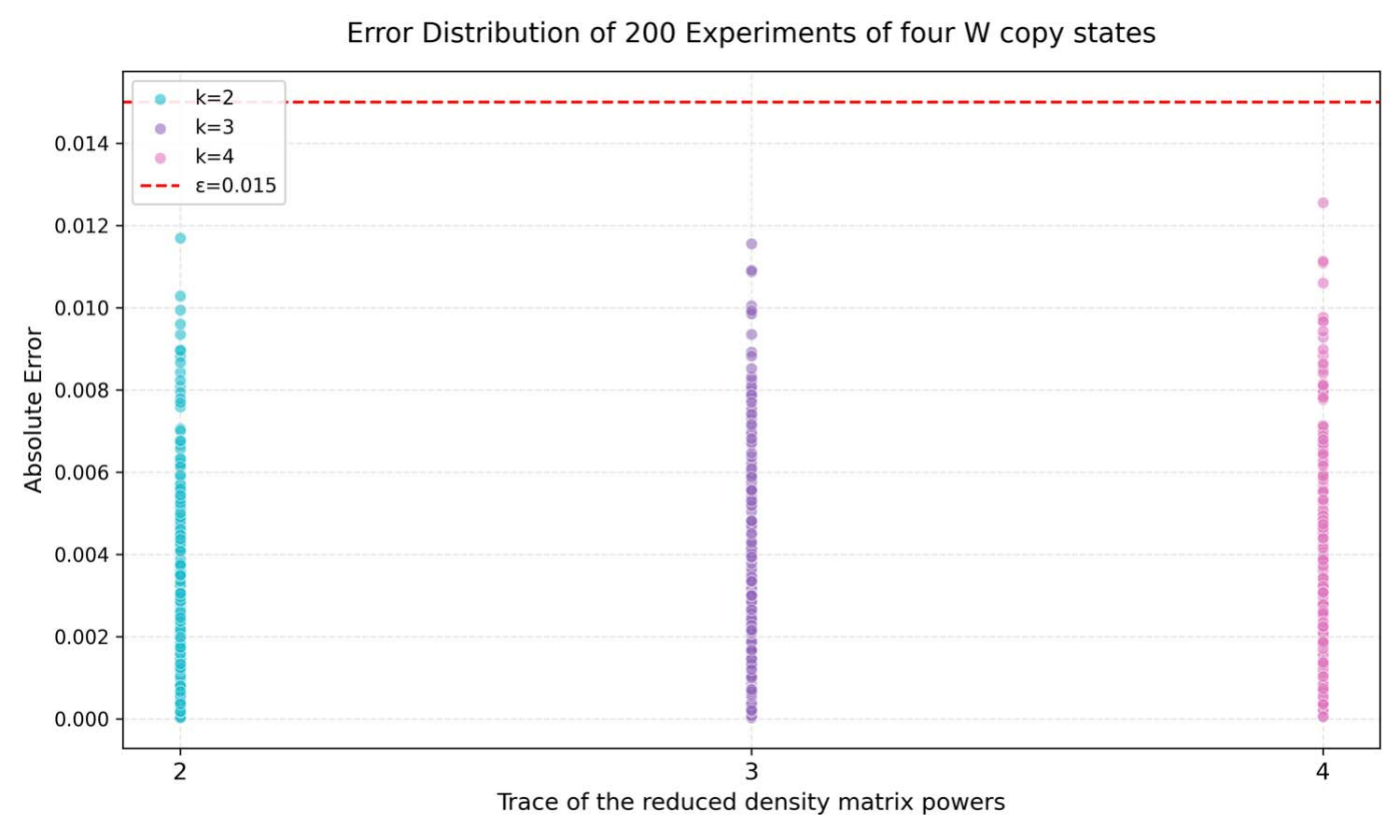}
	\label{fig19d}
}
\subfigure[]{
	\includegraphics[width=0.45\linewidth]{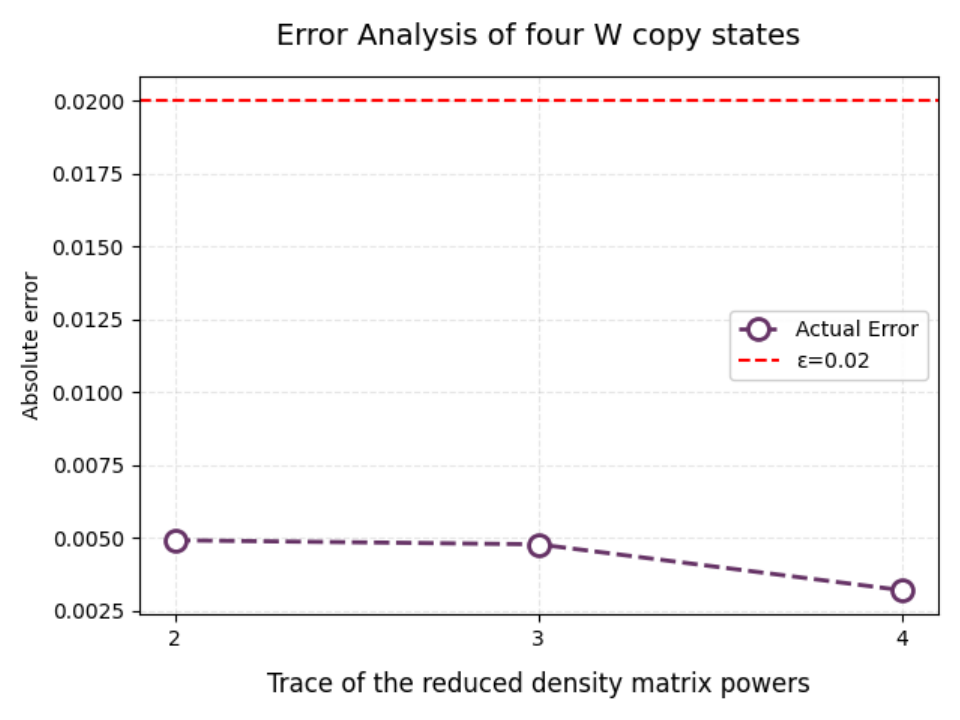}
	\label{fig19e}
}
\subfigure[]{
	\includegraphics[width=0.5\linewidth]{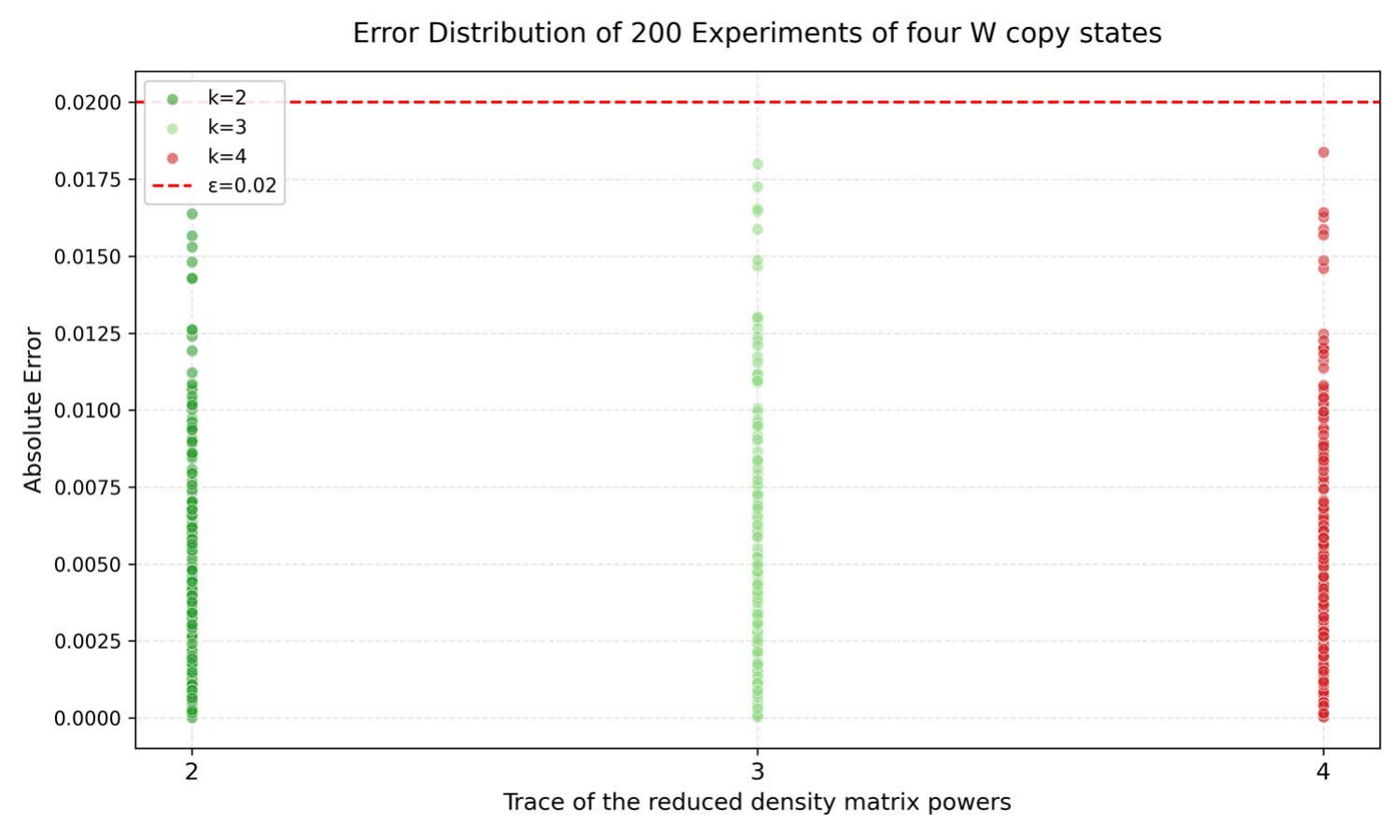}
	\label{fig19f}
}
	\caption{\textbf{Error analysis of the estimation for trace of reduced density matrix of four copy W states under different settings of maximum allowable error and confidence levels.} The red line represents the set maximum allowable error. (a)
		The absolute errors of estimation for trace of reduced density matrix powers when the circuit is run                  $\frac{2}{\epsilon^2} \ln\left(\frac{6}{\delta}\right)$ times  under settings of maximum allowable error $\epsilon=0.01$ and confidence level $\delta=0.05$.  (b) Error distribution of estimation for trace of reduced density matrix powers in 200 experiments with each experiment run $\frac{2}{\epsilon^2} \ln\left(\frac{6}{\delta}\right)$ times. (c) The absolute errors of estimation for trace of reduced density matrix powers when the circuit is run $\frac{2}{\epsilon^2} \ln\left(\frac{6}{\delta}\right)$ times  under settings of maximum allowable error $\epsilon=0.015$ and confidence level $\delta=0.06$. (d) Error distribution of estimation for trace of reduced density matrix powers in 200 experiments with each experiment run $\frac{2}{\epsilon^2} \ln\left(\frac{6}{\delta}\right)$ times. (e) The absolute errors of estimation for trace of reduced density matrix powers when the circuit is run $\frac{2}{\epsilon^2} \ln\left(\frac{6}{\delta}\right)$ times  under settings of maximum allowable error $\epsilon=0.02$ and confidence level $\delta=0.08$. (f) Error distribution of estimation for trace of reduced density matrix powers in 200 experiments with each experiment run $\frac{2}{\epsilon^2} \ln\left(\frac{6}{\delta}\right)$ times.}
\end{figure*}

  To validate the effectiveness of the proposed method, we conducted numerical simulations for two typical maximally entangled states, the GHZ state and the W state, focusing on the error characteristics of estimating traces of reduced density matrix powers using a four-copy quantum circuit. 
  The experiment was set with different maximum allowable errors and confidence levels. According to the sampling complexity theory of Theorem 6, the circuit allows  $\frac{2}{\epsilon^2} \ln\left(\frac{6}{\delta}\right)$ runs to obtain statistical results. Figures \ref{fig18a}, \ref{fig18c}, and \ref{fig18e} show the absolute errors between the estimated values and theoretical values of the traces of the 2nd, 3rd, and 4th powers of the reduced density matrix, respectively. These estimations were obtained by running the quantum circuit with four GHZ copy states $\frac{2}{\epsilon^2} \ln\left(\frac{6}{\delta}\right)$ times under the settings of $\epsilon=0.01, \delta=0.05; \epsilon=0.015, \delta=0.06; and \epsilon=0.02, \delta=0.08,$ respectively.
  Figures \ref{fig18b}, \ref{fig18d}, and \ref{fig18f} display the absolute error distributions between the estimated and theoretical values of the traces of the 2nd, 3rd, and 4th powers of the reduced density matrix. These are derived from 200 experiments, where each experiment run the quantum circuit with four GHZ copy states $\frac{2}{\epsilon^2} \ln\left(\frac{6}{\delta}\right)$ times. Specifically, these figures correspond to the error distributions of 200 repetitions of the experiments shown in Figures \ref{fig18a}, \ref{fig18c}, and \ref{fig18e}.
	Figures \ref{fig19a}, \ref{fig19c}, and \ref{fig19e} illustrate the absolute errors between the estimated and theoretical values of the traces of the 2nd, 3rd, and 4th powers of the reduced density matrix, respectively. These were obtained by running the quantum circuit with four W copy states $\frac{2}{\epsilon^2} \ln\left(\frac{6}{\delta}\right)$ times under the settings of $\epsilon=0.01, \delta=0.05; \epsilon=0.015, \delta=0.06; and \epsilon=0.02, \delta=0.08,$ respectively.
	Figures \ref{fig19b}, \ref{fig19d}, and \ref{fig19f} present the absolute error comparisons between the estimated and theoretical values of the traces of the 2nd, 3rd, and 4th powers of the reduced density matrix from 200 experiments. Each experiment runs the quantum circuit with four W copy states $\frac{2}{\epsilon^2} \ln\left(\frac{6}{\delta}\right)$ times, corresponding to the error distributions of 200 repetitions of the experiments in Figures \ref{fig19a}, \ref{fig19c}, and \ref{fig19e}.\par 
The experimental results show that in the above experiments, the estimation errors of the trace of the reduced density matrix power obtained by running the quantum circuits of four copy quantum states (GHZ states and W states) with $O\left(\frac{1}{\epsilon^2}\log(\frac{n}{\delta})\right)$ experiments are less than the maximum allowable error $\epsilon$ with a very high probability $1-\delta$. This fully verifies the conclusion of Theorem 6, i.e., under the given maximum allowable error and confidence level, by conducting $O\left(\frac{1}{\epsilon^2}\log(\frac{n}{\delta})\right)$ experiments, we can estimate the trace of the reduced density matrix power with an error less than the allowable error $\epsilon$ with a very high probability.\par 
Experimental results demonstrate that multiple power traces can be efficiently and reliably estimated through multiple measurements using a single quantum circuit.\\\par

\section{VII. Conclusions}
	The traces of the reduced density matrix powers play a crucial role in the study of quantum systems and have many important applications. In this Letter, we proposed a unified quantum framework to estimate $\left\lbrace {\rm tr}(\rho_A^k)\right\rbrace _{k=1}^n$ via a single circuit, resolving the redundancy of prior multi-circuit approaches.
	 Specifically, we derive trace expressions for the powers of the reduced density matrix from the measured probabilities of controlled SWAP test, which corrects errors in \cite{Jin}. Estimating the traces of the $2$nd to the $n$th power of the reduced density matrix, we propose two algorithms for estimating them. The first algorithm demands $n$ quantum copy states within a single quantum circuit. By contrast, the second algorithm combines quantum circuits with classical iteration. This approach only requires $r$ quantum copy states, however, it relies on having prior knowledge of the rank of the quantum state. Through the Hoeffding inequality, we rigorously derive a unified upper bound on measurements, which reveals the method’s efficiency and high reliability with finite samples.
	 In addition, we explore various applications including the estimation of nonlinear functions (such as the trace of exponential function, Von Neumann entropy, and a cost function applied in the preparation of variational quantum Gibbs states) and the representation of entanglement measures. We  perform numerical simulations for two maximally entangled states.\\ 
	\indent Future research could focus on deploying the framework in quantum machine learning for kernel estimation and in variational quantum algorithm for Gibbs state preparation. Simultaneously, it is crucial to delve into more significant applications founded on the traces of reduced density matrix powers.
	
	\section*{\bf Acknowledgments} This work is supported by the Shandong Provincial Natural Science Foundation for Quantum Science ZR2021LLZ002, No.ZR2020LLZ003, the Fundamental Research Funds for the Central Universities No.22CX03005A, the NSFC under Grant Nos. 12075159 and 12175147, Beijing Natural Science Foundation (Z190005), and the Academician Innovation Platform of Hainan Province.

\appendix	 
	 
\section{APPENDIX}
In this section, we will present the proof of the lemmas and theorems in the main text. For the convenience of the reader, we restate here the statements of the lemmas and theorems.
{\subsection{A. The proof of lemmas}}
 
\textbf{Lemma 1.} \emph{ Consider $k$-partite quantum systems $\otimes_{i=1}^{k}\mathcal{H}_i$. Let $\rho_i\in \mathcal{H}_i$ be $k$ identical quantum states, i.e., $\rho_i=\rho \ (i=1,2,\cdots ,k)$ and let $S_i:\mathcal{H}_i\otimes\mathcal{H}_{i+1}\longrightarrow \mathcal{H}_i\otimes\mathcal{H}_{i+1}$ denotes the SWAP operator acting on $\mathcal{H}_i$ and $\mathcal{H}_{i+1}$.
	Then we can obtain
	$${\rm tr}(S_{i_1}S_{i_2}\cdots S_{i_{k-1}}\otimes_{i=1}^{k}\rho_i)={\rm tr}( \prod_{i=1}^{k}\rho_i)={\rm tr}( \rho^k), $$
where $i_1,i_2,\cdots,i_{k-1}$ is a rearrangement of $ 1,2,\cdots,k-1$.}\\
\\
\emph{Proof.} Let $\rho_i \ (i=1,2,\cdots ,k)$
be $k$ identical quantum states with spectral decomposition $$\rho_{i}=\sum_{j_i}a_{j_i}|j_i\rangle\langle j_i|,$$ where $\sum_{j_i}a_{j_i}=1$.
First, we consider a special case of $S_{i_1}S_{i_2}\cdots S_{i_{k-1}}$, i.e., $S_{1}S_{2}\cdots S_{k-1}$, then we have 
\begin{widetext}
\begin{align*}
	&{\rm tr}(S_{1}S_{2}\cdots S_{k-1}\rho_1\otimes\rho_2\otimes\cdots\otimes \rho_{k})\notag\\
	=&{\rm tr}(S_{1}S_{2}\cdots S_{k-1}\sum_{j_1}a_{j_1}|j_1\rangle\langle j_1|\otimes\sum_{j_2}a_{j_2}|j_2\rangle\langle j_2|\otimes\cdots\otimes\sum_{j_{k}}a_{j_{k}}|j_{k}\rangle\langle j_{k}|)\notag\\
	=&{\rm tr}(\sum_{j_1,j_2,\cdots,j_{k}}a_{j_1}a_{j_2}\cdots a_{j_{k}}S_{1}S_{2}\cdots S_{k-1}|j_1\rangle\langle j_1|\otimes|j_2\rangle\langle j_2|\otimes\cdots\otimes|j_{k}\rangle\langle j_{k}|)\notag\\
	=&\sum_{j_1,j_2,\cdots,j_{k}}a_{j_1}a_{j_2}\cdots a_{j_{k}}{\rm tr}(S_{1}S_{2}\cdots S_{k-1}|j_1\rangle\langle j_1|\otimes|j_2\rangle\langle j_2|\otimes\cdots\otimes|j_{k}\rangle\langle j_{k}|).
\end{align*}
	\end{widetext}\par 

Since $S_i$ is an operator acting on the adjacent subspaces $\mathcal{H}_i\otimes\mathcal{H}_{i + 1}$, according to the properties of tensor product and trace, we have
\begin{align*}
&{\rm tr}(S_{1}S_{2}\cdots S_{k-1}|j_1\rangle\langle j_1|\otimes|j_2\rangle\langle j_2|\otimes\cdots\otimes|j_{k}\rangle\langle j_{k}|)\\
=&{\rm tr}(|j_{k}\rangle\langle j_{1}|\otimes|j_{1}\rangle\langle j_{2}|\otimes\cdots\otimes|j_{{k-1}}\rangle\langle j_{k}|)\\
=&{\rm tr}(|j_{k}\rangle\langle j_{1}|){\rm tr}(|j_{1}\rangle\langle j_{2}|)\cdots{\rm tr}(|j_{{k-1}}\rangle\langle j_{k}|),
\end{align*}
where for the second equation we have used the expression ${\rm tr}(A\otimes B)={\rm tr}A {\rm tr}B$
 Note that $${\rm tr}(|j_{m}\rangle\langle j_{n}|)=\delta_{j_{m},j_{n}}.$$
Therefore we have
\begin{align*}
&\sum_{j_1,j_2,\cdots,j_{k}}a_{j_1}a_{j_2}\cdots a_{j_{k}}{\rm tr}(|j_{k}\rangle\langle j_{1}|){\rm tr}(|j_{1}\rangle\langle j_{2}|)\\
&\cdots{\rm tr}(|j_{k-1}\rangle\langle j_{k }|)\\
=&\sum_{j_1,j_2,\cdots,j_{k}}a_{j_1}a_{j_2}\cdots a_{j_{k}}\delta_{j_{k},j_{1}}\delta_{j_{1},j_{2}}\cdots\delta_{j_{k-1},j_{k}}\\
=&\sum_{j_{1}}a_{j_{1}}^{k}={\rm tr}(\prod_{i = 1}^{k}\rho_i)={\rm tr}(\rho^k).
\end{align*}
Next we consider the general case $S_{i_1}S_{i_2}\cdots S_{i_{k-1}}$ with an arbitrary order. Actually $S_{i_1}S_{i_2}\cdots S_{i_{k-1}}$ is a permutation of $S_{1}S_{2}\cdots S_{k-1}$ corresponding to an element of the permutation group. Then there exists a corresponding permutation matrix $P$, satisfying $$S_{i_1}S_{i_2}\cdots S_{i_{k-1}}=P^{-1}(S_{1}S_{2}\cdots S_{k-1})P,$$ where the permutation matrix $P$ is orthogonal. Then we have
\begin{align*}
	&{\rm tr}(S_{i_1}S_{i_2}\cdots S_{i_{k-1}}\rho_1\otimes\rho_2\otimes\cdots\otimes \rho_{k})\notag\\
  =&{\rm tr}(P^{-1}(S_{1}S_{2}\cdots S_{k-1})P\rho_1\otimes\rho_2\otimes\cdots\otimes \rho_{k})\\
  =&{\rm tr}(S_{1}S_{2}\cdots S_{k-1}P(\rho_1\otimes\rho_2\otimes\cdots\otimes \rho_{k})P^{-1})\\
  =&{\rm tr}(S_{1}S_{2}\cdots S_{k-1}P(\rho\otimes\rho\otimes\cdots\otimes \rho)P^{-1}).
\end{align*}
Since $P$ is an orthogonal permutation matrix, we get 
\begin{align*}
	&{\rm tr}(S_{i_1}S_{i_2}\cdots S_{i_{k-1}}\rho_1\otimes\rho_2\otimes\cdots\otimes \rho_{k})\notag\\
	=&{\rm tr}(S_{1}S_{2}\cdots S_{k-1}\rho_1\otimes\rho_2\otimes\cdots\otimes \rho_{k})={\rm tr}(\rho^k)
\end{align*}$\hfill\Box$ \par 
Since $i_1,i_2,\cdots,i_{k-1}$ is a rearrangement of $ 1,2,\cdots,k-1$, in fact, when the operator $S_{i_1}S_{i_2}\cdots S_{i_{k-1}}$ acts on multiple copy quantum states, the successive action of all operators is equivalent to a shift in the relative positions of all the original quantum states. For example, when $k=5$, if $i_1=2,i_2=4,i_3=1,i_4=3$, we have
\begin{align*}
	&{\rm tr}(S_2S_4S_1S_3\rho_1\otimes\rho_2\otimes\cdots\otimes\rho_5)\\
	=&\sum_{j_1,j_2,\cdots,j_{5}}a_{j_1}a_{j_2}\cdots a_{j_{5}}{\rm tr}(S_2S_4S_1S_3|j_{1}\rangle\langle j_{1}|\otimes|j_{2}\rangle\langle j_{2}|\\
	&\otimes\cdots\otimes|j_{5}\rangle\langle j_{5}|)\\
	=&\sum_{j_1,j_2,\cdots,j_{5}}a_{j_1}a_{j_2}\cdots a_{j_{5}}{\rm tr}(|j_{2}\rangle\langle j_{1}|\otimes|j_{4}\rangle\langle j_{2}|\otimes|j_{1}\rangle\langle j_{3}|\\
	&\otimes|j_{5}\rangle\langle j_{4}|\otimes|j_{3}\rangle\langle j_{5}|)\\
	=&\sum_{j_1,j_2,\cdots,j_{5}}a_{j_1}a_{j_2}\cdots a_{j_{5}}\delta_{j_{2},j_{1}}\delta_{j_{4},j_{2}}\delta_{j_{1},j_{3}}\delta_{j_{5},j_{4}}\delta_{j_{3},j_{5}}\\
	&\sum_{j_1}a_{j_1}^5={\rm tr}(\rho^5).
\end{align*}$\hfill\Box$\\ \par 
 
\textbf{Lemma 2.} \emph{ Consider $k$-partite quantum systems $\otimes_{i=1}^{k}\mathcal{H}_i$. Let $S_i$ and $S_j$ be two nonadjacent SWAP operators $(i,j\in\left\lbrace 1,2,\cdots,k-1\right\rbrace ,i<j,$ and $i\ne j-1)$ and $\rho_i=\rho \in \mathcal{H}_i, (i=1,2,\cdots,k)$. We can obtain 
\begin{align*}
	{\rm tr}(S_iS_j\otimes_{i=1}^{k}\rho_i)=({\rm tr}\rho^2)^2.
\end{align*}
}\\
\emph{Proof.} We have
\begin{align*}
&{\rm tr}(S_iS_j\otimes_{i=1}^{k}\rho_i)\\
=&{\rm tr}(S_iS_j\rho_1\otimes\cdots\otimes\rho_i\otimes\cdots\otimes\rho_j\otimes\rho_{j+1}\cdots\otimes\rho_{k})\\
=&{\rm tr}(\rho_1\otimes\cdots\otimes\rho_{i-1}\otimes S_i(\rho_i\otimes\rho_{i+1})\otimes\rho_{i+2}\otimes\cdots\\&\otimes\rho_{j-1}\otimes S_{j}(\rho_j\otimes\rho_{j+1})\otimes\rho_{j+2}\otimes\cdots\otimes\rho_{k})\\
=&{\rm tr}(\rho_1)\cdots{\rm tr}(\rho_{i-1}) {\rm tr}(S_i(\rho_i\otimes\rho_{i+1})){\rm tr}(\rho_{i+2})\cdots\\&{\rm tr}(\rho_{j-1})
{\rm tr}(S_{j}(\rho_j\otimes\rho_{j+1})){\rm tr}(\rho_{j+2})\cdots {\rm tr}(\rho_{k})\notag\\
=&{\rm tr}(S_i(\rho_i\otimes\rho_{i+1}))
{\rm tr}(S_{j}(\rho_j\otimes\rho_{j+1}))
\\
=&({\rm tr}\rho^2)^2,
\end{align*}
where for the fifth equation we have used Lemma 1.
$\hfill\Box$\\ \par 
  
\textbf{Lemma 3.} \emph{For $k$-partite quantum systems $\otimes_{i=1}^{k}\mathcal{H}_i$, if there are $k$ copies of the same quantum state $\rho=\rho_i\in\mathcal{H}_{i},(i=1,2,\cdots,k)$, then we have 
\begin{align*}
	&{\rm tr}(\bar{S}_1 \bar{S}_2\cdots\bar{S}_t\otimes_{i=1}^{k}\rho_i)={\rm tr}\rho^{s_1+1}{\rm tr}\rho^{s_2+1}\cdots{\rm tr}\rho^{s_t+1},
\end{align*}
where 
\begin{align*}
	&\bar{S}_1=S_{j_1^{(1)}}S_{j_2^{(1)}}\cdots S_{j_{s_1}^{(1)}},\\
	&\bar{S}_2=S_{j_1^{(2)}}S_{j_2^{(2)}}\cdots S_{j_{s_2}^{(2)}},\cdots,\\ 
    &\bar{S}_t=S_{j_1^{(t)}}S_{j_2^{(t)}}\cdots S_{j_{s_t}^{(t)}}
\end{align*}
 are a series of operators composed of the products of adjacent SWAP operators, and the SWAP operators within operators $\bar{S}_m \ (m=1,2,\cdots, t)$ and those within the other operators  $\bar{S}_n \ (n=1,2,\cdots, t, \ n\ne m)$ have no intersection and are not adjacent. And $s_1+s_2+\cdots+s_t\le k-t$, where $s_m$ denotes the number of all SWAP operators in $\bar{S}_m$.}\\
\\
\emph{Proof.} According to the tensor product property, we have
\begin{align*}
&{\rm tr}(\bar{S}_1 \bar{S}_2\cdots\bar{S}_t\otimes_{i=1}^{k}\rho_i)\\=&{\rm tr}\left( (S_{j_1^{(1)}}S_{j_2^{(1)}}\cdots S_{j_{s_1}^{(1)}}) (S_{j_1^{(2)}}S_{j_2^{(2)}}\right.\\
&\left.\cdots S_{j_{s_2}^{(2)}})\cdots(S_{j_1^{(t)}}S_{j_2^{(t)}}\cdots S_{j_{s_t}^{(t)}}) \otimes_{i={1}}^{k}\rho_i\right) \\=&{\rm tr}\left( S_{j_1^{(1)}}S_{j_2^{(1)}}\cdots S_{j_{s_1}^{(1)}}(\underbrace{\rho\otimes\rho\otimes\cdots\otimes\rho}_{s_1+1})\right.\\
&\left.\otimes S_{j_1^{(2)}}S_{j_2^{(2)}}\cdots S_{j_{s_2}^{(2)}}(\underbrace{\rho\otimes\rho\otimes\cdots\otimes\rho}_{s_2+1})\right.\\
&\left.\cdots\otimes  S_{j_1^{(t)}}S_{j_2^{(t)}}\cdots S_{j_{s_t}^{(t)}} (\underbrace{\rho\otimes\rho\otimes\cdots\otimes\rho}_{s_t+1})\right.\\
&\left. \otimes(\underbrace{\rho\otimes\rho\otimes\cdots\otimes\rho}_{k-(s_1+s_2+\cdots+s_t)-t})\right) \\=&{\rm tr}\left( S_{j_1^{(1)}}S_{j_2^{(1)}}\cdots S_{j_{s_1}^{(1)}}(\underbrace{\rho\otimes\rho\otimes\cdots\otimes\rho}_{s_1+1})\right) \\&{\rm tr} \left( S_{j_1^{(2)}}S_{j_2^{(2)}}\cdots S_{j_{s_2}^{(2)}}(\underbrace{\rho\otimes\rho\otimes\cdots\otimes\rho}_{s_2+1})\right) \\&\cdots{\rm tr}\left( S_{j_1^{(t)}}S_{j_2^{(t)}}\cdots S_{j_{s_t}^{(t)}} (\underbrace{\rho\otimes\rho\otimes\cdots\otimes\rho}_{s_t+1})\right)\\=& 
{\rm tr}\rho^{s_1+1}{\rm tr}\rho^{s_2+1}\cdots{\rm tr}\rho^{s_t+1},
\end{align*}
where for the last equation we have used Lemma 1.$\hfill\Box$\\ \par 

\textbf{Lemma 4.} \emph{Consider $k$-partite quantum systems  $\otimes_{i=1}^{k}\mathcal{H}_i$, let 
\begin{align*}
	&\hat{S}_1=S_{i_1}S_{i_2}\cdots S_{i_m}S_{i_1},\\
	&\hat{S}_2=S_{i_1}\cdots S_{i_{s-1}}S_{i_1}S_{i_{s}}\cdots S_{i_m},\notag\\
	&\hat{S}_3=S_{i_1}\cdots S_{i_{s-1}}S_{i_{s}}\cdots S_{i_m}S_{i_{s}},\\
	&\hat{S}_4=S_{i_1}\cdots S_{i_{s-1}}S_{i_{s}}\cdots S_{i_{t-1}}S_{i_{s}}S_{i_{t}}\cdots S_{i_m}
\end{align*}
be operators composed of some sequentially and continuously adjacent SWAP operators $S_j$ and only one SWAP operator appears twice are repeated in these operators and $m\le k-1$. The indices $i_1, i_2, \cdots, i_m$ denote a sequence of consecutive integers corresponding to adjacent SWAP operators, satisfying $1 \leq i_1 < i_2 < \cdots < i_m \leq k-1$. If there are $k$ copies of the same quantum state $\rho=\rho_i\in\mathcal{H}_{i},(i=1,2,\cdots,k)$, then 
\begin{align*}
	&{\rm tr}(\hat{S}_1\otimes_{i=1}^{k}\rho_i)= {\rm tr}(\hat{S}_2\otimes_{i=1}^{k}\rho_i)={\rm tr}(\hat{S}_3\otimes_{i=1}^{k}\rho_i)\\
	=&{\rm tr}(\hat{S}_4\otimes_{i=1}^{k}\rho_i)={\rm tr}\rho^m.
\end{align*}
}\\
\emph{Proof.} Consider a quantum state with the following form of spectral decomposition
$$\rho_{i}=\sum_{j_i}a_{j_i}|j_i\rangle\langle j_i|.$$ 
First, we have
\begin{align*}
&{\rm tr}( \hat{S}_1\otimes_{i=1}^{k}\rho_i)\\=&{\rm tr}\left( (S_{i_1}S_{i_2}\cdots S_{i_m}S_{i_1}\otimes_{i=i_1}^{i_{m}+1}\rho_i)\otimes\underbrace{\rho\otimes\rho\otimes\cdots\otimes\rho}_{k-m-1}\right)\\
=&{\rm tr}(S_{i_1}S_{i_2}\cdots S_{i_m}S_{i_1}\otimes_{i=i_1}^{i_{m}+1}\rho_i) \notag\\
=&\sum_{j_{i_1},j_{i_2},\cdots,j_{i_{m}+1}}a_{j_{i_1}}a_{j_{i_2}}\cdots a_{j_{i_{m}+1}}{\rm tr}(S_{i_1}S_{i_2}\cdots S_{i_m}S_{i_1}\\&|j_{i_1}\rangle\langle j_{i_1}|\otimes|j_{i_2}\rangle\langle j_{i_2}|\otimes\cdots\otimes|j_{i_m}\rangle\langle j_{i_{m}}|\otimes|j_{i_{m}+1}\rangle\langle j_{i_{m}+1}|).
\end{align*}
Now, we consider the action of the SWAP operators. When applying the SWAP operators successively, we can get
\begin{align*}
&{\rm tr}(S_{i_1}S_{i_2}\cdots S_{i_m}S_{i_1}|j_{i_1}\rangle\langle j_{i_1}|\otimes|j_{i_2}\rangle\langle j_{i_2}|\otimes\cdots\\&\otimes|j_{i_{m}+1}\rangle\langle j_{i_{m}+1}|)\\
=&{\rm tr}(|j_{i_{m} + 1}\rangle\langle j_{i_1}|\otimes|j_{i_2}\rangle\langle j_{i_2}|\otimes|j_{i_1}\rangle\langle j_{i_3}|\otimes|j_{i_3}\rangle\langle j_{i_4}|\\&\otimes|j_{i_4}\rangle\langle j_{i_5}|\otimes\cdots\otimes|j_{i_m}\rangle\langle j_{i_{m}+1}|)
\end{align*}
Then, we can separate the traces of each tensor product component. So we have 
\begin{align*}
&{\rm tr}(\hat{S}_1\otimes_{i=1}^{k}\rho_i)\\=&
\sum_{j_{i_1},j_{i_2},\cdots,j_{i_{m}+1}}a_{j_{i_1}}a_{j_{i_2}}\cdots a_{j_{i_{m}+1}}{\rm tr}((|j_{i_{m }+1}\rangle\langle j_{i_1}|){\rm tr}(|j_{i_2}\rangle\langle j_{i_2}|)\\&{\rm tr}(|j_{i_1}\rangle\langle j_{i_3}|){\rm tr}(|j_{i_3}\rangle\langle j_{i_4}|)\cdots{\rm tr}(|j_{i_m}\rangle\langle j_{i_{m}+1}|)\\=&
\sum_{j_{i_1},j_{i_3},\cdots,j_{i_{m}+1}}a_{j_{i_1}}a_{j_{i_3}}\cdots a_{j_{i_{m}+1}}\delta_{j_{i_{m }+1},j_{i_1}}\delta_{j_{i_{1}},j_{i_3}}\\
&\delta_{j_{i_{3}},j_{i_4}}\cdots\delta_{j_{i_{m}},j_{i_{m}+1}}\\=&\sum_{j_{i_1}}a_{j_{i_1}}^m={\rm tr}\rho^m.
\end{align*}
Similar to the previous case, we have
	\begin{widetext}
\begin{align*}
	&{\rm tr}(\hat{S}_2\otimes_{i=1}^{k}\rho_i)\\=&{\rm tr}\left( (S_{i_1}\cdots S_{i_{s-1}}S_{i_1}S_{i_s}\cdots S_{i_m}\otimes_{i={i_1}}^{{i_{m}+1}}\rho_i)\otimes\underbrace{\rho\otimes\rho\otimes\cdots\otimes\rho}_{k-m-1}\right)\notag\\
	=&{\rm tr}(S_{i_1}\cdots S_{i_{s-1}}S_{i_1}S_{i_s}\cdots S_{i_m}\otimes_{i={i_1}}^{{i_{m}+1}}\rho_i) \\
	=&\sum_{j_{i_1},\cdots,j_{i_{m}+1}}a_{j_{i_1}}\cdots a_{j_{i_{m}+1}}{\rm tr}(S_{i_1}\cdots S_{i_{s-1}}S_{i_1}S_{i_s}\cdots S_{i_m}|j_{i_1}\rangle\langle j_{i_1}|\otimes|j_{i_2}\rangle\langle j_{i_2}|\otimes\cdots\otimes|j_{i_m}\rangle\langle j_{i_m}|\otimes|j_{i_{m}+1}\rangle\langle i_{i_{m}+1}|)\notag\\
	=&\sum_{j_{i_1},\cdots,j_{i_{m}+1}}a_{j_{i_1}}\cdots a_{j_{i_{m}+1}}{\rm tr}(|j_{i_{m}+1}\rangle\langle j_{i_1}|\otimes|j_{i_2}\rangle\langle j_{i_2}|\otimes|j_{i_1}\rangle\langle j_{i_3}|\otimes|j_{i_3}\rangle\langle j_{i_4}|\otimes|j_{i_4}\rangle\langle j_{i_5}|\otimes\cdots\otimes|j_{i_m}\rangle\langle j_{i_{m}+1}|)\\=&
	\sum_{j_{i_1},j_{i_3},\cdots,j_{i_{m}+1}}a_{j_{i_1}}a_{j_{i_3}}\cdots a_{j_{i_{m}+1}}\delta_{j_{i_{m}+1},j_{i_1}}\delta_{j_{i_{1}},j_{i_3}}\delta_{j_{i_{3}},j_{i_4}}\cdots\delta_{j_{i_{m}},j_{i_{m}+1}}\\=&\sum_{j_{i_1}}a_{j_{i_1}}^m={\rm tr}\rho^m,\\
	\\
	&{\rm tr}(\hat{S}_3\otimes_{i=1}^{k}\rho_i)\\=&{\rm tr}\left( (S_{i_1}\cdots S_{i_s}S_{i_{s+1}}\cdots S_{i_m} S_{i_s}\otimes_{i={i_1}}^{{i_{m}+1}}\rho_i)\otimes\underbrace{\rho\otimes\rho\otimes\cdots\otimes\rho}_{k-m-1}\right)\notag\\
	=&{\rm tr} (S_{i_1}\cdots S_{i_s}S_{i_{s+1}}\cdots S_{i_m} S_{i_s}\otimes_{i={i_1}}^{{i_{m}+1}}\rho_i)\notag\\
	=&\sum_{j_{i_1},\cdots,j_{i_{m}+1}}a_{j_{i_1}}\cdots a_{j_{i_{m}+1}}{\rm tr}(S_{i_1}\cdots S_{i_{s-1}}S_{i_s}S_{i_{s+1}}\cdots S_{i_m} S_{i_s}|j_{i_1}\rangle\langle j_{i_1}|\otimes|j_{i_2}\rangle\langle j_{i_2}|\otimes\cdots\otimes|j_{i_m}\rangle\langle j_{i_m}|\otimes|j_{i_{m}+1}\rangle\langle i_{i_{m}+1}|)\notag\\
	=&\sum_{j_{i_1},\cdots,j_{i_{m}+1}}a_{j_{i_1}}\cdots a_{j_{i_{m}+1}}{\rm tr}(|j_{i_{m}+1}\rangle\langle j_{i_1}|\otimes|j_{i_1}\rangle\langle j_{i_2}|\otimes\cdots\otimes|j_{i_{s-1}}\rangle\langle j_{i_s}|\otimes|j_{i_{s+1}}\rangle\langle j_{i_{s+1}}|\otimes|j_{i_s}\rangle\langle j_{i_{s+2}}|\\&
	\otimes|j_{i_{s+2}}\rangle\langle j_{i_{s+3}}|\otimes\cdots\otimes|j_{i_{m}}\rangle\langle j_{i_{m}+1}|)\notag\\	=&\sum_{j_{i_1},\cdots,j_{i_{s}},j_{i_{s+2}},\cdots,j_{i_{m}+1}}a_{j_{i_1}}\cdots a_{j_{i_s}}a_{j_{i_{s+2}}}\cdots a_{j_{i_{m}+1}}\delta_{j_{i_{m}+1},j_{i_1}}\delta_{j_{i_{1}},j_{i_2}}\cdots\delta_{j_{i_{s-1}},j_{i_s}}\delta_{j_{i_{s}},j_{i_{s+2}}}\delta_{j_{i_{s+2}},j_{i_{s+3}}}\cdots\delta_{j_{i_{m}},j_{i_{m}+1}}
	\\
	=&=\sum_{j_{i_1}}a_{j_{i_1}}^m={\rm tr}\rho^m,\\
	\\
	&{\rm tr}(\hat{S}_4\otimes_{i=1}^{k}\rho_i)\\=&{\rm tr}\left( (S_{i_1}\cdots S_{i_s}S_{i_{s+1}}\cdots S_{i_{t-1}} S_{i_s}S_{i_{t}}\cdots S_{i_{m}}\otimes_{i={i_1}}^{{i_{m}+1}}\rho_i)\otimes\underbrace{\rho\otimes\rho\otimes\cdots\otimes\rho}_{k-m-1}\right)\notag\\
	\\=&{\rm tr}(S_{i_1}\cdots S_{i_s}S_{i_{s+1}}\cdots S_{i_{t-1}} S_{i_s}S_{i_{t}}\cdots S_{i_{m}}\otimes_{i={i_1}}^{{i_{m}+1}}\rho_i)\notag\\
	=&\sum_{j_{i_1},\cdots,j_{i_{m}+1}}a_{j_{i_1}}\cdots a_{j_{i_{m}+1}}{\rm tr}(S_{i_1}\cdots S_{i_{s-1}}S_{i_s}S_{i_{s+1}}\cdots S_{i_{t-1}} S_{i_s}S_{i_{t}}\cdots S_{i_{m}}|j_{i_1}\rangle\langle j_{i_1}|\otimes|j_{i_2}\rangle\langle j_{i_2}|\otimes\cdots\otimes|j_{i_m}\rangle\langle j_{i_m}|\otimes|j_{i_{m}+1}\rangle\langle i_{i_{m}+1}|)\notag\\
	=&\sum_{j_{i_1},\cdots,j_{i_{m}+1}}a_{j_{i_1}}\cdots a_{j_{i_{m}+1}}{\rm tr}(|j_{i_{m}+1}\rangle\langle j_{i_1}|\otimes|j_{i_1}\rangle\langle j_{i_2}|\otimes\cdots\otimes|j_{i_{s-1}}\rangle\langle j_{i_s}|\otimes|j_{i_{s+1}}\rangle\langle j_{i_{s+1}}|\otimes|j_{i_s}\rangle\langle j_{i_{s+2}}|
	\\&\otimes|j_{i_{s+2}}\rangle\langle j_{i_{s+3}}|\otimes\cdots\otimes|j_{i_m}\rangle\langle j_{i_{m}+1}|)\notag\\
	=&\sum_{j_{i_1},\cdots,j_{i_s}, j_{i_{s+2}},\cdots ,j_{i_{m}+1}}a_{j_{i_1}}\cdots a_{j_{i_s}}a_{j_{i_{s+2}}}\cdots a_{j_{i_{m}+1}}\delta_{j_{i_{m}+1},j_{i_1}}\delta_{j_{i_{1}},j_{i_2}}\cdots\delta_{j_{i_{s-1}},j_{i_s}}\delta_{j_{i_{s}},j_{i_{s+2}}}\delta_{j_{i_{s+2}},j_{i_{s+3}}}\cdots\delta_{j_{i_{m}},j_{i_{m}+1}}\\
	=&\sum_{j_{i_1}}a_{j_{i_1}}^m={\rm tr}\rho^m.
\end{align*}
	\end{widetext}
$\hfill\Box$\par

In order to give the circuit evolution process, the following lemma is necessary.\\\par 

\textbf{Lemma 5.} \emph{ Any $N$-partite quantum state $\rho \in \mathbb{C}^{d_1}\otimes\mathbb{C}^{d_2}\otimes\cdots\otimes\mathbb{C}^{d_N}$ has the following expression as
	$$	\rho=\sum_{i}p_{i}|\phi_{i}^1\rangle\langle\phi_{i}^1|\otimes|\phi_{i}^2\rangle\langle\phi_{i}^2|\otimes\cdots\otimes|\phi_{i}^N\rangle\langle\phi_{i}^N|,  $$
	where $ p_{i}\in\mathbb{R}$,  $\sum_{i}p_{i}=1,$ and $|\phi_{i}^j\rangle (j=1,2,\cdots,n)$ denote the pure states in the $jth$ subsystem.
}\\
\emph{Proof.} We can prove this lemma through the Bloch representation and spectral decomposition of quantum states.
According to the Bloch representation of quantum states \cite{Vicente,Byrd}, an $N$-partite quantum state $\rho \in \mathbb{C}^{d_1}\otimes\mathbb{C}^{d_2}\otimes\cdots\otimes\mathbb{C}^{d_N}$ can be expressed as follows:
\begin{align*} 	
	\rho&=\frac{1}{d_1d_2\cdots d_N} I_{d_1}\otimes I_{d_2}\otimes\cdots\otimes I_{d_N}\\&+\frac{1}{2d_2d_3\cdots d_N}\sum_{i_1=1}^{d_1^2-1}t_{i_1}^{1}\lambda_{i_1}^{(1)}\otimes I_{d_2}\otimes\cdots\otimes I_{d_N} \\&+\frac{1}{2d_1d_3\cdots d_N}\sum_{i_2=1}^{d_2^2-1}t_{i_2}^{2}I_{d_1}\otimes \lambda_{i_2}^{(2)}\otimes\cdots\otimes I_{d_N}+\cdots\\&+\frac{1}{2^N}\sum_{i_1=1}^{d_1^2-1}\sum_{i_2=1}^{d_2^2-1}\cdots\sum_{i_N=1}^{d_N^2-1}t_{i_1i_2\cdots i_N}^{12\cdots N}\lambda_{i_1}^{(1)}\otimes\cdots\otimes \lambda_{i_N}^{(N)},
\end{align*}
where $I_{d_j} (j=1,2,\cdots,N)$ denote the $d \times d$ identity matrix and $\left\lbrace\lambda_{i_j}^{(j)}\right\rbrace (j=1,2,\cdots,N)$  denote the 
generators of the special unitary Lie group  $SU(d_1),SU(d_2),\cdots,SU(d_N)$, respectively, satisfying the orthogonality relation 
\begin{align*}
	&{\rm tr}(\lambda_{i_1}^{(1)} \lambda_{j_1}^{(1)})=2\delta_{i_1j_1},\\
	& {\rm tr}(\lambda_{i_2}^{(2)} \lambda_{j_2}^{(2)})=2\delta_{i_2j_2},\cdots, \\
	&{\rm tr}(\lambda_{i_N}^{(N)} \lambda_{j_N}^{(N)})=2\delta_{i_Nj_N},
\end{align*}
and
\begin{align*}
	&t_{i_1}^{1}={\rm tr}(\rho\lambda_{i_1}^{(1)}\otimes I_{d_2}\otimes\cdots\otimes I_{d_N}),\\  &t_{i_2}^2={\rm tr}(\rho I_{d_1}\otimes\lambda_{i_2}^{(2)}\otimes\cdots\otimes I_{d_N}),\notag\\
	&t_{i_1i_2\cdots i_N}^{12\cdots N}={\rm tr}(\rho \lambda_{i_1}^{(1)}\otimes\lambda_{i_2}^{(2)}\otimes\cdots\otimes \lambda_{i_N}^{(N)}).
\end{align*}
Performing spectral decomposition on the generators $I_{d_1}, I_{d_2}, \cdots, I_{d_N},\lambda_{i_1}^{(1)},\lambda_{i_2}^{(2)},\cdots,\lambda_{i_N}^{(N)}$ allows us to conclude the result by recombining the coefficients.
$\hfill \Box$\par 

Based on these facts, we can turn to proving our main result. \\\par 
{\subsection{B. The proof of Theorems }}
	
In order to give expressions for arbitrary traces of reduced density matrix powers, we consider an equivalent representation of the evolution of quantum circuits based on the kraus operator. This approach allows us to characterize the results to quantum circuits with multiple copies of quantum states, facilitating the computation of ${\rm tr}\rho_{A_1}^k$ for $k=2,3,\cdots,n$ in the controlled SWAP test for $n$ copy states. \par 
To prove the theorem, next we will first consider the quantum circuits for three copy quantum states and four copy quantum states. According to the Lemmas 1-4, for quantum circuits with three copy states and four copy states, we can give a representation of ${\rm tr}\rho_{A_1}^k,(k=2,3,4)$ based on the kraus operators.\\\par

\textbf{Theorem 1. }\emph{Let $\rho\in \mathbb{C}^{d_{A_1}}\otimes\mathbb{C}^{d_{A_2}}\otimes\cdots\otimes\mathbb{C}^{d_{A_N}}$ be an arbitrary $N$-partite quantum state, and let $S_1$ and $S_2$ denote the SWAP operators acting on the $A_1$ subspace of the first two copies and the last two copies in the quantum circuit of Figure \ref{fig10}, respectively. Passing through the quantum circuit in Figure \ref{fig10}, we obtain the following expressions: 
\begin{align*}
	&{\rm tr}\rho_{A_1}^2=2\sum_{c^\textbf{z}_{1}\ is\ even } p(\textbf{z})-1=2\left( p(00)+p(01)\right) -1,\notag\\
	&{\rm tr}\rho_{A_1}^3=2\sum_{c^\textbf{z}_{1,2}\ is\ even } p(\textbf{z})-1=2\left( p(00)+p(11)\right) -1,
\end{align*}
where  $\rho_{A_1}$ is the reduced density matrix with $A_1$ subsystem.}
\\
\emph{Proof.} 
Let $\emph{\textbf{z}} = z_1z_2$ be a bitstring of the two control qubits and $z_1,z_2\in\left\lbrace 0,1\right\rbrace $. To give a probabilistic representation, we first consider 
	\begin{widetext}
\begin{align*}
	&K_\emph{\textbf{z}}^{\dagger }K_\emph{\textbf{z}}\\=&\frac{1}{2^{4}}( [I+(-1)^{z_1}S_1][I+(-1)^{z_2}S_2][I+(-1)^{z_2}S_2][I+(-1)^{z_1}S_1] )\notag\\
	=&\frac{1}{2^{4}}\left( 2^2I+2^2(-1)^{z_1}S_1+2(-1)^{z_2}S_2+2(-1)^{z_1+z_2}S_1S_2+2(-1)^{z_1+z_2}S_2S_1+2(-1)^{2z_1+z_2}S_1S_2S_1\right) \notag\\
	=&\frac{1}{2^{4}}\left( 2^2I+2^2(-1)^{z_1}S_1+2(-1)^{z_2}S_2+2(-1)^{z_1+z_2}S_1S_2+2(-1)^{z_1+z_2}S_2S_1+2(-1)^{z_2}S_1S_2S_1\right) .
\end{align*}
	\end{widetext}
Therefore, we can get the probability of obtaining a string of $\emph{\textbf{z}}$ as
	\begin{widetext} 
\begin{align}
	&p(\emph{\textbf{z}})={\rm tr}(K_\emph{\textbf{z}}^{\dagger }K_\emph{\textbf{z}}\rho\otimes\rho\otimes\rho)\notag\\
	=&\frac{1}{2^{4}}{\rm tr}[\left( 2^2I+2^2(-1)^{z_1}S_1+2(-1)^{z_2}S_2+2(-1)^{z_1+z_2}S_1S_2
	+2(-1)^{z_1+z_2}S_2S_1+2(-1)^{z_2}S_1S_2S_1\right) \rho\otimes\rho\otimes\rho]\notag\\
	=&\frac{1}{2^{4}}[2^2+2^2(-1)^{z_1}{\rm tr}(\rho_{A_1}^2)+2(-1)^{z_2}{\rm tr}(\rho_{A_1}^2)+2(-1)^{z_1+z_2}{\rm tr}(\rho_{A_1}^3)+2(-1)^{z_1+z_2}{\rm tr}(\rho_{A_1}^3)+2(-1)^{z_2}{\rm tr}(\rho_{A_1}^2)]\notag\\
	=&\frac{1}{2^{2}}[1+(-1)^{z_1}{\rm tr}(\rho_{A_1}^2)+(-1)^{z_2}{\rm tr}(\rho_{A_1}^2)+(-1)^{z_1+z_2}{\rm tr}(\rho_{A_1}^3)],
\end{align}
	\end{widetext}
where the third equation is obtained by Lemma 1 and Lemma 4.
Due to $\mathcal{S}^\emph{\textbf{z}}=\left\lbrace i\in\mathcal{S}|z_{i}=1\right\rbrace $ and $c^\emph{\textbf{z}}_{ x}=\left|\mathcal{S}^\emph{\textbf{z}}\cap x\right|$, for any  $x\in\mathcal{P}(\mathcal{S})$, from the above probability representation, we get
\begin{align*}
&2\sum_{c^\emph{\textbf{z}}_{1}\ is\ even } p(\emph{\textbf{z}})-1=2(p(00)+p(01))-1={\rm tr}(\rho_{A_{1}}^2),\notag\\
&2\sum_{c^\emph{\textbf{z}}_{1,2}\ is\ even } p(\emph{\textbf{z}})-1=2(p(00)+p(11))-1={\rm tr}(\rho_{A_{1}}^3).
\end{align*}
$\hfill\Box$\\ 

\textbf{Theorem 2.} \emph{Let $\rho$ in $ \mathbb{C}^{d_{A_1}}\otimes\mathbb{C}^{d_{A_2}}\otimes\cdots\otimes\mathbb{C}^{d_{A_N}}$ be any $N$-partite quantum state, and $S_i\ (i=1,2,3)$  denotes the SWAP operator acting on the subsystem $A_1$ in the space where the $i${\rm th} and $(i+1)${\rm th} copy states are located in Figure \ref{fig11}. Passing the quantum circuit in Figure \ref{fig11}, we can also get 
\begin{align*}
	&{\rm tr}\rho_{A_1}^2=2\sum_{c^\textbf{z}_{1}\ is\ even } p(\textbf{z})-1\\=&2[ p(000)+p(001)+p(010)+p(011)] -1,\notag\\
	&{\rm tr}\rho_{A_1}^3=2\sum_{c^\textbf{z}_{1,2}\ is\ even } p(\textbf{z})-1\\=&2[ p(000)+p(001)+p(110)+p(111)] -1,\notag\\
	&{\rm tr}\rho_{A_1}^4=2\sum_{c^\textbf{z}_{1,2,3}\ is\ even } p(\textbf{z})-1\\=&2[ p(000)+p(011)+p(101)+p(110)] -1.
\end{align*}
}
\emph{Proof.} Let $\emph{\textbf{z}} = z_1z_2z_3$ be a bitstring of the three control qubits and $z_1,z_2,z_3\in\left\lbrace 0,1\right\rbrace $. To give a probabilistic representation, we first consider \\
	\begin{widetext}
\begin{align*}
	&K_\emph{\textbf{z}}^{\dagger }K_\emph{\textbf{z}}\\=&\frac{1}{2^{6}}\left(  [I+(-1)^{z_1}S_1][I+(-1)^{z_2}S_2][I+(-1)^{z_3}S_3][I+(-1)^{z_3}S_3][I+(-1)^{z_2}S_2][I+(-1)^{z_1}S_1] \right) \notag\\
	=&\frac{1}{2^{6}}\left( 2^3I+2^3(-1)^{z_1}S_1+2^2(-1)^{z_2}S_2+2(-1)^{z_3}S_3+2^2(-1)^{z_1+z_2}S_1S_2
	+2^2(-1)^{z_1+z_2}S_2S_1\right.\\
	&\left.+ 2(-1)^{z_1+z_3}S_1S_3+2(-1)^{z_1+z_3}S_3S_1+2(-1)^{z_2+z_3}S_2S_3
	+2(-1)^{z_2+z_3}S_3S_2+2^2(-1)^{2z_1+z_2}S_1S_2S_1\right.\\
	&\left.+2(-1)^{2z_1+z_3}S_1S_3S_1+2(-1)^{2z_2+z_3}S_2S_3S_2+2(-1)^{z_1+z_2+z_3}S_1S_2S_3+2(-1)^{z_1+z_2+z_3}S_3S_2S_1\right.\\
	&\left.+2(-1)^{z_1+z_2+z_3}S_2S_3S_1+2(-1)^{z_1+z_2+z_3}S_1S_3S_2+2(-1)^{2z_1+z_2+z_3}S_1S_2S_3S_1+2(-1)^{z_1+2z_2+z_3}S_1S_2S_3S_2\right.\\
	&\left.+2(-1)^{2z_1+z_2+z_3}S_1S_3S_2S_1+2(-1)^{z_1+2z_2+z_3}S_2S_3S_2S_1+2(-1)^{2z_1+2z_2+z_3}S_1S_2S_3S_2S_1\right)\\
	=&\frac{1}{2^{6}}\left(2^3I+2^3(-1)^{z_1}S_1+2^2(-1)^{z_2}S_2+2(-1)^{z_3}S_3+2^2(-1)^{z_1+z_2}S_1S_2+2^2(-1)^{z_1+z_2}S_2S_1+2(-1)^{z_1+z_3}S_1S_3\right.\\
	&\left.+2(-1)^{z_1+z_3}S_3S_1+2(-1)^{z_2+z_3}S_2S_3+2(-1)^{z_2+z_3}S_3S_2+2^2(-1)^{z_2}S_1S_2S_1+2-1)^{z_3}S_1S_3S_1+2(-1)^{z_3}S_2S_3S_2\right.\\
	&\left.+2(-1)^{z_1+z_2+z_3}S_1S_2S_3+2(-1)^{z_1+z_2+z_3}S_3S_2S_1+2(-1)^{z_1+z_2+z_3}S_2S_3S_1+2(-1)^{z_1+z_2+z_3}S_1S_3S_2\right.\\
	&\left.+2(-1)^{z_2+z_3}S_1S_2S_3S_1+2(-1)^{z_1+z_3}S_1S_2S_3S_2+2(-1)^{z_2+z_3}S_1S_3S_2S_1+2(-1)^{z_1+z_3}S_2S_3S_2S_1+2(-1)^{z_3}S_1S_2S_3S_2S_1\right)\notag\\
\end{align*}
	\end{widetext}

And we can get the probability of obtaining a string of $\emph{\textbf{z}}$ as 
	\begin{widetext}
\begin{align}
	&p(\emph{\textbf{z}})={\rm tr}(K_\emph{\textbf{z}}^{\dagger }K_\emph{\textbf{z}}\rho\otimes\rho\otimes\rho\otimes\rho)\notag\\
	=&\frac{1}{2^{6}}{\rm tr}[(2^3I+2^3(-1)^{z_1}S_1+2^2(-1)^{z_2}S_2+2(-1)^{z_3}S_3+2^2(-1)^{z_1+z_2}(S_1S_2+S_2S_1)+2(-1)^{z_1+z_3}(S_1S_3+S_3S_1)\notag\\&+2(-1)^{z_2+z_3}(S_2S_3+S_3S_2)+2^2(-1)^{z_2}S_1S_2S_1+2(-1)^{z_3}S_1S_3S_1+2(-1)^{z_3}S_2S_3S_2\notag\\&+2(-1)^{z_1+z_2+z_3}(S_1S_2S_3+S_3S_2S_1+S_2S_3S_1+S_1S_3S_2)+2(-1)^{z_2+z_3}(S_1S_2S_3S_1+S_1S_3S_2S_1)\notag\\&+2(-1)^{z_1+z_3}(S_1S_2S_3S_2+S_2S_3S_2S_1)+2(-1)^{z_3}S_1S_2S_3S_2S_1)\rho\otimes\rho\otimes\rho\otimes\rho]\notag\\
	=&\frac{1}{2^{3}}[1+(-1)^{z_1}{\rm tr}(\rho_{A_1}^2)+(-1)^{z_2}{\rm tr}(\rho_{A_1}^2)+(-1)^{z_3}{\rm tr}(\rho_{A_1}^2)+(-1)^{z_1+z_2}{\rm tr}(\rho_{A_1}^3)+(-1)^{z_2+z_3}{\rm tr}(\rho_{A_1}^3)\notag\\&+(-1)^{z_1+z_2+z_3}{\rm tr}(\rho_{A_1}^4)]+\frac{1}{2^{6}}[2^2(-1)^{z_1+z_3}({\rm tr}(\rho_{A_1}^2))^2+2^2(-1)^{z_1+z_3}{\rm tr}(\rho_{A_1}^3)],\label{p1}
\end{align}
	\end{widetext}
where the third equation is obtained by Lemma 1, Lemma 2 and Lemma 4.
According to $\mathcal{S}^\emph{\textbf{z}}=\left\lbrace i\in\mathcal{S}|z_{i}=1\right\rbrace $ and $c^\emph{\textbf{z}}_{ x}=\left|\mathcal{S}^\emph{\textbf{z}}\cap x\right|$ for any  $x\in\mathcal{P}(\mathcal{S})$, from the above probability representation (\ref{p1}), we get
\begin{align}
&2\sum_{c^\emph{\textbf{z}}_{1}\ is\ even } p(\textbf{z})-1\notag\\=&2\left( p(000)+p(001)+p(010)+p(011)\right) -1={\rm tr}(\rho_{A_{1}}^2),\notag\\
&2\sum_{c^\emph{\textbf{z}}_{1,2}\ is\ even } p(\textbf{z})-1\notag\\=&2\left( p(000)+p(001)+p(110)+p(111)\right) -1={\rm tr}(\rho_{A_{1}}^3),\notag\\
&2\sum_{c^\emph{\textbf{z}}_{1,2,3}\ is\ even } p(\emph{\textbf{z}})-1\notag\\=&2\left( p(000)+p(011)+p(101)+p(110)\right) -1={\rm tr}(\rho_{A_{1}}^4).
\end{align}
$\hfill\Box$\\ 
\par 
And the proof of above discussion reveal that only the terms where $``c^\emph{\textbf{z}}_{1,\cdots,s}\ is\ even$'' $(s=1,2,\cdots,n-1)$ are relevant and affect the outcome, while the terms in other cases, i.e., those not satisfying the $``c^\emph{\textbf{z}}_{1,\cdots,s}\ is\ even$'' $(s=1,2,\cdots,n-1)$ condition, cancel out during the summation process. This cancellation occurs due to the parity characteristics of these terms. For instance, certain terms contain summation forms of $z_i$ that differ from those in the terms meeting the condition. When performing the summation operation, the combined value of these terms amounts to 0, thereby having no substantial contribution to the final result. Consequently, certain terms in the expressions for $K_\emph{\textbf{z}}^{\dagger}K_\emph{\textbf{z}}$ and $p(\emph{\textbf{z}})$ are irrelevant to the proof's process and result. Given that the expansion of $K_\emph{\textbf{z}}^{\dagger}K_\emph{\textbf{z}}$ can reach up to $2^{2n-2}$ terms, it is impractical and unnecessarily complex to specify an exact general expression for this expansion. For the sake of simplicity, we will focus on the essential terms, and for those that are insignificant, we will denote them with the symbol $\mathcal{S}_{waste}$, as they do not influence the proof's process and outcome. Next we consider circuit of $n$ copy quantum states via kraus operator representation. \\\par
\textbf{Theorem 3.} \emph{For any $N$-partite quantum state $\rho_{A_1A_2\cdots A_N}$, consider $n$ copy states, and passing through the controlled SWAP test with $n-1$ control qubits as shown in Figure \ref{fig1}, for any $k=2,3,\cdots,n,$  we obtain 
\begin{align*}
	{\rm tr}\rho_{A_1}^k=2\sum_{c^\textbf{z}_{1,2\cdots,k-1}\ is \ even }p(\textbf{z})-1.
\end{align*}}
\emph{Proof.} In order to consider the probabilistic representation of $n-1$ qubits, we can first consider the kraus operator
\begin{widetext}
\begin{align}
&K_\emph{\textbf{z}}^{\dagger}K_\emph{\textbf{z}}\notag\\=&\frac{1}{2^{2n-2}}( \prod_{k=1}^{n-1} [I+(-1)^{z_k}S_k])^{\dagger }(\prod_{k=1}^{n-1} [I+(-1)^{z_k}S_k]) \notag\\=& \frac{1}{2^{2n-2}}\left([I+(-1)^{z_1}S_1]\cdots[I+(-1)^{z_{n-1}}S_n][I+(-1)^{z_{n-1}}S_n]\cdots[I+(-1)^{z_1}S_1] \right) \notag\\=&  \frac{1}{2^{2n-2}}\left(2^{n-1}I+2^{n-1}(-1)^{z_1}S_1+2^{n-2}(-1)^{z_1+z_2}(S_1S_2+S_2S_1)+2^{n-3}(-1)^{\sum_{i=1}^3z_i}(S_1S_2S_3+S_2S_3S_1+S_3S_2S_1+S_1S_3S_2)\right.\notag\\& \left.+\cdots+2^2(-1)^{\sum_{i=1}^{n-2}z_i}\sum_{i_1,\cdots,i_{n-2}\in\sigma_{n-2}^*}S_{i_1}S_{i_2}\cdots S_{i_{n-1}}+2(-1)^{\sum_{i=1}^{n-1}z_i}\sum_{j_1,\cdots,j_{n-1}\in\sigma_{n-1}^*}S_{j_1}S_{j_2}\cdots S_{j_{n-1}}+\mathcal{S}_{waste}\right),  \label{1}
\end{align}
	\end{widetext} 
where the second and the third equations use the relations $S_k^{\dagger}=S_k$ and $S_k^{\dagger}S_k=I (k=1,2,\cdots,n)$  respectively \cite{Beckey}. $\sigma_{n-1}$ and $\sigma_{n}$ denote the set of permutations of $1,2,\cdots, n-2$ and $1,2,\cdots, n-1$, respectively, and $\sigma_{i}^*$ is denoted as the subset of $\sigma_{i}$ that satisfies the order of the permutations in the original equation (\ref{1}). The summation term $\sum_{i_1,i_2,\cdots,i_{n-2}\in\sigma_{n-2}^*}S_{i_1}S_{i_2}\cdots S_{i_{n-2}}$ contains $2^{n-3}$ terms and the summation term  $\sum_{j_1,j_2,\cdots,j_{n-1}\in\sigma_{n-1}^*}S_{j_1}S_{j_2}\cdots S_{j_{n-1}}$ contains $2^{n-2}$ terms, and the $\mathcal{S}_{waste}$ denotes the remaining summation term containing the product of a series of $S_i$ with the product of $(-1)^{\hat{z}}$, where $\hat{z}$ is a series of expressions for the summation of  $z_i\ (i=1,2,\cdots,n-1)$ cases and does not contain expressions of the form $\sum_{k=1}^lz_k (l=1,2,3,\cdots,n-1)$.\par  
Then we have 
	\begin{widetext}
\begin{align}
&p(\emph{\textbf{z}})={\rm tr}(K_\emph{\textbf{z}}^{\dagger}K_\emph{\textbf{z}} \underbrace{\rho\otimes\rho\otimes\cdots\otimes\rho}_{n})\\
=&\frac{1}{2^{2n-2}} 
\Bigg\{  {\rm tr}(2^{n-1}I+2^{n-1}(-1)^{z_1}S_1+2^{n-2}(-1)^{z_1+z_2}(S_1S_2+S_2S_1)+2^{n-3}(-1)^{z_1+z_2+z_3}(S_1S_2S_3+S_2S_3S_1+S_3S_2S_1+S_1S_3S_2)\notag\\&\cdots+2^2(-1)^{z_1\cdots z_{n-2}}\sum_{i_1,\cdots,i_{n-2}\in\sigma_{n-2}^*}S_{i_1}S_{i_2}\cdots S_{i_{n-2}} +2(-1)^{z_1\cdots z_{n-1}}\sum_{j_1,\cdots,j_{n-1}\in\sigma_{n-1}^*}S_{j_1}\cdots S_{j_{n-1}}
+\mathcal{S}_{waste})\rho\otimes\rho\otimes\cdots\otimes\rho\Bigg\} \notag \\ 
=&\frac{1}{2^{2n-2}}\Bigg\{2^{n-1}+2^{n-1}(-1)^{z_1}{\rm tr}\rho_{A_{1}}^2+2^{n-1}(-1)^{z_1+z_2}{\rm tr}\rho_{A_{1}}^3+2^{n-1}(-1)^{z_1+z_2+z_3}{\rm tr}\rho_{A_{1}}^4+\cdots+2^{n-1}(-1)^{z_1+\cdots +z_{n-2}}{\rm tr}\rho_{A_{1}}^{n-1}\notag\\&+2^{n-1}(-1)^{z_1+z_2+\cdots +z_{n-1}}{\rm tr}\rho_{A_{1}}^{n}+{\rm tr}(\mathcal{S}_{waste}\rho\otimes\rho\otimes\cdots\otimes\rho)\Bigg\}\notag\\ 
=&\frac{1}{2^{n-1}}\Bigg\{1+(-1)^{z_1}{\rm tr}\rho_{A_{1}}^2+(-1)^{z_1+z_2}{\rm tr}\rho_{A_{1}}^3+(-1)^{z_1+z_2+z_3}{\rm tr}\rho_{A_{1}}^4+\cdots +(-1)^{z_1+\cdots +z_{n-2}}{\rm tr}\rho_{A_{1}}^{n-1}+(-1)^{z_1+\cdots +z_{n-1}}{\rm tr}\rho_{A_{1}}^{n}\Bigg\}\notag\\&+\frac{1}{2^{2n-2}}{\rm tr}(\mathcal{S}_{waste}\rho\otimes\rho\otimes\cdots\otimes\rho), \label{pz}
\end{align}
	\end{widetext}
where  we have used Lemma 1 and Lemma 4 in the third equation.\par
Given that $z_i\in\left\lbrace 0,1\right\rbrace $, for a fixed $k$, we then consider the parity of $c^\emph{\textbf{z}}_{1,2,\cdots ,k-1}$. It is worth mentioning that when considering $c^\emph{\textbf{z}}_{1,2,\cdots,k-1}$ as an even number, in the summation $$\sum_{c^\emph{\textbf{z}}_{1,2\cdots,k-1}\ is \ even }{\rm tr}(\mathcal{S}_{waste}\rho\otimes\rho\otimes\cdots\otimes\rho),$$ each term in ${\rm tr}(\mathcal{S}_{waste}\rho\otimes\rho\otimes\cdots\otimes\rho)$  is canceled out, since each term contains a series of $z_i (i=1,2,\cdots,n-1)$ summation cases other than those of the form $\sum_{s=1}^{k-1}z_s $ with the same parity. \par 
For the case $k = 2$, we can get  
\begin{align*}
&2\sum_{c^\emph{\textbf{z}}_{1}\ is \ even }p(\emph{\textbf{z}})-1\\=&\frac{2}{2^{n-1}}(2^{n-2}(1+{\rm tr}\rho_{A_{1}}^2))-1={\rm tr}\rho_{A_{1}}^2,
\end{align*}
where the summation term contains $2^{n-2}$ terms, and these summation terms $(-1)^{\sum_{l=1}^s{z_l}}{\rm tr}\rho_{A_1}^{l+1}$, $\sum_{l=1}^s{z_l} \ (s=2,3,\cdots,n) $ have the same parity. So the value of them is 0.
And when $k=n$, we can obtain
\begin{align*}
&2\sum_{c^\emph{\textbf{z}}_{1,2,\cdots,n-1}\ is \ even }p(\emph{\textbf{z}})-1\\=&\frac{2}{2^{n-1}}(2^{n-2}(1+{\rm tr}\rho_{A_{1}}^{n}))-1={\rm tr}\rho_{A_{1}}^{n},
\end{align*}
where the summation term contains $2^{n-2}$ terms, and other summation terms have the same parity. So the value of them is 0.
By a similar discussion, we can obtain that for any $k=2,3,\cdots,n,$ 
\begin{align*}
{\rm tr}\rho_{A_1}^k=2\sum_{c^\emph{\textbf{z}}_{1,2\cdots,k-1}\ is \ even }p(\emph{\textbf{z}})-1,
\end{align*}
which completes the proof of the theorem. $\hfill\Box$\\ \par
Actually the above discussion is equivalent to evolution of quantum circuit. In the main text we mention that the kraus operator representation gives estimates of the traces of the reduced density matrix powers for mixed states. Subsequently, we present the evolution process of the mixed state quantum circuit, and prove that the result is consistent with that based on the kraus operator representation. And the following results reproduce the above results based on the kraus operator representation.
\\
\par
\textbf{Theorem 4.} \emph{Let $\rho$ in $ \mathbb{C}^{d_{A_1}}\otimes\mathbb{C}^{d_{A_2}}\otimes\cdots\otimes\mathbb{C}^{d_{A_N}}$ be any quantum state, and let the initial state be $|0\rangle\langle 0|\otimes|0\rangle\langle 0|\otimes\rho\otimes\rho\otimes\rho$.  After the quantum circuit in Figure \ref{fig10} is applied, we have 
\begin{align*}
&	{\rm tr}\rho_{A_1}^2=2\left( p(00)+p(01)\right) -1,\\
&	{\rm tr}\rho_{A_1}^3=2\left( p(00)+p(11)\right) -1.
\end{align*}
}\\
\emph{Proof.} According to Lemma 5, any quantum state $\rho$ can be expressed as $$
\rho=\sum_{i}q_{i}|\phi_{i}^1\rangle\langle\phi_{i}^1|\otimes|\phi_{i}^2\rangle\langle\phi_{i}^2|\otimes\cdots\otimes|\phi_{i}^N\rangle\langle\phi_{i}^N|,
$$
where $ q_{i}\in\mathbb{R}, \sum_iq_i=1$. The initial state of the circuit is then
\begin{align*}
&|00\rangle\langle00|\otimes\sum_{i}q_i|\phi_{i}^1\rangle|\phi_{i}^2\rangle\cdots|\phi_{i}^N\rangle\langle\phi_{i}^1|\langle\phi_{i}^2|\cdots\langle\phi_{i}^N|\\&\otimes\sum_{j}q_j |\phi_{j}^1\rangle|\phi_{j}^2\rangle\cdots|\phi_{j}^N\rangle\langle\phi_{j}^1|\langle\phi_{j}^2|\cdots\langle\phi_{j}^N|\notag\\
&\otimes\sum_{k}q_k|\phi_{k}^1\rangle|\phi_{k}^2\rangle\cdots|\phi_{k}^N\rangle\langle\phi_{k}^1|\langle\phi_{k}^2|\cdots\langle\phi_{k}^N|\notag\\
=&\sum_{i,j,k}q_iq_jq_k|00\rangle|\phi_{i}^1\rangle\cdots|\phi_{i}^N\rangle|\phi_{j}^1\rangle\cdots|\phi_{j}^N\rangle|\phi_{k}^1\rangle\cdots|\phi_{k}^N\rangle\\& \langle00|\langle\phi_{i}^1|\cdots\langle\phi_{i}^N|\langle\phi_{j}^1|\cdots\langle\phi_{j}^N|\langle\phi_{k}^1|\cdots\langle\phi_{k}^N|.
\end{align*}
For simplicity, we first consider the initial input state 
\begin{equation*}
|00\rangle|\phi_{i}^1\rangle\cdots|\phi_{i}^N\rangle|\phi_{j}^1\rangle\cdots|\phi_{j}^N\rangle|\phi_{k}^1\rangle\cdots|\phi_{k}^N\rangle.
\end{equation*}\par
Passing through the circuit in Figure \ref{fig10} yields the final state 
\begin{widetext}
\begin{align*}
&|\Psi_{ijk}\rangle\\=&\frac{1}{4}\left( |00\rangle
(|\phi_{i}^1\rangle\cdots|\phi_{i}^N\rangle|\phi_{j}^1\rangle\cdots|\phi_{j}^N\rangle|\phi_{k}^1\rangle\cdots|\phi_{k}^N\rangle 
+|\phi_{i}^1\rangle \cdots
|\phi_{i}^N\rangle|\phi_{k}^1\rangle\cdots|\phi_{j}^N\rangle|\phi_{j}^1\rangle\cdots|\phi_{k}^N\rangle
\right.\\
&\left.+|\phi_{j}^1\rangle\cdots|\phi_{i}^N\rangle|\phi_{i}^1\rangle  \cdots  |\phi_{j}^N\rangle|\phi_{k}^1\rangle\cdots|\phi_{k}^N\rangle+|\phi_{j}^1\rangle\cdots|\phi_{i}^N\rangle|\phi_{k}^1\rangle  \cdots|\phi_{j}^N\rangle|\phi_{i}^1\rangle\cdots|\phi_{k}^N\rangle)
\right.\\
&\left.+|01\rangle
(|\phi_{i}^1\rangle\cdots |\phi_{i}^N\rangle|\phi_{j}^1\rangle\cdots|\phi_{j}^N\rangle|\phi_{k}^1\rangle\cdots|\phi_{k}^N\rangle-
|\phi_{i}^1\rangle\cdots|\phi_{i}^N\rangle|\phi_{k}^1\rangle\cdots|\phi_{j}^N\rangle|\phi_{j}^1\rangle\cdots|\phi_{k}^N\rangle\right.\\
&\left.+|\phi_{j}^1\rangle\cdots|\phi_{i}^N\rangle|\phi_{i}^1\rangle\cdots|\phi_{j}^N\rangle|\phi_{k}^1\rangle\cdots|\phi_{k}^N\rangle
-|\phi_{j}^1\rangle\cdots|\phi_{i}^N\rangle|\phi_{k}^1\rangle\cdots|\phi_{j}^N\rangle|\phi_{i}^1\rangle\cdots|\phi_{k}^N\rangle)\right.\\
&\left.+|10\rangle
(|\phi_{i}^1\rangle\cdots|\phi_{i}^N\rangle|\phi_{j}^1\rangle\cdots|\phi_{j}^N\rangle|\phi_{k}^1\rangle\cdots|\phi_{k}^N\rangle+
|\phi_{i}^1\rangle\cdots|\phi_{i}^N\rangle|\phi_{k}^1\rangle\cdots|\phi_{j}^N\rangle|\phi_{j}^1\rangle\cdots|\phi_{k}^N\rangle\right.\\
&\left.-|\phi_{j}^1\rangle  \cdots|\phi_{i}^N\rangle|\phi_{i}^1\rangle\cdots|\phi_{j}^N\rangle|\phi_{k}^1\rangle\cdots|\phi_{k}^N\rangle
-|\phi_{j}^1\rangle\cdots|\phi_{i}^N\rangle|\phi_{k}^1\rangle\cdots|\phi_{j}^N\rangle|\phi_{i}^1\rangle \cdots|\phi_{k}^N\rangle)\right.\\
&\left.+|11\rangle
(|\phi_{i}^1\rangle\cdots|\phi_{i}^N\rangle|\phi_{j}^1\rangle\cdots|\phi_{j}^N\rangle|\phi_{k}^1\rangle\cdots|\phi_{k}^N\rangle-
|\phi_{i}^1\rangle\cdots|\phi_{i}^N\rangle|\phi_{k}^1\rangle\cdots|\phi_{j}^N\rangle|\phi_{j}^1\rangle\cdots|\phi_{k}^N\rangle\right.\\
&\left.-|\phi_{j}^1\rangle\cdots|\phi_{i}^N\rangle|\phi_{i}^1\rangle\cdots|\phi_{j}^N\rangle|\phi_{k}^1\rangle\cdots|\phi_{k}^N\rangle
+|\phi_{j}^1\rangle\cdots|\phi_{i}^N\rangle|\phi_{k}^1\rangle\cdots|\phi_{j}^N\rangle|\phi_{i}^1\rangle\cdots|\phi_{k}^N\rangle)\right).
\end{align*}
	\end{widetext}
After measuring the control qubits, the probabilities of observing $|00\rangle,|01\rangle,|10\rangle,|11\rangle$ are
	\begin{widetext}
\begin{align*}
&p_{ijk}(00)= {\rm tr}(|0\rangle\langle0|\otimes|0\rangle\langle0|\otimes I_{d_{A_1}d_{A_2}\cdots d_{A_n}}\otimes I_{d_{B_1}d_{B_2}\cdots d_{B_n}}\otimes I_{d_{C_1}d_{C_2}\cdots d_{C_n}}\cdot	|\Psi_{ijk}\rangle\langle\Psi_{ijk}|)\\=&\frac{1}{16}\left( 4+2{\rm tr}(|\phi_{k}^1\rangle\langle\phi_{k}^1| \cdot|\phi_{j}^1\rangle\langle\phi_{j}^1|) +4{\rm tr}(|\phi_{j}^1\rangle\langle\phi_{j}^1|\cdot|\phi_{i}^1\rangle\langle\phi_{i}^1|)+2{\rm tr}(|\phi_{k}^1\rangle\langle\phi_{k}^1|\cdot|\phi_{i}^1\rangle\langle\phi_{i}^1|)+4{\rm tr}(|\phi_{j}^1\rangle\langle\phi_{j}^1|\cdot|\phi_{k}^1\rangle\langle\phi_{k}^1|\cdot|\phi_{i}^1\rangle\langle\phi_{i}^1|)\right),\notag\\
&p_{ijk}(01) ={\rm tr}(|0\rangle\langle1|\otimes|0\rangle\langle1|\otimes I_{d_{A_1}d_{A_2}\cdots d_{A_n}}\otimes I_{d_{B_1}d_{B_2}\cdots d_{B_n}}\otimes I_{d_{C_1}d_{C_2}\cdots d_{C_n}}\cdot	|\Psi_{ijk}\rangle\langle\Psi_{ijk}|)\\=&\frac{1}{16}\left( 4-2{\rm tr}(|\phi_{k}^1\rangle\langle\phi_{k}^1|\cdot|\phi_{j}^1\rangle\langle\phi_{j}^1|)+4{\rm tr}(|\phi_{j}^1\rangle\langle\phi_{j}^1|\cdot|\phi_{i}^1\rangle\langle\phi_{i}^1|)-2{\rm tr}(|\phi_{k}^1\rangle\langle\phi_{k}^1|\cdot|\phi_{i}^1\rangle\langle\phi_{i}^1|)-4{\rm tr}(|\phi_{j}^1\rangle\langle\phi_{j}^1|\cdot|\phi_{k}^1\rangle\langle\phi_{k}^1|\cdot|\phi_{i}^1\rangle\langle\phi_{i}^1|)\right),\notag\\
&p_{ijk}(10) = {\rm tr}(|1\rangle\langle0|\otimes|1\rangle\langle0|\otimes I_{d_{A_1}d_{A_2}\cdots d_{A_n}}\otimes I_{d_{B_1}d_{B_2}\cdots d_{B_n}}\otimes I_{d_{C_1}d_{C_2}\cdots d_{C_n}}\cdot	|\Psi_{ijk}\rangle\langle\Psi_{ijk}|)\\=&\frac{1}{16}\left( 4+2{\rm tr}(|\phi_{k}^1\rangle\langle\phi_{k}^1|\cdot|\phi_{j}^1\rangle\langle\phi_{j}^1|)-4{\rm tr}(|\phi_{j}^1\rangle\langle\phi_{j}^1|\cdot|\phi_{i}^1\rangle\langle\phi_{i}^1|)+2{\rm tr}(|\phi_{k}^1\rangle\langle\phi_{k}^1|\cdot|\phi_{i}^1\rangle\langle\phi_{i}^1|)-4{\rm tr}(|\phi_{j}^1\rangle\langle\phi_{j}^1|\cdot|\phi_{k}^1\rangle\langle\phi_{k}^1|\cdot|\phi_{i}^1\rangle\langle\phi_{i}^1|)\right),\notag\\
&p_{ijk}(11)={\rm tr}(|0\rangle\langle0|\otimes|0\rangle\langle0|\otimes I_{d_{A_1}d_{A_2}\cdots d_{A_n}}\otimes I_{d_{B_1}d_{B_2}\cdots d_{B_n}}\otimes I_{d_{C_1}d_{C_2}\cdots d_{C_n}}\cdot	|\Psi_{ijk}\rangle\langle\Psi_{ijk}|)\\=& \frac{1}{16}\left( 4-2{\rm tr}(|\phi_{k}^1\rangle\langle\phi_{k}^1|\cdot|\phi_{j}^1\rangle\langle\phi_{j}^1|) -4{\rm tr}(|\phi_{j}^1\rangle\langle\phi_{j}^1|\cdot|\phi_{i}^1\rangle\langle\phi_{i}^1|)-2{\rm tr}(|\phi_{k}^1\rangle\langle\phi_{k}^1|\cdot|\phi_{i}^1\rangle\langle\phi_{i}^1|)+4{\rm tr}(|\phi_{j}^1\rangle\langle\phi_{j}^1|\cdot|\phi_{k}^1\rangle\langle\phi_{k}^1|\cdot|\phi_{i}^1\rangle\langle\phi_{i}^1|)\right).
\label{pij}
\end{align*}
	\end{widetext}
Therefore, for the any state $  |0\rangle\langle 0|\otimes|0\rangle\langle 0|\otimes\rho\otimes \rho\otimes\rho$ passes through the circuit, the probabilities of measuring $|00\rangle,|01\rangle,|10\rangle,|11\rangle$ are
	\begin{widetext}
\begin{align*}
&p(00)=\sum_{i,j,k}q_{i}q_{j}q_kp_{ijk}(00)\\=&\frac{1}{16}\left(4+2\sum_{i,j,k}q_{i}q_{j}q_k{\rm tr}(|\phi_{k}^1\rangle\langle\phi_{k}^1|\cdot|\phi_{j}^1\rangle\langle\phi_{j}^1|)+4\sum_{i,j,k}q_{i}q_{j}q_k{\rm tr}(|\phi_{j}^1\rangle\langle\phi_{j}^1|\cdot|\phi_{i}^1\rangle\langle\phi_{i}^1|)+2\sum_{i,j,k}q_{i}q_{j}q_k{\rm tr}(|\phi_{k}^1\rangle\langle\phi_{k}^1|\cdot|\phi_{i}^1\rangle\langle\phi_{i}^1|)\right.\\
&\left.+4\sum_{i,j,k}q_{i}q_{j}q_k{\rm tr}(|\phi_{j}^1\rangle\langle\phi_{j}^1|\cdot|\phi_{k}^1\rangle\langle\phi_{k}^1|\cdot|\phi_{i}^1\rangle\langle\phi_{i}^1|)\right)=\frac{1}{4}(1+2{\rm tr}\rho_{A_1}^2+{\rm tr}\rho_{A_1}^3),\\
&p(01)=\sum_{i,j,k}q_{i}q_{j}q_kp_{ijk}(01)\\=&\frac{1}{16}\left( 4-2\sum_{i,j,k}q_{i}q_{j}q_k{\rm tr}(|\phi_{k}^1\rangle\langle\phi_{k}^1|\cdot|\phi_{j}^1\rangle\langle\phi_{j}^1|) +4\sum_{i,j,k}q_{i}q_{j}q_k{\rm tr}(|\phi_{j}^1\rangle\langle\phi_{j}^1|\cdot|\phi_{i}^1\rangle\langle\phi_{i}^1|)-2\sum_{i,j,k}q_{i}q_{j}q_k{\rm tr}(|\phi_{k}^1\rangle\langle\phi_{k}^1|\cdot|\phi_{i}^1\rangle\langle\phi_{i}^1|)\right.\\
&\left.-4\sum_{i,j,k}q_{i}q_{j}q_k{\rm tr}(|\phi_{j}^1\rangle\langle\phi_{j}^1|\cdot|\phi_{k}^1\rangle\langle\phi_{k}^1|\cdot|\phi_{i}^1\rangle\langle\phi_{i}^1|)\right)=\frac{1}{4}(1-{\rm tr}\rho_{A_1}^3),\notag\\   
&p(10)=\sum_{i,j,k}q_{i}q_{j}q_kp_{ijk}(10)\\=&\frac{1}{16}\left( 4+2\sum_{i,j,k}q_{i}q_{j}q_k{\rm tr}(|\phi_{k}^1\rangle\langle\phi_{k}^1|\cdot|\phi_{j}^1\rangle\langle\phi_{j}^1|) -4\sum_{i,j,k}q_{i}q_{j}q_k{\rm tr}(|\phi_{j}^1\rangle\langle\phi_{j}^1|\cdot|\phi_{i}^1\rangle\langle\phi_{i}^1|)+2\sum_{i,j,k}q_{i}q_{j}q_k{\rm tr}(|\phi_{k}^1\rangle\langle\phi_{k}^1|\cdot|\phi_{i}^1\rangle\langle\phi_{i}^1|)\right.\\
&\left.-4\sum_{i,j,k}q_{i}q_{j}q_k{\rm tr}(|\phi_{j}^1\rangle\langle\phi_{j}^1|\cdot|\phi_{k}^1\rangle\langle\phi_{k}^1|\cdot|\phi_{i}^1\rangle\langle\phi_{i}^1|)\right)=\frac{1}{4}(1-{\rm tr}\rho_{A_1}^3),\notag\\
&p(11)=\sum_{i,j,k}q_{i}q_{j}q_kp_{ijk}(00)\\&=\frac{1}{16}\left( 4-2\sum_{i,j,k}q_{i}q_{j}q_k{\rm tr}(|\phi_{k}^1\rangle\langle\phi_{k}^1|\cdot|\phi_{j}^1\rangle\langle\phi_{j}^1|) -4\sum_{i,j,k}q_{i}q_{j}q_k{\rm tr}(|\phi_{j}^1\rangle\langle\phi_{j}^1|\cdot|\phi_{i}^1\rangle\langle\phi_{i}^1|)-2\sum_{i,j,k}q_{i}q_{j}q_k{\rm tr}(|\phi_{k}^1\rangle\langle\phi_{k}^1|\cdot|\phi_{i}^1\rangle\langle\phi_{i}^1|)\right.\\
&\left.+4\sum_{i,j,k}q_{i}q_{j}q_k{\rm tr}(|\phi_{j}^1\rangle\langle\phi_{j}^1|\cdot|\phi_{k}^1\rangle\langle\phi_{k}^1|\cdot|\phi_{i}^1\rangle\langle\phi_{i}^1|)\right)=\frac{1}{4}(1-2{\rm tr}\rho_{A_1}^2+{\rm tr}\rho_{A_1}^3). 
\end{align*}
	\end{widetext}
Therefore from above equations, we can obtain 
\begin{align*}
{\rm tr}\rho_{A_1}^2=2(p(00)+p(01))-1,\\ {\rm tr}\rho_{A_1}^3=2(p(00)+p(11))-1.
\end{align*}
$\hfill\Box$\\ \par

Furthermore, using a circuit with four copies of the quantum state as shown in Figure \ref{fig11}, we give estimates of ${\rm tr}\rho^k_{A_1}$, for $k=2,3,4$.\\\par 
\textbf{Theorem 5. }\emph{ Let $\rho$ in $ \mathbb{C}^{d_{A_1}}\otimes\mathbb{C}^{d_{A_2}}\otimes\cdots\otimes\mathbb{C}^{d_{A_N}}$ be any $N$-partite quantum state, and consider the initial state $|0\rangle\langle 0|\otimes|0\rangle\langle 0|\otimes|0\rangle\langle 0|\otimes\rho\otimes\rho\otimes\rho\otimes\rho$. After passing through the quantum circuit in Figure \ref{fig11}, we can obtain 
\begin{align*}
{\rm tr}\rho_{A_1}^2=2\left( p(000)+p(001)+p(010)+p(011)\right) -1,\\
{\rm tr}\rho_{A_1}^3=2\left( p(000)+p(001)+p(110)+p(111)\right) -1.\\
{\rm tr}\rho_{A_1}^4=2\left( p(000)+p(011)+p(101)+p(110)\right) -1.
\end{align*}
}\\
\emph{Proof.}  First the initial state of the circuit can be expressed as
\begin{align*}
&|000\rangle\langle000|\otimes\sum_{i}q_i|\phi_{i}^1\rangle|\phi_{i}^2\rangle\cdots|\phi_{i}^N\rangle\langle\phi_{i}^1|\langle\phi_{i}^2|\cdots	\\&\langle\phi_{i}^N|
\otimes\sum_{j}q_j |\phi_{j}^1\rangle|\phi_{j}^2\rangle\cdots|\phi_{j}^N\rangle\langle\phi_{j}^1|\langle\phi_{j}^2|\cdots\langle\phi_{j}^N|\\&\otimes\sum_{k}q_k
|\phi_{k}^1\rangle|\phi_{k}^2\rangle
\cdots|\phi_{k}^N\rangle\langle\phi_{k}^1|\langle\phi_{k}^2|\cdots\langle\phi_{k}^N|	\\&\otimes\sum_{l}q_l|\phi_{l}^1\rangle\cdots|\phi_{l}^N\rangle\langle\phi_{l}^1|\cdots\langle\phi_{l}^N|\notag\\
=&\sum_{i,j,k}q_iq_jq_kq_l|000\rangle|\phi_{i}^1\rangle\cdots|\phi_{i}^N\rangle|\phi_{j}^1\rangle\cdots|\phi_{j}^N\rangle|\phi_{k}^1\rangle	\\&\cdots|\phi_{k}^N\rangle|\phi_{l}^1\rangle\cdots|\phi_{l}^N\rangle\langle000|\langle\phi_{i}^1|\cdots\langle\phi_{i}^N|\langle\phi_{j}^1|\\&\cdots\langle\phi_{j}^N|\langle\phi_{k}^1|\cdots\langle\phi_{k}^N|\langle\phi_{l}^1|\cdots\langle\phi_{l}^N|.
\end{align*}
Since the quantum circuit only involves operations on the $A_1$ subsystem, we can focus solely on the states of this subsystem, disregarding the others. Thus, the initial state simplifies to
\begin{equation*}
|000\rangle|\phi_{i}^1|\phi_{j}^1\rangle|\phi_{k}^1\rangle\rangle|\phi_{l}^1\rangle.
\end{equation*}\par
After the circuit in Figure \ref{fig11} is applied, the final state is
	\begin{widetext}
\begin{align*}
&|\Psi_{ijkl}\rangle\\=&\frac{1}{8}[|000\rangle
(|\phi_{i}^1\rangle|\phi_{j}^1\rangle|\phi_{k}^1\rangle|\phi_{l}^1\rangle+|\phi_{i}^1\rangle|\phi_{j}^1\rangle|\phi_{l}^1\rangle|\phi_{k}^1\rangle+|\phi_{i}^1\rangle|\phi_{k}^1\rangle|\phi_{j}^1\rangle|\phi_{l}^1\rangle+|\phi_{i}^1\rangle|\phi_{k}^1\rangle|\phi_{l}^1\rangle|\phi_{j}^1\rangle\\&+|\phi_{j}^1\rangle|\phi_{i}^1\rangle|\phi_{k}^1\rangle|\phi_{l}^1\rangle+|\phi_{j}^1\rangle|\phi_{i}^1\rangle|\phi_{l}^1\rangle|\phi_{k}^1+|\phi_{j}^1\rangle|\phi_{k}^1\rangle|\phi_{i}^1\rangle|\phi_{l}^1\rangle+|\phi_{j}^1\rangle|\phi_{k}^1\rangle|\phi_{l}^1\rangle|\phi_{i}^1\rangle)\\&+|001\rangle
(|\phi_{i}^1\rangle|\phi_{j}^1\rangle|\phi_{k}^1\rangle|\phi_{l}^1\rangle-|\phi_{i}^1\rangle|\phi_{j}^1\rangle|\phi_{l}^1\rangle|\phi_{k}^1\rangle+|\phi_{i}^1\rangle|\phi_{k}^1\rangle|\phi_{j}^1\rangle|\phi_{l}^1\rangle-|\phi_{i}^1\rangle|\phi_{k}^1\rangle|\phi_{l}^1\rangle|\phi_{j}^1\rangle\\&+|\phi_{j}^1\rangle|\phi_{i}^1\rangle|\phi_{k}^1\rangle|\phi_{l}^1\rangle-|\phi_{j}^1\rangle|\phi_{i}^1\rangle|\phi_{l}^1\rangle|\phi_{k}^1\rangle+|\phi_{j}^1\rangle|\phi_{k}^1\rangle|\phi_{i}^1\rangle|\phi_{l}^1\rangle-|\phi_{j}^1\rangle|\phi_{k}^1\rangle|\phi_{l}^1\rangle|\phi_{i}^1\rangle)\\&+|010\rangle(|\phi_{i}^1\rangle|\phi_{j}^1\rangle|\phi_{k}^1\rangle|\phi_{l}^1\rangle+|\phi_{i}^1\rangle|\phi_{j}^1\rangle|\phi_{l}^1\rangle|\phi_{k}^1\rangle-|\phi_{i}^1\rangle|\phi_{k}^1\rangle|\phi_{j}^1\rangle|\phi_{l}^1\rangle-|\phi_{i}^1\rangle|\phi_{k}^1\rangle|\phi_{l}^1\rangle|\phi_{j}^1\rangle\\&+|\phi_{j}^1\rangle|\phi_{i}^1\rangle|\phi_{k}^1\rangle|\phi_{l}^1\rangle+|\phi_{j}^1\rangle|\phi_{i}^1\rangle|\phi_{l}^1\rangle|\phi_{k}^1\rangle-|\phi_{j}^1\rangle|\phi_{k}^1\rangle|\phi_{i}^1\rangle|\phi_{l}^1\rangle-|\phi_{j}^1\rangle|\phi_{k}^1\rangle|\phi_{l}^1\rangle|\phi_{i}^1\rangle)\\&
+|011\rangle
(|\phi_{i}^1\rangle|\phi_{j}^1\rangle|\phi_{k}^1\rangle|\phi_{l}^1\rangle-|\phi_{i}^1\rangle|\phi_{j}^1\rangle|\phi_{l}^1\rangle|\phi_{k}^1\rangle-|\phi_{i}^1\rangle|\phi_{k}^1\rangle|\phi_{j}^1\rangle|\phi_{l}^1\rangle+|\phi_{i}^1\rangle|\phi_{k}^1\rangle|\phi_{l}^1\rangle|\phi_{j}^1\rangle\\&+|\phi_{j}^1\rangle|\phi_{i}^1\rangle|\phi_{k}^1\rangle|\phi_{l}^1\rangle-|\phi_{j}^1\rangle|\phi_{i}^1\rangle|\phi_{l}^1\rangle|\phi_{k}^1\rangle-|\phi_{j}^1\rangle|\phi_{k}^1\rangle|\phi_{i}^1\rangle|\phi_{l}^1\rangle+|\phi_{j}^1\rangle|\phi_{k}^1\rangle|\phi_{l}^1\rangle|\phi_{i}^1\rangle)
\\&+|100\rangle
(|\phi_{i}^1\rangle|\phi_{j}^1\rangle|\phi_{k}^1\rangle|\phi_{l}^1\rangle+|\phi_{i}^1\rangle|\phi_{j}^1\rangle|\phi_{l}^1\rangle|\phi_{k}^1\rangle+|\phi_{i}^1\rangle|\phi_{k}^1\rangle|\phi_{j}^1\rangle|\phi_{l}^1\rangle+|\phi_{i}^1\rangle|\phi_{k}^1\rangle|\phi_{l}^1\rangle|\phi_{j}^1\rangle\\&-|\phi_{j}^1\rangle|\phi_{i}^1\rangle|\phi_{k}^1\rangle|\phi_{l}^1\rangle-|\phi_{j}^1\rangle|\phi_{i}^1\rangle|\phi_{l}^1\rangle|\phi_{k}^1\rangle-|\phi_{j}^1\rangle|\phi_{k}^1\rangle|\phi_{i}^1\rangle|\phi_{l}^1\rangle-|\phi_{j}^1\rangle|\phi_{k}^1\rangle|\phi_{l}^1\rangle|\phi_{i}^1\rangle)
\\&	+|101\rangle
(|\phi_{i}^1\rangle|\phi_{j}^1\rangle|\phi_{k}^1\rangle|\phi_{l}^1\rangle-|\phi_{i}^1\rangle|\phi_{j}^1\rangle|\phi_{l}^1\rangle|\phi_{k}^1\rangle+|\phi_{i}^1\rangle|\phi_{k}^1\rangle|\phi_{j}^1\rangle|\phi_{l}^1\rangle-|\phi_{i}^1\rangle|\phi_{k}^1\rangle|\phi_{l}^1\rangle|\phi_{j}^1\rangle\\&-|\phi_{j}^1\rangle|\phi_{i}^1\rangle|\phi_{k}^1\rangle|\phi_{l}^1\rangle+|\phi_{j}^1\rangle|\phi_{i}^1\rangle|\phi_{l}^1\rangle|\phi_{k}^1\rangle-|\phi_{j}^1\rangle|\phi_{k}^1\rangle|\phi_{i}^1\rangle|\phi_{l}^1\rangle+|\phi_{j}^1\rangle|\phi_{k}^1\rangle|\phi_{l}^1\rangle|\phi_{i}^1\rangle)\\&
+|110\rangle
(|\phi_{i}^1\rangle|\phi_{j}^1\rangle|\phi_{k}^1\rangle|\phi_{l}^1\rangle+|\phi_{i}^1\rangle|\phi_{j}^1\rangle|\phi_{l}^1\rangle|\phi_{k}^1\rangle-|\phi_{i}^1\rangle|\phi_{k}^1\rangle|\phi_{j}^1\rangle|\phi_{l}^1\rangle-|\phi_{i}^1\rangle|\phi_{k}^1\rangle|\phi_{l}^1\rangle|\phi_{j}^1\rangle\\&-|\phi_{j}^1\rangle|\phi_{i}^1\rangle|\phi_{k}^1\rangle|\phi_{l}^1\rangle-|\phi_{j}^1\rangle|\phi_{i}^1\rangle|\phi_{l}^1\rangle|\phi_{k}^1\rangle+|\phi_{j}^1\rangle|\phi_{k}^1\rangle|\phi_{i}^1\rangle|\phi_{l}^1\rangle+|\phi_{j}^1\rangle|\phi_{k}^1\rangle|\phi_{l}^1\rangle|\phi_{i}^1\rangle)\\&
+|111\rangle
(|\phi_{i}^1\rangle|\phi_{j}^1\rangle|\phi_{k}^1\rangle|\phi_{l}^1\rangle-|\phi_{i}^1\rangle|\phi_{j}^1\rangle|\phi_{l}^1\rangle|\phi_{k}^1\rangle-|\phi_{i}^1\rangle|\phi_{k}^1\rangle|\phi_{j}^1\rangle|\phi_{l}^1\rangle+|\phi_{i}^1\rangle|\phi_{k}^1\rangle|\phi_{l}^1\rangle|\phi_{j}^1\rangle\\&-|\phi_{j}^1\rangle|\phi_{i}^1\rangle|\phi_{k}^1\rangle|\phi_{l}^1\rangle+|\phi_{j}^1\rangle|\phi_{i}^1\rangle|\phi_{l}^1\rangle|\phi_{k}^1\rangle+|\phi_{j}^1\rangle|\phi_{k}^1\rangle|\phi_{i}^1\rangle|\phi_{l}^1\rangle-|\phi_{j}^1\rangle|\phi_{k}^1\rangle|\phi_{l}^1\rangle|\phi_{i}^1\rangle)].
\end{align*}
	\end{widetext}
After the mixed state $  |0\rangle\langle 0|\otimes|0\rangle\langle 0|\otimes|0\rangle\langle 0|\otimes\rho\otimes \rho\otimes\rho\otimes\rho$ passes through the circuit, the probabilities of being in $|000\rangle,|001\rangle,|010\rangle,|011\rangle,|100\rangle,|101\rangle,|110\rangle,|111\rangle$ are given by
\begin{align*}
p(000)=&\frac{1}{16}\left( 2+6{\rm tr}\rho_{A_1}^2+5{\rm tr}\rho_{A_1}^3+2{\rm tr}\rho_{A_1}^4+({\rm tr}\rho_{A_1}^2)^2\right), \notag\\
p(001)=&\frac{1}{16}\left( 2+2{\rm tr}\rho_{A_1}^2-{\rm tr}\rho_{A_1}^3-2{\rm tr}\rho_{A_1}^4-({\rm tr}\rho_{A_1}^2)^2\right), \notag\\
p(010)=&\frac{1}{16}\left( 2+2{\rm tr}\rho_{A_1}^2-3{\rm tr}\rho_{A_1}^3-2{\rm tr}\rho_{A_1}^4+({\rm tr}\rho_{A_1}^2)^2\right), \notag\\
p(011)=&\frac{1}{16}\left( 2-2{\rm tr}\rho_{A_1}^2-{\rm tr}\rho_{A_1}^3+2{\rm tr}\rho_{A_1}^4-({\rm tr}\rho_{A_1}^2)^2\right), \notag\\
p(100)=&\frac{1}{16}\left( 2+2{\rm tr}\rho_{A_1}^2-{\rm tr}\rho_{A_1}^3-2{\rm tr}\rho_{A_1}^4-({\rm tr}\rho_{A_1}^2)^2\right), \notag\\
p(101)=&\frac{1}{16}\left( 2-2{\rm tr}\rho_{A_1}^2-3{\rm tr}\rho_{A_1}^3+2{\rm tr}\rho_{A_1}^4+({\rm tr}\rho_{A_1}^2)^2\right), \notag\\
p(110)=&\frac{1}{16}\left( 2-2{\rm tr}\rho_{A_1}^2-{\rm tr}\rho_{A_1}^3+2{\rm tr}\rho_{A_1}^4-({\rm tr}\rho_{A_1}^2)^2\right), \notag\\
p(111)=&\frac{1}{16}\left( 2-6{\rm tr}\rho_{A_1}^2+5{\rm tr}\rho_{A_1}^3-2{\rm tr}\rho_{A_1}^4+({\rm tr}\rho_{A_1}^2)^2\right). 
\end{align*}
From the above expressions, we can derive the following expressions:
\begin{align*}
&{\rm tr}\rho_{A_1}^2=2(p(000)+p(001)+p(010)+p(011))-1,\notag\\
& {\rm tr}\rho_{A_1}^3=2(p(000)+p(001)+p(110)+p(111))-1,\notag\\
&{\rm tr}\rho_{A_1}^4=2(p(000)+p(011)+p(101)+p(110))-1.
\end{align*}
$\hfill\Box$\\ 
\par

\textbf{Theorem 6.}\emph{ For any $N$-qubit quantum state $\rho$ with reduced density matrix $\rho_A$ on subsystem $A$, consider performing a controlled SWAP test in Figure \ref{fig1} with $n$ copies of $\rho$, yielding measurement results $\textbf{z} \in \left\lbrace 0,1\right\rbrace^{n-1}$. Let $\left\lbrace {\rm tr}\rho_A^k\right\rbrace_{k=2}^n$ denote the set of power traces to be estimated. The following statistical guarantees hold:\\ 
	(1) There exist random variables $\left\lbrace \hat{T}_k\right\rbrace_{k=2}^n$ such that with $M = O\left(\frac{1}{\epsilon^2}\log(\frac{n}{\delta})\right)$ independent experiments, the following holds simultaneously:
	\begin{align*}
		{\rm P}\left(\bigcap_{k=2}^n \left|\hat{T}_k - {\rm tr}(\rho_A^k)\right| \leq \epsilon \right) \geq 1-\delta,
	\end{align*}
	where $\epsilon>0$ is the maximum allowable error and $\delta \in (0,1)$ is the confidence level. \\
	(2) For any $ k \in \left\lbrace 2, 3, \ldots, n\right\rbrace $, the variance of the estimator $\hat{T}_k$ satisfies 
		\begin{align*}
			{\rm Var}(\hat{T}_k)=\frac{4}{M}p_k(1 - p_k),
		\end{align*}
		where $ p_k = \frac{{\rm tr}(\rho_A^k)+1}{2}$.} \\
\\
\emph{Proof.} By Theorem 3, ${\rm tr}\rho_A^k$ can be expressed as a linear combination of measurement probabilities:
\begin{align*}
	{\rm tr}\rho_{A}^k=2\sum_{c^\emph{\textbf{z}}_{1,2\cdots,k-1}\ is \ even }p(\emph{\textbf{z}})-1.
\end{align*}
Suppose we perform $M$ independent experiments where measuring the control qubits in each experiment yields a bitstring  $\emph{\textbf{z}}=z_{1} z_{2}\cdots z_{n-1}$. For each power $k\in\left\lbrace 2, 3,\dots, n\right\rbrace $, define the indicator function:
\begin{widetext}
\begin{align*}
Y_i^{(k)} = \begin{cases} 
	2 & \text{if the $i$-th measurement result $\emph{\textbf{z}}$ satisfies that $c_{1,2,\dots,k-1}^{\emph{\textbf{z}}}$ is even}, \\
	0 & \text{otherwise}.
\end{cases}
	\end{align*}
\end{widetext}
The estimator is given by:
\begin{align*}
    \hat{T}_k=\frac{1}{M}\sum_{i=1}^M Y_i^{(k)}-1.
\end{align*}
For each $k$, the expectation of $Y_i^{(k)} \in \left\lbrace 0,2\right\rbrace$ is
\begin{align*}
	\mathbb{E}[Y_i^{(k)}]=2 \sum_{c^\emph{\textbf{z}}_{1,2\cdots,k-1}\ is \ even } p(\emph{\textbf{z}})={\rm tr}\rho_A^k+1,
\end{align*}
 thus it follows that
\begin{align*}
  \mathbb{E}\left[\hat{T}_k\right] =&{\rm tr}\rho_{A}^k.
\end{align*}
Because $Y_i^{(k)}\in\left\lbrace 0,2\right\rbrace $ holds, it follows that $\hat{T}_k\in\left[ -1,1\right].$ Let $Z_i^{(k)}=\frac{Y_i^{(k)}}{2}$, then $Z_i^{(k)}\in\left[ 0,1\right]$, and 
\begin{align*}
	  \hat{T}_k=2\left(\frac{1}{M}\sum_{i=1}^{M}Z_i^{(k)}\right) -1.
\end{align*}
Define $S=\sum_{i=1}^{M}Z_i^{(k)}$, and its mean is \begin{align*}
	\mu=\mathbb{E}\left[ S\right] =\frac{M}{2}\left( {\rm tr}\rho_{A}^k+1\right).
\end{align*}

 According to Hoeffding inequality:
\begin{align*}
	{\rm P}(\left|S-\mu\right|\geq t)\le 2\exp(-\frac{2t^2}{M}),
\end{align*}
where $t=\frac{M\epsilon}{2}$, then we have 
\begin{align*}
	{\rm P}\left(\left|\hat{T}_k-{\rm tr}\rho_{A}^k\right|\ge\epsilon\right)\le  2\exp\left(-\frac{M\epsilon^2}{2}\right). 
\end{align*}
To ensure all $n-1$ estimators simultaneously satisfy the error bound, apply the union bound, we have
 \begin{align*}
 	&{\rm P}\left(\bigcup_{k=2}^n \left|\hat{T}_k - {\rm tr}\rho_A^k\right| \geq \epsilon \right) \\
 	\leq &\sum_{k=2}^n {\rm P}\left(\left|\hat{T}_k - {\rm tr}\rho_A^k\right| \geq \epsilon \right)\\
 	 \leq &2(n-1) \exp\left(-\frac{M\epsilon^2}{2}\right).
 \end{align*}
Setting $2(n-1) \exp\left(-\frac{M\epsilon^2}{2}\right) \leq \delta$ yields
\begin{align*}
	M \geq \frac{2}{\epsilon^2} \ln\left(\frac{2(n-1)}{\delta}\right).
\end{align*}
The random variable $Y_i^{(k)}\in\left\lbrace 0,2\right\rbrace $ has the following expectation
\begin{align*}
	\mathbb{E}[Y_i^{(k)}]=2 \sum_{c^\emph{\textbf{z}}_{1,2\cdots,k-1}\ is \ even } p(\emph{\textbf{z}})={\rm tr}\rho_A^k+1.
\end{align*}
Let $p_k =\frac{{\rm tr}\rho_A^k+1}{2}$, then we have $\mathbb{E}\left[Y_i^{(k)}\right]=2p_k$.
Since the value of $Y_i^{(k)}$ is either 0 or 2, its square value is
\begin{align*}
	(Y_i^{(k)})^2=
	\begin{cases} 
		4 & \text{if $Y_i^{(k)}$ is 2}, \\
		0 & \text{if $Y_i^{(k)}$ is 0}.
	\end{cases}
\end{align*}

Therefore variance of $Y_i^{(k)}$ is
\begin{align*}
	 {\rm Var}\left(Y_i^{(k)}\right)&= \mathbb{E}\left[\left(Y_i^{(k)}\right)^2\right]-\left(\mathbb{E}\left[Y_i^{(k)}\right]\right)^2 \\
	 &=4p_k-(2p_k)^2 \\
	 &=4p_k(1-p_k).
\end{align*}
Since $\hat{T}_k = \frac{1}{M} \sum_{i=1}^M Y_i^{(k)}-1$, its variance is
\begin{align*}
	 \text{Var}(\hat{T}_k)&= \text{Var}\left(\frac{1}{M}\sum_{i=1}^M Y_i^{(k)}\right)\\
	 &=\frac{1}{M^2}\sum_{i=1}^M \text{Var}\left(Y_i^{(k)}\right)\\
	 &=\frac{4p_k(1-p_k)}{M}.
\end{align*}$\hfill\Box$\\ \par


\begin{thebibliography}{99}
		
		\bibitem{Horodecki} R. Horodecki, P. Horodecki, M. Horodecki, K. Horodecki, Quantum entanglement. Reviews of modern physics \textbf{81(2)}, 865-942 (2009).
		\bibitem{Vedral} V. Vedral, Quantum entanglemen. Nature Physics \textbf{10(4)}, 256-258 (2014).
		\bibitem{Barrett} J. Barrett, Nonsequential positive-operator-valued measurements on entangled mixed states do not always violate a Bell inequality. Phys. Rev. A \textbf{65(4)}, 042302 (2002).
		\bibitem{Gisin} N. Gisin, R. Thew, Quantum communication. Nature photonics \textbf{1(3)}, 165-171 (2007).		
		\bibitem{Bose} S. Bose, Quantum communication through an unmodulated spin chain. Phys. Rev. Lett. \textbf{91(20)}, 207901 (2003).
		\bibitem{Ribordy} N. Gisin, G. Ribordy, W. Tittel, H. Zbinden, Quantum cryptography. Reviews of modern physics \textbf{74(1)}, 145 (2002).
		\bibitem{Bernstein} D. J. Bernstein, T. Lange, Post-quantum cryptography. Nature \textbf{549(7671)}, 188-194 (2017).
		\bibitem{DiVincenzo} D. P. DiVincenzo, Quantum computation. Science \textbf{270(5234)}, 255-261 (1995).
		\bibitem{Briegel} H. J. Briegel, D. E. Browne,  W. D\"{u}r, R. Raussendorf, M. Van den Nest, Measurement-based quantum computation.  Nature Physics \textbf{5(1)}, 19-26 (2009).
		\bibitem{Horodecki1} M. Horodecki, Entanglement measures. Quantum Inf. Comput. \textbf{1(1)}, 3-26 (2001).
		\bibitem{Guhne}  O. G\"{u}hne, M. Reimpell, R. F. Werner, Estimating entanglement measures in experiments. Phys. Rev. Lett. \textbf{98(11)}, 110502 (2007).
		\bibitem{Szalay} S. Szalay, Multipartite entanglement measures. Phys. Rev. A  \textbf{92(4)}, 042329 (2015).
		\bibitem{Wootters} W. K. Wootters, Entanglement of formation and concurrence. Quantum Inf. Comput. \textbf{1(1)}, 27-44 (2001).
		\bibitem{Fan} H. Fan, K. Matsumoto, H. Imai, Quantify entanglement by concurrence hierarchy. Journal of Physics A: Mathematical and General \textbf{36(14)}, 4151 (2003).
		\bibitem{Wootters1}  W. K. Wootters, Entanglement of formation of an arbitrary state of two qubits. Phys. Rev. Lett. \textbf{80(10)}, 2245 (1998).
		\bibitem{Kai} K. Chen, S. Albeverio, S. M. Fei, Entanglement of formation of bipartite quantum states. Phys. Rev. Lett. \textbf{95(21)}, 210501 (2005).
		\bibitem{Henderson} L. Henderson, V. Vedral, Information, relative entropy of entanglement, and irreversibility. Phys. Rev. Lett. \textbf{84(10)}, 2263 (2000).
		\bibitem{Moitra} U. Moitra, R. M. Soni, S. P. Trivedi, Entanglement entropy, relative entropy and duality. Journal of High Energy Physics \textbf{2019(8)}, 1-24 (2019).
		\bibitem{Cullen} A. R. Cullen, P. Kok, Calculating concentratable entanglement in graph states. Phys. Rev. A  \textbf{106(4)}, 042411 (2022).		
		\bibitem{Schatzki}  L. Schatzki, G. Liu, M. Cerezo, E. Chitambar, Hierarchy of multipartite correlations based on concentratable entanglement. Physical Review Research \textbf{6(2)}, 023019 (2024).
		\bibitem{Beckey1} J. L. Beckey,  G. Pelegr\'{i},  S. Foulds, N. J. Pearson, Multipartite entanglement measures via Bell-basis measurements. Phys. Rev. A \textbf{107(6)}, 062425 (2023).
		\bibitem{Beckey} J. L. Beckey, N. Gigena, P. J. Coles, M. Cerezo, Computable and operationally meaningful multipartite entanglement measures. Phys. Rev. Lett. \textbf{127(14)}, 140501 (2021).	
		\bibitem{Li} H. Li, T. Gao, F. Yan, Parametrized multipartite entanglement measures. Phys. Rev. A \textbf{109(1)}, 012213 (2024).
		\bibitem{Zhou} W. Zhou, Z.X. Shen, D.P. Xuan,  Z.X. Wang, S.M. Fei, Parameterized bipartite entanglement measures and entanglement constraints. Adv. Quantum Technol. 2400707 (2025).
		\bibitem{Jin} Z. X. Jin, X. Li-Jost, S. M. Fei, C. F. Qiao, Entanglement measures based on the complete information of reduced states. Phys. Rev. A \textbf{107(1)}, 012409 (2023).	
		\bibitem{Foulds} S. Foulds, V. Kendon, T. Spiller, The controlled SWAP test for determining quantum entanglement. Quantum Science and Technology \textbf{6(3)}, 035002 (2021).
		\bibitem{Zhang} R. Q. Zhang, Y.D. Qu,  S.Q. Shen, M. Li, J. Wang, The controlled SWAP test for entanglement of mixed quantum states. Europhysics Letters \textbf{146(1)}, 18001 (2024).
		\bibitem{Ekert} A. K. Ekert, C. M. Alves, D. K. Oi, M. Horodecki, P. Horodecki, L. C. Kwek, Direct estimations of linear and nonlinear functionals of a quantum state. Phys. Rev. Lett. \textbf{88(21)}, 217901 (2002).
		\bibitem{Johri} S. Johri,  D. S. Steiger, M. Troyer, Entanglement spectroscopy on a quantum computer. Phys. Rev. B \textbf{96(19)}, 195136 (2017).
		\bibitem{Suba}Y. Suba\c{s}\i, L. Cincio, P. J. Coles, Entanglement spectroscopy with a depth-two quantum circuit. J. Phys. A \textbf{52(4)}, 044001 (2019).
		\bibitem{Yirka} J. Yirka, Y. Suba\c{s}\i, Qubit-efficient entanglement spectroscopy using qubit resets. Quantum \textbf{5}, 535 (2021).
		\bibitem{Shin} M. Shin, J. Lee, S. Lee,  K. Jeong, Rank Is All You Need: Estimating the Trace of Powers of Density Matrices.  arXiv:2408.00314 (2024).
		\bibitem{Huang} H. Y. Huang, R. Kueng, J. Preskill, Predicting many properties of a quantum system from very few measurements.  Nature Physics \textbf{16(10)}, 1050-1057 (2020).
		\bibitem{YangR} R. Yang, W. Zhao, T. Chen, Scalable algorithms for calculating power functions of random quantum states in the noisy intermediate-scale quantum era. Phys. Rev. A \textbf{109(6)}, 062421 (2024).
		\bibitem{Liu} Z. Liu, Y. Tang, H. Dai, P. Liu, S. Chen, X. Ma, Detecting entanglement in quantum many-body systems via permutation moments. Phys. Rev. Lett. \textbf{129(26)}, 260501 (2022).
		\bibitem{Neven} A. Neven, J. Carrasco,  V. Vitale, C. Kokail, A. Elben, M. Dalmonte, P. Calabrese, P. Zoller, B. Vermersch, R. Kueng, B. Kraus, Symmetry-resolved entanglement detection using partial transpose moments. npj Quantum Information, \textbf{7(1)}, 152 (2021).
		\bibitem{Natalini} P. Natalini, P. E. Ricci, Newton sum rules of polynomials defined by a three-term recurrence relation. Computers $\&$ Mathematics with Applications \textbf{42(6-7)}, 767-771 (2001).
		\bibitem{Hoeffding}W. Hoeffding,  Probability inequalities for sums of bounded random variables. J. Am. Stat. Assoc. \textbf{58}, 13(1963).
		\bibitem{Rotter}  I. Rotter, Dynamics of quantum systems. Phys. Rev. E \textbf{64(3)}, 036213 (2001).
		\bibitem{Mohseni} S. Lloyd, M. Mohseni, P. Rebentrost, Quantum principal component analysis. Nature physics \textbf{10(9)}, 631-633 (2014).
		\bibitem{Boes} P. Boes, J. Eisert, R. Gallego, M. P. M\"{u}ller, H. Wilming, Von Neumann entropy from unitarity. Phys. Rev. Lett. \textbf{122(21)}, 210402 (2019).
		\bibitem{wang1} Y. Wang, G. Li, X. Wang, Variational quantum Gibbs state preparation with a truncated Taylor series. Phys. Rev. Appl. \textbf{16(5)}, 054035 (2021).
		\bibitem{Consiglio} M. Consiglio, J. Settino, A. Giordano, C. Mastroianni, F. Plastina, S. Lorenzo,  S. Maniscalco, J. Goold, T. JG Apollaro, Variational Gibbs state preparation on noisy intermediate-scale quantum devices. Phys. Rev. A \textbf{110(1)}, 012445 (2024).
		\bibitem{SanKim}J. San Kim, Generalized entanglement constraints in multi-qubit systems in terms of Tsallis entropy. Annals of Physics \textbf{373}, 197-206 (2016).
		\bibitem{Luo}Y. Luo, T. Tian, L. H. Shao, Y. Li, General monogamy of Tsallis q-entropy entanglement in multiqubit systems. Phys. Rev. A \textbf{93(6)}, 062340 (2016).
		\bibitem{Yang}X. Yang, M. X. Luo, Y. H. Yang, S. M. Fei, Parametrized entanglement monotone. Phys. Rev. A \textbf{103(5)}, 052423 (2021).
		\bibitem{Vicente}J. I. De Vicente, Separability criteria based on the bloch representation of density matrices. Quantum Inf Comput \textbf{7(7)}, 624-638 (2007).
		\bibitem{Byrd} M. S. Byrd, N. Khaneja, Characterization of the positivity of the density matrix in terms of the coherence vector representation. Phys. Rev. A \textbf{68(6)}, 062322 (2003).
		

	\end{thebibliography}
\end{document}